\documentclass[preprint,preprintnumbers,aps,superscriptaddress,nofootinbib,tightenlines,floatfix]{revtex4}
\usepackage{amsmath,amssymb}
\usepackage{graphicx}
\usepackage{bm}
\usepackage{comment}

\newcommand{\beq}{\begin{equation}}
\newcommand{\eeq}{\end{equation}}
\newcommand{\bea}{\begin{eqnarray}}
\newcommand{\eea}{\end{eqnarray}}

\def\Nprops{292,500} 
\def\Ncfgs{1194}
\def\PropsperCFG{245}
\def\TotalCost{$7\times 10^6$}

\def\mpi{M_\pi}
\def\mK{M_K}

\def\OMIT#1{{}}

\newcount\hour \newcount\hourminute \newcount\minute 
\hour=\time \divide \hour by 60
\hourminute=\hour \multiply \hourminute by 60
\minute=\time \advance \minute by -\hourminute
\newcommand{\mydate}{\ \today \ - \number\hour :\number\minute}

\begin{document}
\begin{figure}[!t]

  \vskip -1.5cm
  \leftline{\includegraphics[width=0.25\textwidth]{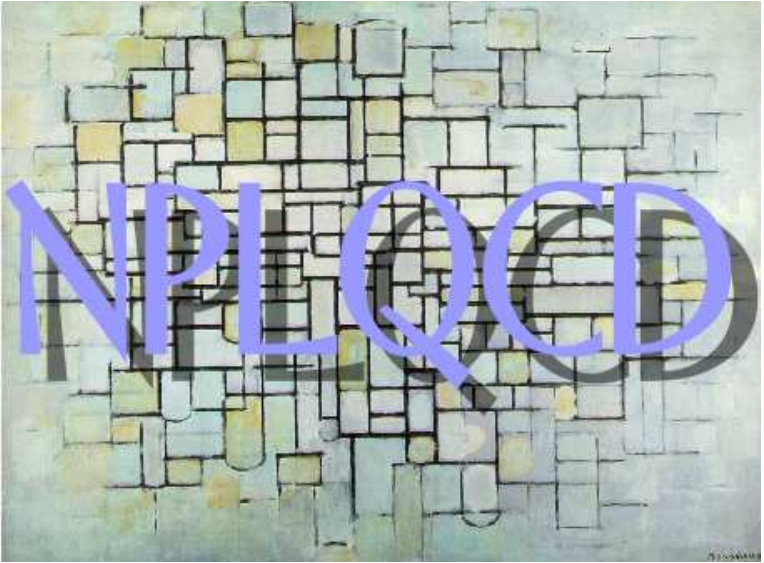}}
\end{figure}

\preprint{\vbox{ \hbox{UNH-09-01} \hbox{JLAB-THY-09-960}
    \hbox{NT@UW-09-08} \hbox{ICCUB-09-184}
    \hbox{ATHENA-PUB-09-012}}}

\vskip .5cm

\title{High Statistics Analysis using Anisotropic Clover Lattices: (I)
  Single Hadron Correlation Functions}

\vskip .5cm \author{Silas R.~Beane} \affiliation{Department of
  Physics, University of New Hampshire, Durham, NH 03824-3568.}
\author{William Detmold} \affiliation{Department of Physics, College
  of William and Mary, Williamsburg, VA 23187-8795.}
\affiliation{Jefferson Laboratory, 12000 Jefferson Avenue, Newport
  News, VA 23606.}  \author{Thomas C.~Luu} \affiliation{N Division,
  Lawrence Livermore National Laboratory, Livermore, CA 94551.}
\author{Kostas Orginos} \affiliation{Department of Physics, College of
  William and Mary, Williamsburg, VA 23187-8795.}
\affiliation{Jefferson Laboratory, 12000 Jefferson Avenue, Newport
  News, VA 23606.}  \author{Assumpta Parre\~no}
\affiliation{Departament d'Estructura i Constituents de la Mat\`{e}ria
  and Institut de Ci\`encies del Cosmos, Universitat de Barcelona,
  E--08028 Barcelona, Spain.}  \author{Martin J.~Savage}
\affiliation{Department of Physics, University of Washington, Seattle,
  WA 98195-1560.}  \author{Aaron Torok} \affiliation{Department of
  Physics, University of New Hampshire, Durham, NH 03824-3568.}
\author{Andr\'e Walker-Loud} \affiliation{Department of Physics,
  College of William and Mary, Williamsburg, VA 23187-8795.}
\collaboration{ NPLQCD Collaboration } \noaffiliation \vphantom{}

\date{\mydate}

\vskip 0.8cm
\begin{abstract}
  \noindent
  We present the results of high-statistics calculations of
  correlation functions generated with single-baryon interpolating
  operators on an ensemble of dynamical anisotropic gauge-field
  configurations generated by the Hadron Spectrum Collaboration using
  a tadpole-improved clover fermion action and Symanzik-improved gauge
  action. A total of $\Nprops$ sets of measurements are made using
  $\Ncfgs$ gauge configurations of size $20^3\times 128$ with an
  anisotropy parameter $\xi= b_s/b_t = 3.5$, a spatial lattice spacing
  of $b_s=0.1227\pm 0.0008~{\rm fm}$, and pion mass of $\mpi\sim
  390~{\rm MeV}$.  Ground state baryon masses are extracted with
  fully quantified uncertainties that are at or below the $\sim
  0.2\%$-level in lattice units.  The lowest-lying negative-parity
  states are also extracted albeit with a somewhat lower level of
  precision.  In the case of the nucleon, this negative-parity state
  is above the $N\pi$ threshold and, therefore, the isospin-${1\over
    2}$ $\pi N$ s-wave scattering phase-shift can be extracted using
  L\"uscher's method.  The disconnected contributions to this process
  are included indirectly in the gauge-field configurations and do not
  require additional calculations.  The signal-to-noise ratio in the
  various correlation functions is explored and is found to degrade
  exponentially faster than naive expectations on many
  time-slices. This is due to backward propagating states arising from
  the anti-periodic boundary conditions imposed on the
  quark-propagators in the time-direction.  We explore how best to
  distribute computational resources between configuration generation
  and propagator measurements in order to optimize the extraction of
  single baryon observables.
\end{abstract}
\maketitle

% \vfill\eject

\tableofcontents

\vfill\eject

%%%%%%%%%%%%%%%%%%%%%%%%%%%%%%%%%%%%%%%%%%%%%%%%%%%
%
% Intro
%
%%%%%%%%%%%%%%%%%%%%%%%%%%%%%%%%%%%%%%%%%%%%%%%%%%%
\section{Introduction \label{sec:Intro} }

\noindent
One of the primary goals of lattice QCD (LQCD) is to calculate the
properties and interactions of nucleons and, more generally, systems
comprised of multiple hadrons.  Precise exploration of the simplest
multi-hadron systems has recently become possible with significant
advances in computing resources, as well as through algorithmic and
theoretical developments.  The two-pion system $\pi^+\pi^+$ is the
simplest of such multi-hadron systems to calculate in LQCD, and
current computational resources have allowed for a precise
determination of the $\pi^+\pi^+$ scattering
length~\cite{Beane:2005rj,Beane:2007xs} at the $\sim 1\% $ level.
Recently, we have explored systems comprised of up to twelve
$\pi^+$'s~\cite{Beane:2007es,Detmold:2008fn} and also systems
comprised of up to twelve $K^+$'s~\cite{Detmold:2008yn} for the first
time, allowing a determination of the three-$\pi^+$ and three-$K^+$
interactions.  In general, a determination of the two-particle
scattering amplitude, or multi-body interactions, with LQCD requires
calculating the energy-eigenvalues of the system in the
finite-volume~\cite{Hamber:1983vu,Maiani:1990ca,Luscher:1986pf,Luscher:1990ux}.
The energy differences between the multi-particle energy-levels in the
finite-volume and the sum of the particle masses determines the
scattering amplitude or interaction.  As processes of interest to
low-energy nuclear physics are in the MeV energy-regime, while the
masses of the baryons and nuclei are in the GeV regime, the
energy-levels in the volume must be determined to high precision to
yield useful constraints and predictions for scattering amplitudes,
phase-shifts and electroweak properties.  Consequently, correlation
functions of systems comprised of more than one hadron must be
calculated with small statistical and systematic uncertainties ($\ll
1$~\%) in order to provide useful information about low-energy nuclear
interactions and nuclei.

The correlation functions associated with systems of baryons (and,
more generally, states other than the pion) suffer from an exponential
degradation of the signal-to-noise ratio as a function of time as
argued by Lepage~\cite{Lepage:1989hd}.  The scale that dictates this
degradation is the difference between the total energy of the baryons
in the system and half of the total energy of hadrons that contribute
to the correlation function associated with the square of the
interpolating operator for the baryon system. An example is provided
by the two point nucleon correlator ($N$ is an interpolating field
with the quantum numbers of the nucleon) where,
\begin{eqnarray}
  % \label{eq:1}
  {\rm signal} &\sim& \langle N(t)\ \overline{N}(0) \rangle\ 
  \stackrel{t\to\infty}{\longrightarrow}\  Z\, e^{-M_N t}\,, \\
  \nonumber
  {\rm noise} &\sim& \sqrt{\langle N(t)\ \overline{N}(t)N(0)\
    \overline{N}(0)\rangle}\  
  \stackrel{t\to\infty}{\longrightarrow} \ Z^\prime\,
  e^{-{3\over 2} \mpi t}\,,
\end{eqnarray}
neglecting effects of the finite temporal extent which we discuss
below (here $Z$ and $Z^\prime$ are overlap factors).  Since
$M_N-\frac{3}{2}\mpi>0$, the signal degrades exponentially in time
with this exponent. Further, for multi-baryon systems, this exponent
is scaled by the baryon number, $B$, and it consequently requires
exponentially larger computational resources to calculate the
properties of systems containing $B>1$ baryons than a single
baryon. In many regards, it is this signal-to-noise problem that
distinguishes LQCD calculations of quantities typically of importance
to nuclear physics from those typically of importance to particle
physics.

The main motivation for our present work is to explore very high
statistics calculations of the energy spectrum of $B=0,1$ correlation
functions, quantifying the statistical scaling and identifying any
issues that appear in the regime of precision calculations.  More
generally, we aim to assess the feasibility of extracting precise
phase-shifts and multi-nucleon interactions from multi-baryon systems
but we leave these discussions to subsequent work. Our focus is on the
statistical scaling behavior of these measurements instead of on
measuring physically relevant quantities. Consequently, we work with a
single ensemble of gauge configurations that was produced by the
Hadron Spectrum Collaboration~\cite{Lin:2008pr} (the details of these
configurations are discussed below).  The analysis presented here
enables us to identify a number of issues that will be important to
LQCD calculations of quantities where exponentially degrading
signal-to-noise ratios are a dominant concern:
\begin{enumerate}
    \item While the classic argument of Lepage \cite{Lepage:1989hd}
  concerning the behavior of the signal-to-noise ratios of baryon
  correlation functions seems to be on a solid theoretical footing, it
  has yet to be explored and verified through direct calculation.  We
  examine the signal-to-noise ratios of the single hadron correlation
  functions in detail and present a modified version of the Lepage
  argument that incorporates the finite extent of the temporal
  direction of the gauge-field configuration, focusing on the case of
  quark propagators subject to anti-periodic temporal boundary
  conditions (BCs). Over large regions of the temporal extent of the
  lattice, the signal-to-noise ratio degrades exponentially faster
  than expected from the original Lepage argument, see
  Sec.~\ref{sec:ston}.

    \item At present, and even more so in the past, the generation of
  gauge-field ensembles consumes most of the computational resources
  of LQCD calculations.\footnote{For example, the USQCD collaboration
    used $\sim60$\% of its resources for ensemble generation in
    2008/9.} However, it is not clear what the optimal distribution of
  computational resources between gauge-field production and
  measurements (propagator calculations and contractions) is when one
  is interested in noisy quantities.  To address this, we explore what
  can be accomplished by performing hundreds of measurements per
  configuration, and how precisely the baryon ground-state masses can
  be determined from an ensemble of $\Ncfgs$ configurations. We also
  study whether the measurements ``saturate'' after some critical
  number have been performed on one configuration (that is, exhibit
  little or no improvement in uncertainties after a critical number of
  measurements), finding for baryons that they do not, even up to
  $\sim200$ measurements per configuration.

    \item Correlation functions that are determined to high precision
  are amenable to analysis with a variety of techniques, beyond those
  typically used successfully with low statistics data.  On these
  anisotropic configurations, multiple (five or more) exponential fits
  to such correlation functions become stable as statistical
  fluctuations decrease, and the ground state energies can be
  extracted with high precision. We show that the generalized
  effective mass (EM) method, in which multiple energies are extracted
  from a linear system (a method developed by Gaspard Riche de Prony
  in 1795) also becomes useful for correlation functions with small
  uncertainties.  As two (different but correlated) correlation
  functions are computed per species of hadron, this method is
  extended to construct the matrix-Prony method, which is found to be
  a very clean and effective tool for determining the ground-state
  energies.

    \item While the correlation function generated by a single baryon
  interpolating operator will be dominated by the baryon ground-state
  at large times, it also contains contributions from all states that
  can couple to the operator.  This includes multi-hadron states.  The
  backward propagating component of the nucleon correlation function
  is dominated by the lowest energy negative-parity $I={1\over 2}$
  state for the projection-operator we have applied to the correlation
  functions.  By measuring the energy of this state, which is above
  the $\pi N$ threshold and therefore is a continuum state, the
  phase-shift associated with the $s$-wave $\pi N$ interaction is
  determined at this energy.  The important point here is that this
  process contains disconnected diagrams, which are encoded in the
  gauge-field configurations, and do not require additional (of order
  the volume in number) calculations.
  
    \item We also note that thermal states, while strongly suppressed,
  are seen in our high precision data. In these states, some part of
  the hadronic state propagates backward in time and can consequently
  manifest itself in the correlation function as an exponential with
  energy less than that of the zero temperature ground state.  These
  contributions have amplitudes that are exponentially suppressed by
  the temporal extent of the configuration, but they can be extracted
  in certain temporal regions of the correlation function(s) where
  other components are also small.  They can lead to pollution of the
  ground state signal.

\end{enumerate}

The structure of this work is as
follows. Section~\ref{sec:lattice-calculations} introduces the details
of the lattice calculations we perform, and in
section~\ref{sec:expectations} we discuss our expectations for the
hadron spectrum on this ensemble. Section~\ref{sec:methods} introduces
the tools used in our analysis and presents detailed comparisons of
the different methods we utilize.  Following this,
sections~\ref{sec:mesons}, \ref{sec:baryons} and \ref{sec:odds}
present our main results for the pseudo-scalar mesons, ground-state
baryons and negative-parity excited states, respectively.  In
sections~\ref{sec:ston} and \ref{sec:saturation} we discuss the
behaviour of noise in our measurements and investigate the scaling of
uncertainties in hadron masses for varying numbers of gauge
configurations and measurements. We conclude in section
\ref{sec:conclusions}. In subsequent works, we will address states
with baryon number, $B>1$.

%%%%%%%%%%%%%%%%%%%%%%%%
\section{Lattice QCD Calculations}
\label{sec:lattice-calculations}

\subsection{Calculational Details}
\label{sec:calc-deta}
\noindent
In this study, we employ an ensemble of the $n_f=2+1$-flavor
anisotropic clover gauge-field configurations that are currently being
produced by the Hadron Spectrum Collaboration~\cite{Lin:2008pr}.
These ensembles are being generated with dynamical tadpole-improved
clover fermions and a Symanzik-improved gauge action (see Ref
\cite{Lin:2008pr} for full details).  All of the calculations that we
present here were performed on a single ensemble of gauge-field
configurations of size $20^3\times 128$ with an anisotropy parameter
of $\xi= b_s/b_t= 3.5$, a spatial lattice spacing of $b=b_s =
0.1227\pm 0.0008~{\rm fm}$, a pion mass of $\mpi\sim 390~{\rm MeV}$
and a kaon mass of $\mK\sim 546$~MeV.  The ensemble used in this study
contains \Ncfgs\ configurations taken at intervals of 10 trajectories,
after allowing 1000 trajectories for thermalization.
Ref.~\cite{Lin:2008pr} provides a comprehensive analysis of
autocorrelation times and thermalization. Some correlation is seen to
be present between configurations separated by 30 trajectories.

The light and strange quark-propagators were computed using the same
fermion action used in the gauge-field generation. We use the clover
discretization of the fermion action as it requires significantly less
computational resources than, for instance, the Domain-Wall
discretization, in both the production of gauge-field configurations
and in the calculation of quark-propagators, while retaining an ${\cal
  O}(b)$-improved spectrum. Unlike the Domain-Wall discretization, the
Clover discretization does not have a lattice chiral symmetry.  At
moderate lattice spacings, this may significantly impact the
extraction of the properties and interactions of pions and kaons, but
it is not expected to produce systematic uncertainties that are as
significant in the properties and interactions of baryons(this remains
to be verified and will not be addressed here).

\begin{figure}
  \centering
  \includegraphics[width=0.99\columnwidth]{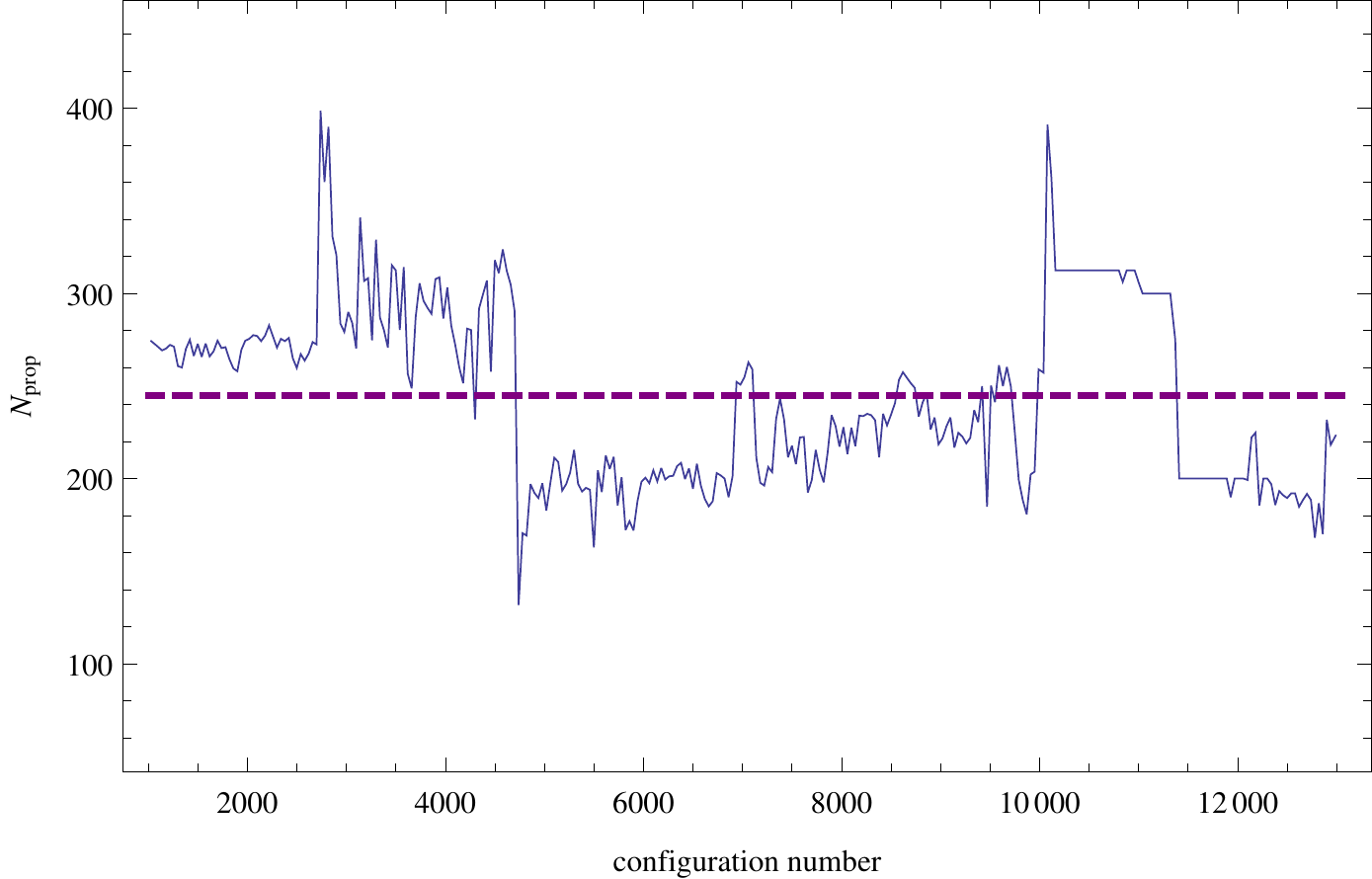}
  \caption{The number of propagators, $N_{\rm prop}$, used in
    measurements of correlation functions on each configuration used
    in this study.  For the purpose of clarity, bins of four
    configurations (40 trajectories) have been averaged.  The
    ensemble-average of the number of propagators calculated per
    configuration, $\PropsperCFG$, is indicated by the dashed
    horizontal line.}
  \label{fig:srcpercfg}
\end{figure}

The quark propagators are calculated with anti-periodic BC's imposed
on the time-direction and periodic BCs imposed on the spatial
directions.  As multiple propagators are calculated on each
configuration, iterative solvers beyond the simple conjugate gradient
algorithm can provide significant speed improvements.  In particular,
we employ the deflated conjugate gradient algorithm proposed in
Ref.~\cite{Stathopoulos:2007zi}, and implemented in the {\tt Chroma}
lattice field theory library \cite{Edwards:2004sx} as the {\tt EigCG}
inverter. In our typical production runs, we compute from 30 to 100
propagators in sequence, observing a factor of $\sim 7$ improvement in
inversion speed after the first few solves are used to deflate
low-lying eigenvalues from subsequent inversions.
Figure~\ref{fig:srcpercfg} shows the details of the propagators
computed in this work. The histogram indicates the number of
propagators computed on each of the \Ncfgs\ configurations (averaged
over four adjacent configurations for clarity) . The total number of
propagators computed in this dataset is \Nprops, an average of
\PropsperCFG\ propagators per configuration (we note that the maximum
number of point-to-all propagators that could be computed on each of
the configurations is $\sim10^6$)

\begin{figure}
  \centering
  \includegraphics[width=0.99\columnwidth]{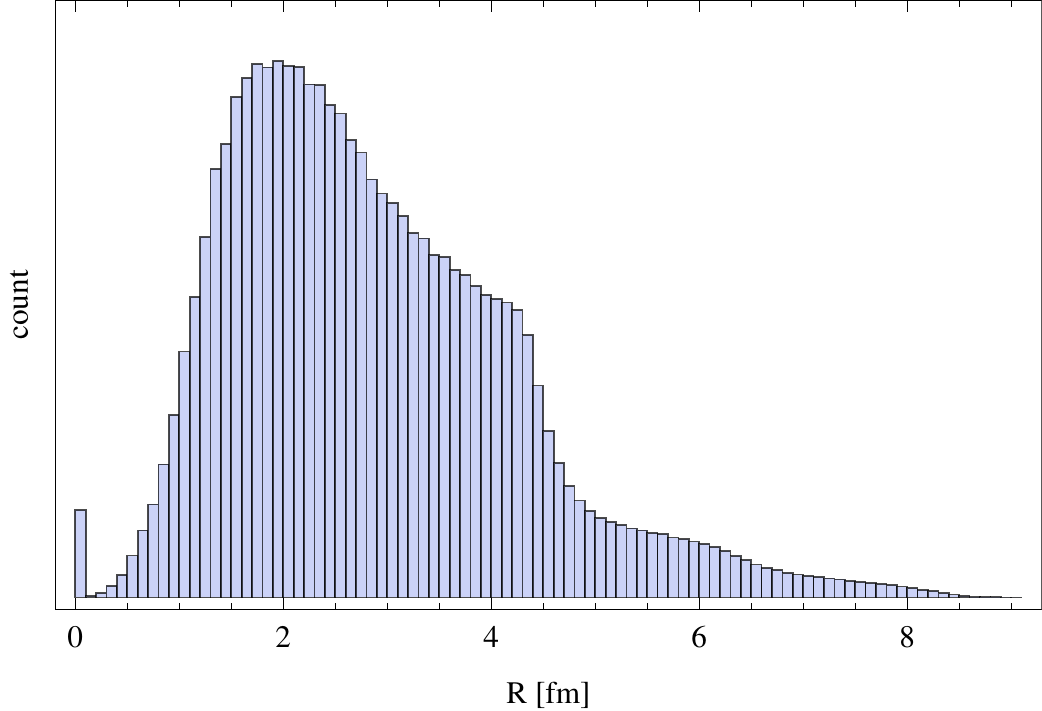}
  \caption{The separation between pairs of sources on a given
    configuration, defined to be the minimum distance between two
    sources, including the effect of the (anti-) periodic boundary
    conditions.  The height of the bar at $R=0$ corresponds to the
    total number of propagators.  }
  \label{fig:sourcesep}
\end{figure}

Each propagator is generated from a gauge-invariantly Gaussian-smeared
source~\cite{Teper:1987wt,Albanese:1987ds}, on a stout-smeared
gauge-field in order to optimize the overlap onto the ground-state
hadrons.  On each configuration, the locations of the propagator
source points are chosen randomly throughout the configuration.  In
fig.~\ref{fig:sourcesep}, we show a histogram of the 4d-separation,
$R$, between the each pair of sources on each configuration. The
shoulder at $R\sim 4$~fm appears because of the (anti) periodic
boundary conditions. The average source separation is $\langle
R\rangle\sim 2.9~{\rm fm}$ and the source density is $3.43\ {\rm
  fm}^{-4}$.

%%%%%%%%%%%%%%%%%%%%%%%%%%%%%%%%%%%%%%%%%%%%%%
\subsection{Correlation Functions}
\label{sec:corr-funct}

The propagators are used to compute two-point correlation functions
which, for baryons, take the form
\begin{equation}
  \label{eq:2}
  C_{{\cal H}; \Gamma }({\bf p};t)= \sum_{{\bf x}}e^{i{\bf p \cdot x}}\ 
  \Gamma^\alpha_\beta\ 
  \langle\  {\cal H}^\beta({\bf x},t) \overline{\cal H}_\alpha({\bf
    x_0},0)\  \rangle
  \ \ \ ,
\end{equation}
where ${\cal H}^\alpha({\bf x},t)$ is an interpolating operator for
the appropriate baryon state, {\it e.g.}, for the proton ${\cal
  H}^\alpha({\bf x},t) = \epsilon_{abc} \left(u^{a,T}\ C\ \gamma_5
  d^b\right) u^{c,\alpha}({\bf x},t)$ where $C$ is the charge
conjugation matrix. The Dirac matrix $\Gamma$ is an arbitrary
particle-spin-projector and the point ${\bf x_0}$ is the propagator
source point.  Similar correlation functions are used for the mesons.
The interpolating-operator at the source, $\overline{\cal H}$, is
constructed from gauge-invariantly-smeared quark field operators,
while at the sink, the interpolating operator is constructed from
either local quark field operators, or from the same smeared quark
field operators used at the source, leading to two sets of correlation
functions.  For brevity, we refer to the two sets of correlation
functions that result from these source and sink operators as {\it
  smeared-point} (SP) and {\it smeared-smeared} (SS) correlation
functions, respectively.

We calculate the smeared-point and smeared-smeared correlation
functions associated with the $\pi^+$, $K^+$ ($J^\pi=0^-$) mesons, and
the N, $\Lambda$, $\Sigma$, $\Xi$ ($J^\pi={1\over 2}^+$) baryons.  For
the baryons, the energy projectors $\Gamma_\pm = {1\over 2}(1\pm
\gamma_0)$ are used to project separately onto either the positive- or
negative-energy (parity) states (one of which can be time-reversed and
added to the other to improve statistics).  Correlation functions
associated with a given pair of interpolating fields are averaged over
all sources on each configuration, producing one correlation function
per interpolating operator pair per configuration.

%%%%%%%%%%%%%%%%%%%%%%%%%%%%%%%%%%%%%%
\subsection{Statistical Behavior}
\label{sec:stat-behav}

Before extracting results for observables, we analyze the statistical
behavior of the measured correlators.  As the computational cost of
each measurement is much less than the computational cost of
generating each configuration, performing multiple, ${\cal O}(10)$,
measurements on each configuration is a practical way to cheaply
reduce uncertainties and is an approach that has been used by many
groups. Averaging the measurements on a given configuration produces a
more accurate estimation of the correlation function on that
configuration. {\it A priori}, one might argue that performing a
significantly larger number, say ${\cal O}$(100--1000), of
measurements on a given configuration is an inefficient use of
computing resources as the additional measurements will contain little
or no new information and will not decrease the statistical
uncertainty in the measurements of interest.  This argument is likely
true for configurations extending over small volumes.  Physically, one
expects there are a number of length scales associated with the
possible ``saturation'' density of the measurements on a given
configuration.  As the lightest hadron is the pion, one expects the
critical saturation density of measurements to depend parametrically
upon the dimensionless quantity $\rho/\mpi^4$, where $\rho = N_{\rm
  src}/V$ where $V$ is the four-volume, and $ N_{\rm src}$ is the
number of measurements on the configuration.  For a simple quantity
such as the energy, $E$, of an eigenstate in the volume, one also
expects to find a dependence upon $\rho/E^4$.  For instance, we expect
a dependence upon the dimensionless quantity $N_{\rm src}/(V \ M_N^4)$
for the nucleon.
%
%%%%%%%%%%%%%%%%%%%%%%%%%%%%%%%%%%%%%%%%%%%%%%%%%%%
%
% FIGURE: sh-sh and sh_pt signal-to-noise
%
%%%%%%%%%%%%%%%%%%%%%%%%%%%%%%%%%%%%%%%%%%%%%%%%%%%
\begin{figure}[!ht]
  \vskip0.5in \center
  \begin{tabular}{c}
    \includegraphics[width=0.99\columnwidth]{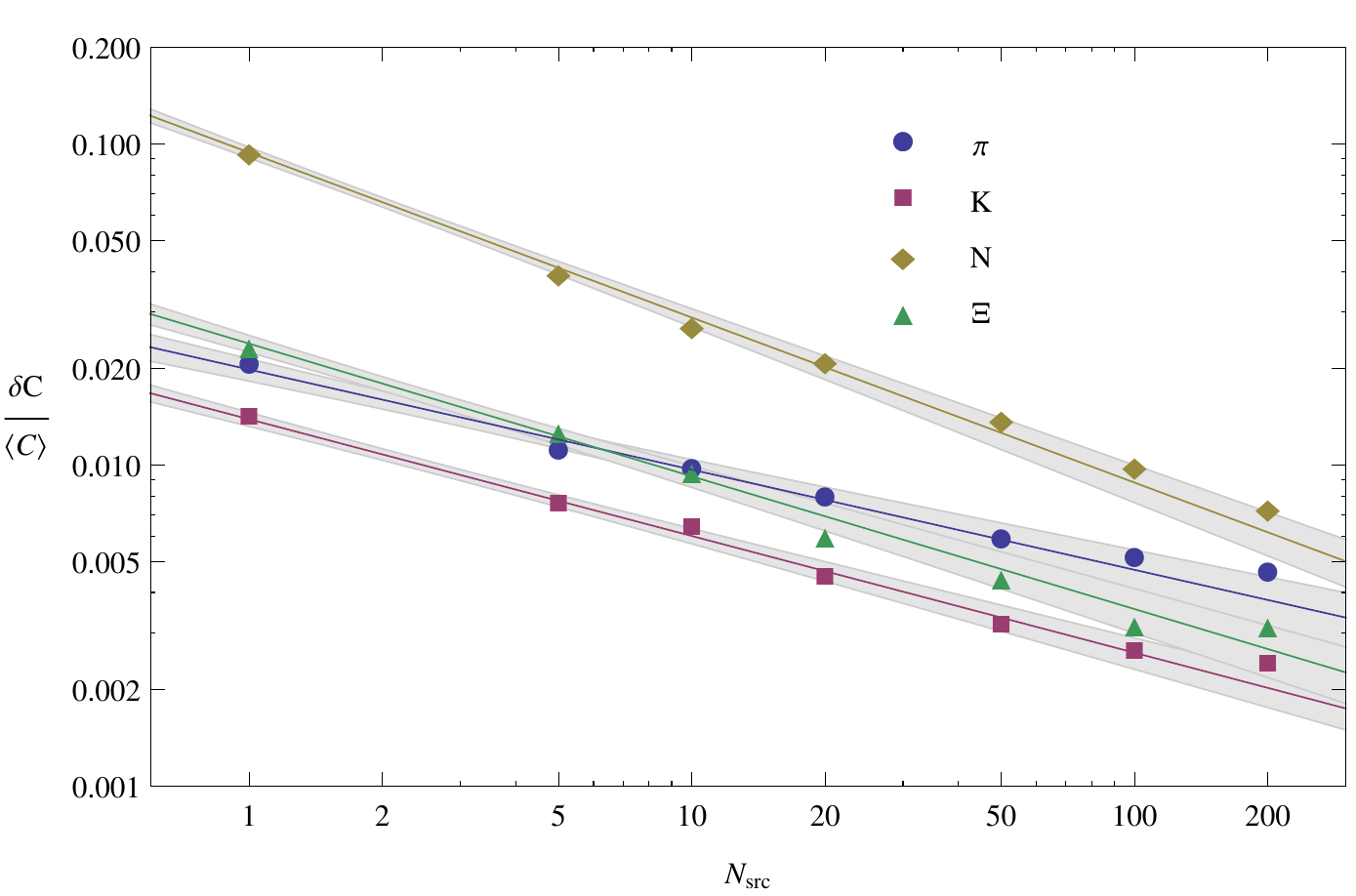}\ \ \ \ \ 
  \end{tabular}
  \caption{\label{fig:saturation} A $\log$-$\log$ plot of the
    normalized uncertainty in the mean value of the effective mass of
    the $\pi^+$, $K^+$, N, and $\Xi$ at time-slice $t=34$, $t=34$,
    $t=39$, and $t=49$, respectively, as a function of the number of
    sources on 664 of the gauge-field configurations (those with more
    than 200 measurements).  The fits correspond to a power-law of the
    form $\delta C/\langle C\rangle = A\ \left(N_{\rm src}\right)^b$.
    The best fit values for the exponent are $b= -0.31(2), -0.36(1),
    -0.51(9)$, and $-0.41(5)$ for the $\pi^+$, $K^+$, $N$, and $\Xi$,
    respectively.  (Statistically independent measurements would
    produce $b=-{1\over 2}$.)  }
\end{figure}
Figure~\ref{fig:saturation} shows the scaling of the uncertainty in
the effective mass (the logarithm of the ratio of the correlator on
adjacent time-slices) at one particular time-slice for the $\pi^+$,
$K^+$, N and $\Xi$ as a function of the number of measurements per
configuration.  This calculation was performed on 664 of the 1194
configurations in the ensemble, those for which we have made more than
200 measurements.  The correlation functions, after being averaged
over the sources on each configuration, were blocked in units of 10
configurations (100 trajectories), and the uncertainties in the
effective mass (EM) on each time-slice were generated with the single
omission Jackknife procedure.  The $\pi^+$ and $K^+$ correlation
functions clearly show deviations from statistical independence beyond
$\sim 10$ sources per configuration, and by 200 sources per
configuration there is little to be gained by performing additional
measurements on a configuration.  In contrast, measurements of the
baryon correlation functions are behaving as if they are statistically
independent even with 200 sources per configuration.  It is clear that
the statistical uncertainties in the baryon correlators can be further
reduced by performing even more measurements per configuration.  These
observations are consistent with the arguments regarding the critical
source densities.

An alternative way to investigate this question is to consider the
correlation between measurements of a correlation function from
different sources on the same configuration. A natural quantity to
consider is an extension of the standard autocorrelator to a
source-to-source autocorrelator, $\chi_{\rm src}$, defined by
\begin{equation}
  \label{eq:srctosrccorrelation}
  \chi_{\rm src}(R;t_0) = \left[\sum_{c,s}
    C(t_0;c,s)\right]^{-2}
  \left[\sum_c\sum_{s_1} \sum_{s_2} C(t_0;s_1,c) C(t_0;s_2,c)
    \theta(s_1,s_2:R)\right] -1
\end{equation}
where $C(t,c,s)$ is the correlation function of interest evaluated on
time-slice $t$, configuration $c$ and from source $s$ and the function
$\theta(s_1,s_2:R)$ is unity if the two sources are separated by a
4d-distance $|s_1-s_2|<R$. A nonzero value of $\chi_{\rm src}(R)$
indicates the presence of significant correlations over distances
shorter than $R$. We have calculated $\chi_{\rm src}$ for a number of
the correlation functions we analyze but find no sign of deviation
from zero even for the case of the $\pi^+$. This may in part be due to
the poor statistics at small source--source separations (see
Fig.~\ref{fig:sourcesep}).

\begin{figure}
  \centering
  \includegraphics[width=0.99\columnwidth]{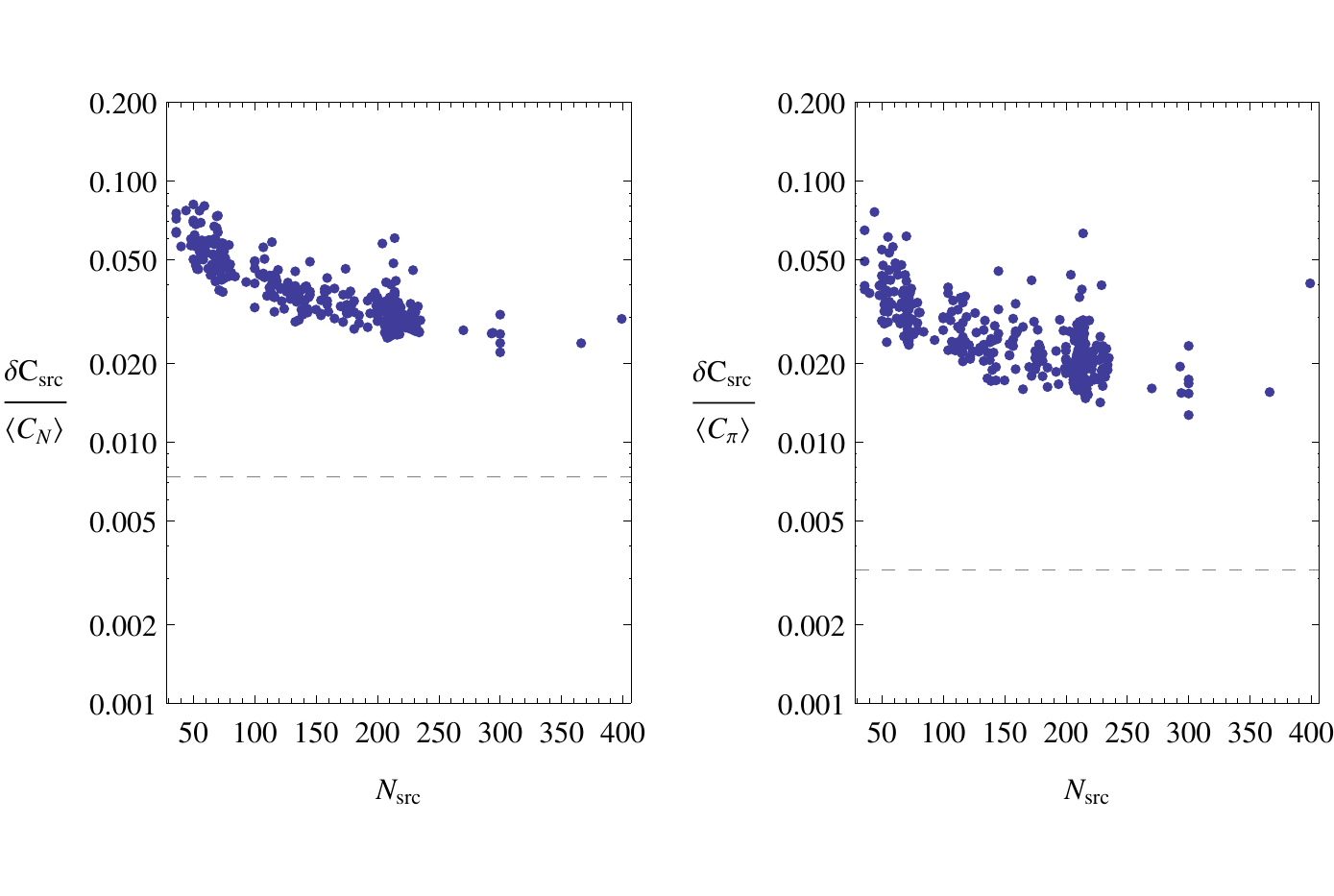}
  \caption{The normalized uncertainties in the measurements of $N$
    (left panel) and $\pi^+$ (right panel) correlation functions for
    time-slice $t=10$ are shown for some individual configurations as
    a function of the number of measurements on that
    configuration. The dashed lines significantly below the data are
    the normalized uncertainties on our full ensemble.}
  \label{fig:srcvscfg}
\end{figure}

A further consideration is that for a given number of configurations,
at some value of $N_{\rm src}$, the uncertainty in the measurements of
a correlation function on a given configuration will become smaller
than the uncertainty in the measurements over the entire
ensemble. Once this limit is reached, it is pointless to perform
further measurements without also increasing the ensemble size. Our
measurements are far from this limit as is illustrated by
fig.~\ref{fig:srcvscfg} where the uncertainties in the measurements of
$\pi^+$ and $N$ correlation functions on some individual
configurations are shown as a function of the number of sources and
compared to the overall uncertainty attained with the full ensemble.

%%%%%%%%%%%%%%%%%%%%%%%%%%%%%%%%%%%%%%%%%%%%%%%%%%%
%
% FIGURE: cfg saturation ?  signal-to-noise
%
%%%%%%%%%%%%%%%%%%%%%%%%%%%%%%%%%%%%%%%%%%%%%%%%%%%
\begin{figure}[!ht]
  \centering
  \includegraphics[width=0.99\columnwidth]{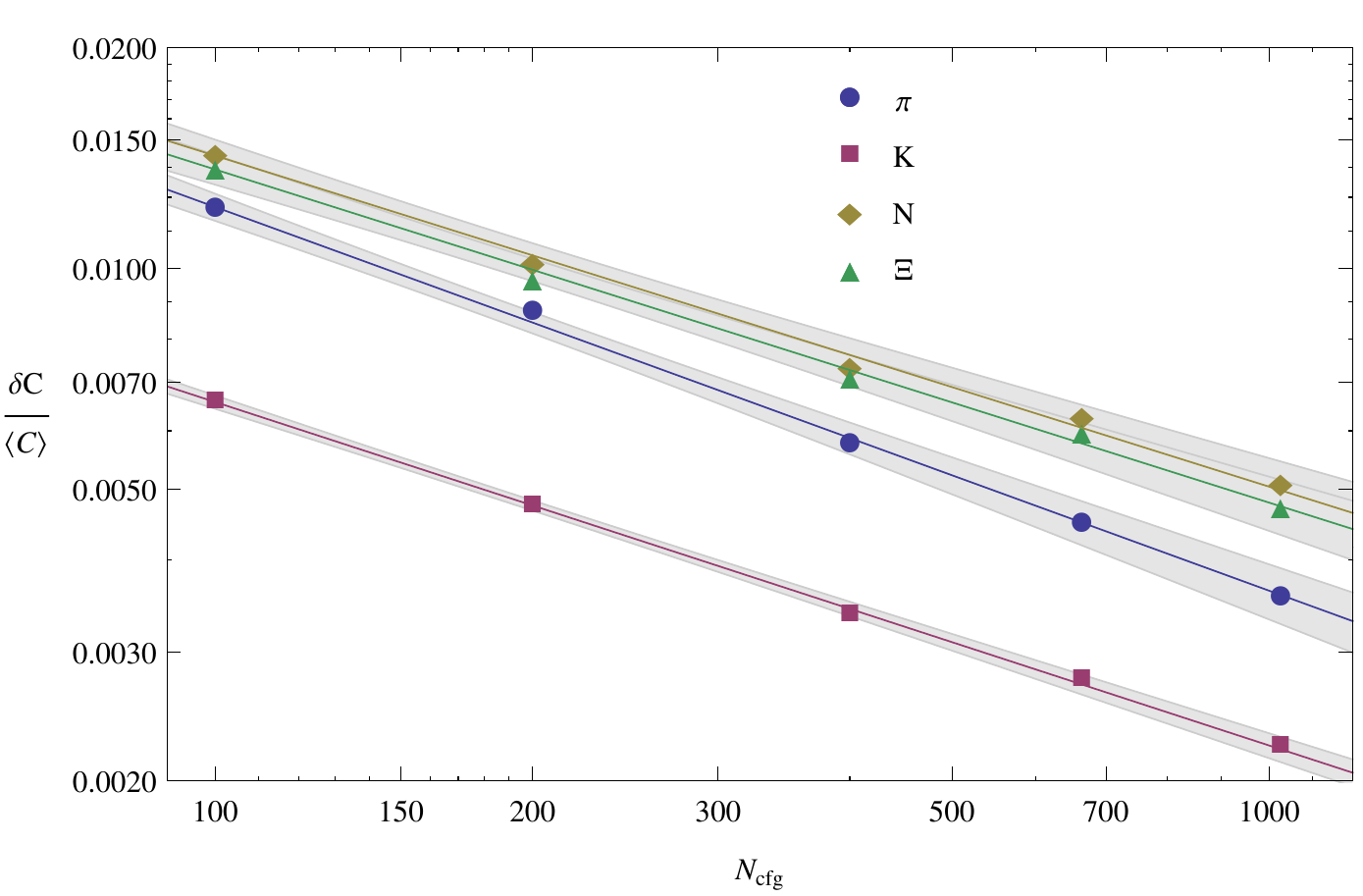}
  \caption{\label{fig:saturationlatts} A $\log$-$\log$ plot of the
    normalized uncertainty in time-slice $t=34$, $t=34$ $t=29$ and
    $t=39$ of the EM of the $\pi^+$, $K^+$, N and $\Xi$, respectively,
    as a function of the number of gauge-field configurations, each
    with 50 measurements.  The fits correspond to a power-law of the
    form $\delta C/\langle C\rangle = A\ \left(N_{\rm cfg}\right)^b$.
    The best fit values are $b= -0.52(1), -0.47(1), -0.45(2)$, and
    $-0.45(1)$ for the $\pi^+$ $K^+$, N, and $\Xi$, respectively.
    (Statistically independent measurements would produce $b=-{1\over
      2}$.) }
\end{figure}
An important consideration in generating high statistics measurements
is the correlation between configurations.  Ideally, enough
trajectories separate each gauge-field in the ensemble so that they
are statistically independent to the precision of the calculation of
interest.  The degree of correlation between configurations dictates
the number of measurements that should be performed on a given set of
configurations before it is more cost effective to enlarge the
ensemble.  In fig.~\ref{fig:saturationlatts} we show the uncertainty
at given time-slices of the EM for the $\pi^+$, $K^+$, N and $\Xi$ as
a function of the number of gauge-fields on which 50 measurements are
performed.  The configurations are maximally separated in Monte-Carlo
time, but an increasing number of configurations means a reduced
separation in Monte-Carlo time between each configuration.  The curves
in fig.~\ref{fig:saturationlatts} correspond to what is expected for
statistically independent configurations.  The 100 maximally separated
configurations are separated by 80 trajectories, the configurations
separated by 20 trajectories appear to be contributing as one expects
for statistically independent configurations (assuming that those
separated by 80 trajectories are statistically independent). This is
consistent with the hadronic auto-correlation times measured on sets
of configurations similar to this ensemble in Ref.~\cite{Lin:2008pr},
$\hat\tau_{\pi}\sim \hat\tau_{N}\sim 40$.

%%%%%%%%%%%%%%%%%%%%%%%%%%%%%%%%%%%%%%%%%%%%
\subsection{Computational Costs}
\label{sec:computational-costs}

These calculations required significant computing resources to
perform; the total cost of the measurements was approximately
\TotalCost\ JLab {\tt 6n} cluster node hours (this is an older machine
with a dual core 3 GHz Pentium D processor per node) distributed over
various computational facilities.  To put this into context, the
generation of the gauge-field configurations required approximately
one-third of this time~\cite{EdwardsPC}.  Our calculational method
makes use of hadronic building blocks (partly contracted sets of
propagators) which are extremely useful for contracting multi-particle
correlation functions but inefficient for single hadron correlation
functions; only $4\times 10^6$ JLab {\tt 6n} cluster node hours were
directly relevant to the calculations presented herein.  Nevertheless,
it seems that in order to achieve the level of precision of the $B=1$
correlation functions presented here, propagator generation rather
than gauge-field generation is the most computationally intensive
component of the LQCD calculation.  However, even this will be
superseded by the calculation of the contractions that are required
for systems involving more than two baryons (the subject of future
work).

%%%%%%%%%%%%%%%%%%%%%%%%
\section{Expected Spectra \label{sec:expectations}}
\label{sec:expected-spectra}

\noindent
The form of the correlation functions that are expected to emerge from
these calculations is a textbook discussion, but is now becoming more
relevant as advances in the field are enabling more complicated
processes to be explored, such as scattering, excited states and
multi-hadron interactions. Additionally, the accurate statistical
sampling we perform in the current work brings to light features that
have been safely neglected in the past.  A discussion of the impact of
the boundary conditions (here we use anti-periodic temporal BCs for
the quark fields) on multiple meson correlation functions that were
used in a recent calculation of $K^+K^+$ scattering can be found in
Ref.~\cite{Detmold:2008yn}, and a more detailed derivation for a two
particle system can be found in Ref.~\cite{Prelovsek:2008rf}.

For interpolating functions ${\cal O}_{A,B}$, the correlation function
that is calculated with anti-periodic BCs on the quark-fields is
\begin{eqnarray}
  G_{\cal O}(t) & = & {1\over Z} {\rm Tr}\left[ e^{-\hat H T}\ \hat {\cal O}_A^\dagger (t) \  \hat
    {\cal O}_B(0) \right] \nonumber \\
  &=& 
  {1\over Z} \sum_{j,k} e^{-E_j T}\ e^{(E_j-E_k)t}\ \langle j|\  \hat {\cal
    O}_A^\dagger(0)\  |k\rangle  \langle k|\  \hat {\cal
    O}_B(0)\  |j\rangle
  \ \ \ ,
  \label{eq:thermal}
\end{eqnarray}
where $T$ is the length of the time-direction and $Z={\rm Tr}\left[
  e^{-\hat H T}\right]$ is the partition function.\footnote{Typically,
  ${\cal O}_A$ and ${\cal O}_B$ are closely related; in our
  calculations, they differ only in the type of smearing of the quark
  fields and in the momentum injection.}

\begin{figure}
  \centering
  \includegraphics[width=0.5\columnwidth]{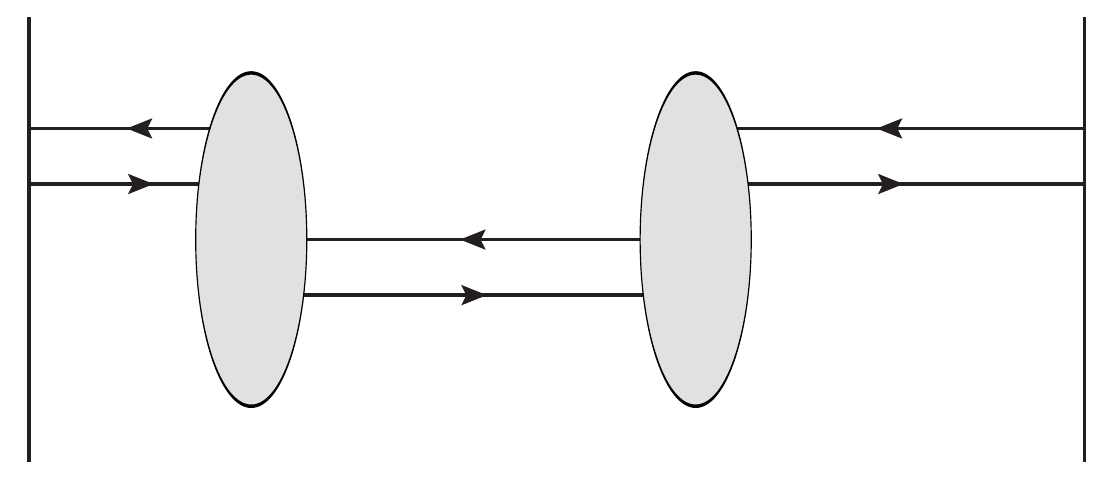}
  \caption{A depiction of the thermal contribution to $\pi\pi$
    correlation function.  The vertical lines indicate the
    anti-periodic temporal boundaries of the configuration and the
    grey regions represent the $\pi^+\pi^+$ source and sink. The solid
    lines correspond to valence quark propagators.  }
  \label{fig:thermalpipidiagram}
\end{figure}

As an example, consider the interpolating operator with baryon number
zero, strangeness zero ($S=0$), and isospin equal to two ($I=2$) that
couples to the $\pi^+\pi^+$-state.  This state can be written in terms
of hadronic field operators as $ \hat {\cal O}(0)\ = Z_{\pi^+\pi^+}\
\pi^+\pi^+\ +\ Z_{\pi^+\pi^+\pi^0\pi^0}\ {\pi^+\pi^+\pi^0\pi^0}\
+...$, where the ellipses denote all other possible hadronic field
operators with the same quantum numbers and the $Z$'s are unknown
overlap factors.  In Eq.~(\ref{eq:thermal}), this operator thus gives
non-zero values for $\langle \pi^-\pi^-|\ \hat {\cal O}(0)\
|0\rangle$, $\langle \pi^-|\ \hat {\cal O}(0)\ |\pi^+\rangle$,
$\langle 0|\ \hat {\cal O}(0)\ |\pi^+\pi^+\rangle$, plus all other
states with the same quantum numbers as the $\pi^+\pi^+$
source. Consequently the corresponding correlation function contains
exponentials $e^{-M\ t}$ with energies $M=E_{\pi^+\pi^+}$,
$M=E_{\pi^+} -E_{\pi^+}=0$, $M=-E_{\pi^+\pi^+}$,
$M=E_{\pi^+\pi^+\pi^0\pi^0}$, $M=-E_{\pi^+\pi^+\pi^0\pi^0}$,
$M=E_{\pi^+\pi^+K^+K^-}$,\ldots. In the zero temperature limit, only
those exponentials with $M\ge E_{\pi^+\pi^+}$ survive. States with
energies less than $E_{\pi^+\pi^+}$ are thermal excitations, for
instance arising from the process shown in
fig.~\ref{fig:thermalpipidiagram},
and are quite apparent in the measured $I=2$ $\pi\pi$ correlation
functions and have also been observed in hadronic systems involving a
static quark \cite{Detmold:2008ww}.

\begin{figure}
  \centering
  \includegraphics[width=0.5\columnwidth]{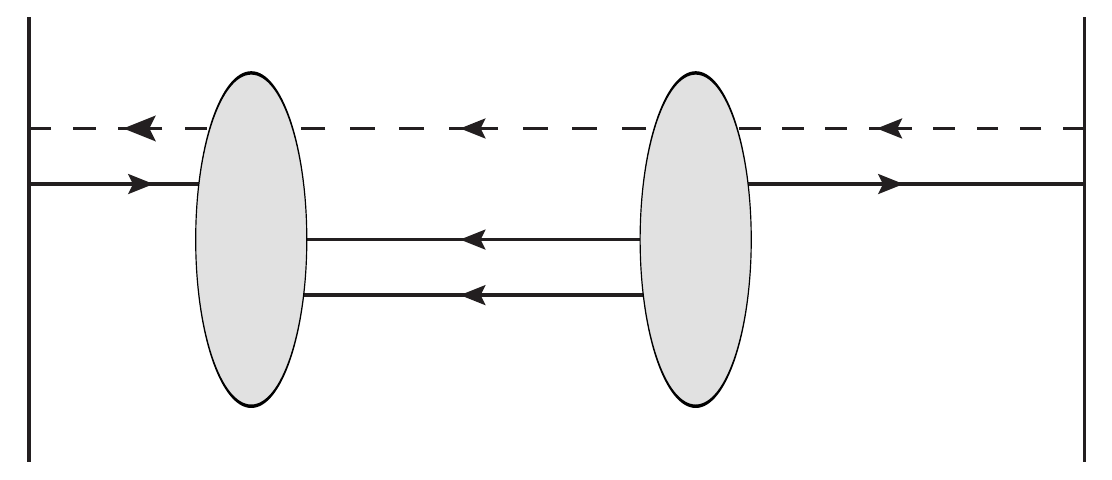}
  \caption{A depiction of the thermal $\overline{N}\pi$ system
    produced by the single-nucleon interpolating field. The vertical
    lines indicate the anti-periodic temporal boundaries of the
    configuration and the grey regions represent the single nucleon
    source and sink. The solid lines correspond to valence quark
    propagators, while the dashed lines correspond to a sea quark loop
    from the gauge-field.  }
  \label{fig:thermaldiagram}
\end{figure}

Baryon correlation functions are somewhat different, as the
interpolating operator for the single nucleon, for instance, can
couple not only to the N, but also to a state containing the N and an
even number of $\pi$'s, to a p-wave $\Lambda K$ state, to a p-wave
$\Sigma K$ state,to a p-wave $N\pi$ state, and to any other state with
the same quantum numbers as the nucleon.  Further, it can also couple
to backward propagating negative-parity states, such as an s-wave
$N\pi$.  Finally, the single nucleon interpolating operator can couple
forward and backward propagating hadronic states (these are thermal
states as they exist only because of the finite temporal extent
(temperature) of the configuration), an example being a forward
propagating N and a backward propagating $\pi$ or vice versa.  These
states are simply illustrated by an example shown in
fig.~\ref{fig:thermaldiagram}, a $N\pi$ thermal state. Here the finite
temporal extent of the configuration is indicated by the vertical
lines (these should be (anti-)identified). The two grey regions
correspond to the source and sink interpolating field. In the case
depicted, the interpolating field at the source is
$\overline{N}=(\overline{u} C\gamma_5 \overline{d^T})\overline{u}$ and
that at the sink is $N=(u^T C\gamma_5 d)u$, suppressing spin and color
indices. For the usual zero temperature ground state, the source
produces three valence quarks and the sink annihilates three valence
quarks. In the thermal state depicted, the source (right grey region)
produces two valence anti-quarks and a valence quark (solid lines)
while also producing, via gluonic interactions, a sea quark-antiquark
pair (dashed line). The three anti-quarks between the grey regions
combine to form an anti-nucleon propagating as $\exp(-M_N (T-t))$
where $t$ is the separation between the source and sink and we ignore
excited states for simplicity. The quark--anti-quark pair propagating
around the temporal boundary (since two quarks propagate, the boundary
appears periodic at the hadronic level) contribute a factor of
$\exp(-\mpi t)$ where $T$ is the temporal extent of the
configuration. The resulting contribution to the two point correlator
is then
\begin{eqnarray}
  G(t) & \sim & 
  Z_{N\overline{\pi}}\  e^{-\mpi T}\  e^{-(M_N-\mpi)t}
  \  ,
\end{eqnarray}
corresponding to a state with energy $M_N - \mpi$ in the observed
spectrum.

In the case of the correlation function resulting from a single
nucleon interpolating operator in the $A_1$ representation of the
cubic group, one expects to see a state with energy equal to the N
mass, $M_N$, and also states (a subset of all states) with energy
$M_N-2 \mpi - \delta E_{\pi\pi}$, $M_N + \delta E_{N\pi}$, and $M_N+2
\mpi + \delta E_{N\pi\pi} $, where $\delta E_{\pi\pi}$ is the
interaction energy of two $\pi$'s in an $I=0$ state, $\delta E_{N\pi}$
is the interaction energy of the $N\pi$ system, while $\delta
E_{N\pi\pi}$ is the interaction energy of the $N\pi\pi$ system.
Particularly disturbing is the state with energy $M_N + \delta
E_{N\pi}$ corresponding to $N\pi$ moving forward in time and a $\pi$
moving backwards in time, that conspire to produce a state with an
energy that differs from the nucleon mass only by the $N\pi$
interaction energy. Such states will be exponentially suppressed by
the temporal extent of the configuration, however, accurately
disentangling such states from the zero temperature ground state will
ultimately require calculations on ensembles of gauge-field
configurations with different temporal extents.

It is important to realize that thermal states are not simply a
curiosity that can be safely ignored.  As we shall see in
Section~\ref{sec:ston}, they dominate the statistical uncertainty of
baryon correlation functions at large times, providing deviations from
the naive form of the signal-to-noise ratio.  The amplitudes of these
states are exponentially suppressed by the temporal extent of the
configuration times the mass of the backward going hadronic state.
Consequently, the most important thermal states involve backward
propagating pions, and, to suppress these states, the product $\mpi T$
must be large. As the chiral limit is approached this will become more
and more difficult since, in the limit, it is impossible to separate
any particular state from itself and any number of pions.

%%%%%%%%%%%%%%%%%%%%%%%%
\section{Analysis Methods \label{sec:methods}}

%%%%%%%%%%%%%%%%%%%%%%%%%%%%%%%%%%%%%%%%%%%%%%%%%%%
\subsection{Multi-Exponential Fits}
\label{sec:unbi-expon-fits}

The high statistics accumulated for this work allows us to perform
stable multi-exponential fits using a standard $\chi^2$ minimization.
In this section, we explore the determination of the ground and
excited states as a function of several variables; the number of
exponentials used in the fit function, $N_{\rm exp}$, the range of the
fit, $R$, the number of sources per configuration, the number of
configurations, the blocking time $\tau_{\rm block}$, and the
(effective) anisotropy, $\xi_{\rm eff}$.  We present details of our
fits for the $\Xi$, using a correlated fit to the smeared-smeared and
smeared-point correlation functions.

To begin, we performed combined multi-exponential fits by minimizing
\begin{equation}
  \chi^2 = \sum_{t,t',s,s'} \left[ y_s(t) - C_s(t) \right]
  \left( Cov^{-1} \right)_{t,t'}^{s,s'} \left[ y_{s'}(t') - C_s'(t')\right]
\end{equation}
where $y_s(t)$ are the lattice measured correlation functions,
$s=[SS,SP]$, and $Cov$ is the covariance matrix between both
time-slices and correlation functions.  The fitting functions used
are,
\begin{align}
  &C_{SS}(t) = \sum_n Z^S_n Z^S_n\, e^{-E_n t}\, ,&
  % \nonumber\\
  &C_{SP}(t) = \sum_n Z^S_n Z^P_n\, e^{-E_n t}\, ,&
\end{align}
where $C_{SS}$ ($C_{SP}$) denotes the smeared-smeared (smeared-point)
correlation function.
%%%%%%%%%%%%%%%%
%% Table
\begin{table}[b]
  \caption{\label{tab:nExpFits}{Multi-exponential fits to
      smeared-smeared and smeared-point $\Xi$--correlation functions as a function
      of the number of exponentials, $N_{\rm exp}$.  The number of successful
      fits, $N_{\rm suc}$ has been defined to be fits with  $Q> 0.1$ for a fixed time-length $R$ with a fixed set of initial parameter values.  The time window listed is taken as a representative example of a good fit.  The first
      uncertainty is statistical while the second is taken from the standard
      deviation of successful fits, as defined above.}}
  \begin{ruledtabular}
    \begin{tabular}{lcccccccccc}
      &&&&$\Xi$ \\
      $b_t M$ & $b_t M^\prime$& $N_{\rm exp}$& R: $t_{\rm
        min},..,t_{\rm max}$& $N_{\rm suc}$& $\tau_{\rm block}$& $\xi_{\rm eff}$& $N_{\rm src}$ & $N_{\rm cfg}$& $\chi^2/{\rm dof}$& Q  \\
      \hline
      0.24138(33)(44)&--& 1&20: 47--67 & 16& 1& 3.5 & 245 & 1194& 0.50& 1.00\\
      0.24108(28)(10)& 0.377(6)(8)& 2&45: 24--69& 12& 1& 3.5& 245 & 1194& 0.70& 0.99\\
      0.24115(24)(07)& 0.371(8)(5)& 3&50: 14--64 & 10& 1& 3.5& 245 & 1194& 0.77& 0.95\\
      0.24115(25)(07)& 0.368(9)(4)& 4&50: 10--60& 12& 1& 3.5& 245 & 1194& 0.84& 0.86\\
      % 0.24120(20)(?)& 0.374(2)(?)& 5& 63: 1--64& 1& 1& 3.5& 245 &
      % 1194& 1.04& 0.35\\
      % & & 6&&& 1& 3.5& 245 & 1194&&\\
    \end{tabular}
  \end{ruledtabular}
\end{table}
%%%%%%%%%%%%%%%%
To perform these fits we start with a single exponential and perform
the correlated fit to $Z^S_0$, $Z^P_0$ and $E_0$.  A selected set of
best fit parameters from this fit are used as initial estimates for
the two exponential fits.  This is performed recursively by taking the
best fit results from the N exponential fit as an initial estimates to
the N+1 exponential fit.  With this strategy, successful minimizations
with up to six exponentials have been performed.  However, with the
inclusion of the fifth and higher exponentials, the minimizer performs
poorly, and often returns two masses that are degenerate within their
uncertainties.  Furthermore, as discussed in detail in the previous
Section, the expected spectrum of states on these anisotropic
configurations is such that the resulting masses for the excited
states are likely averages of nearby energy levels, see also
Section~\ref{sec:prony-histograms} below for demonstrations of this.
For these reasons, we are only confident in the ground state energies
extracted in these fits.  However, the number of exponentials used in
a successful minimization plays an important role in minimizing the
fitting systematic uncertainty.

\begin{figure}[t]
  \begin{tabular}{cc}
    \includegraphics[width=0.9\textwidth]{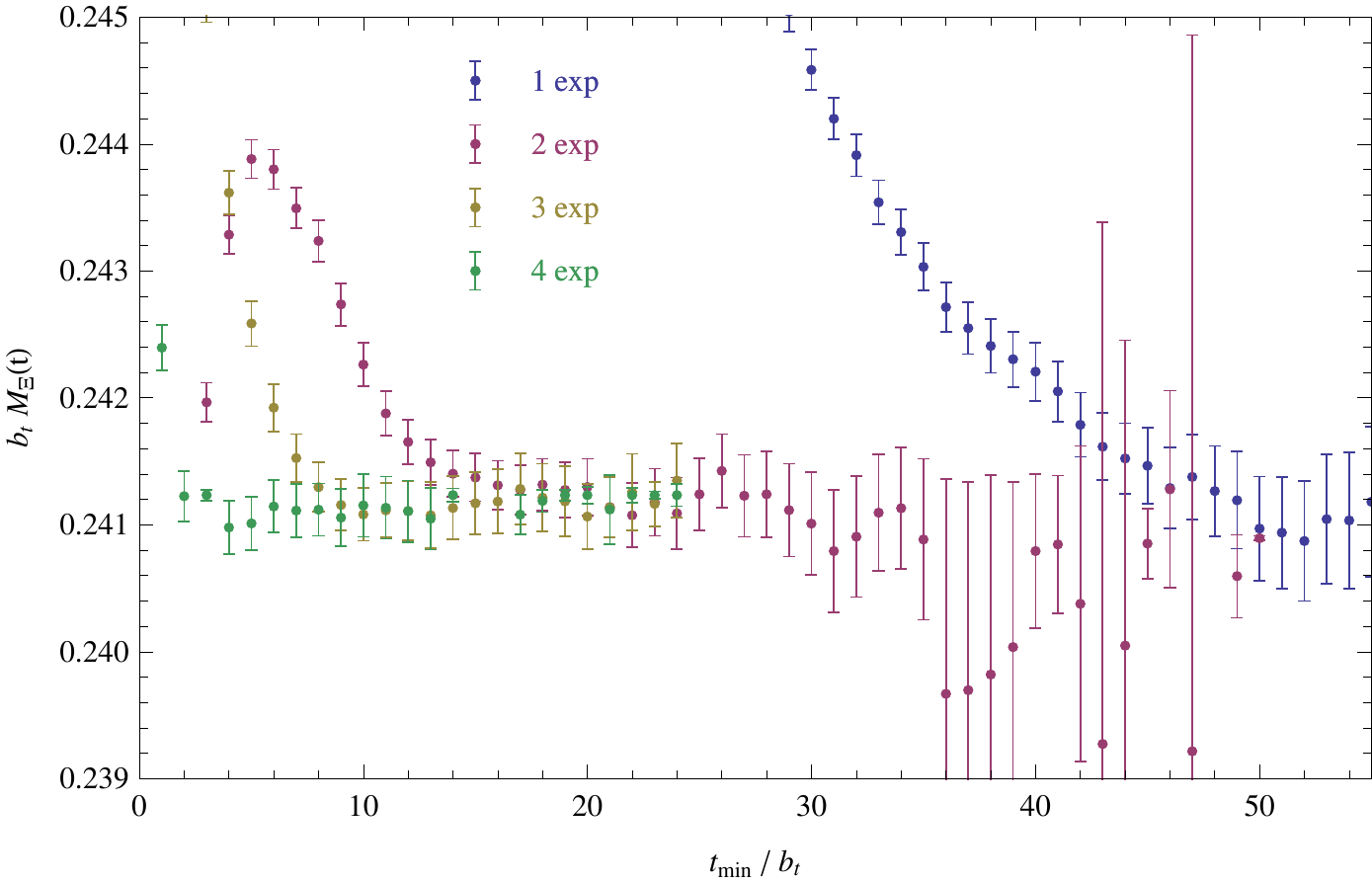}
    \\ \vspace*{3mm}
    \includegraphics[width=0.9\textwidth]{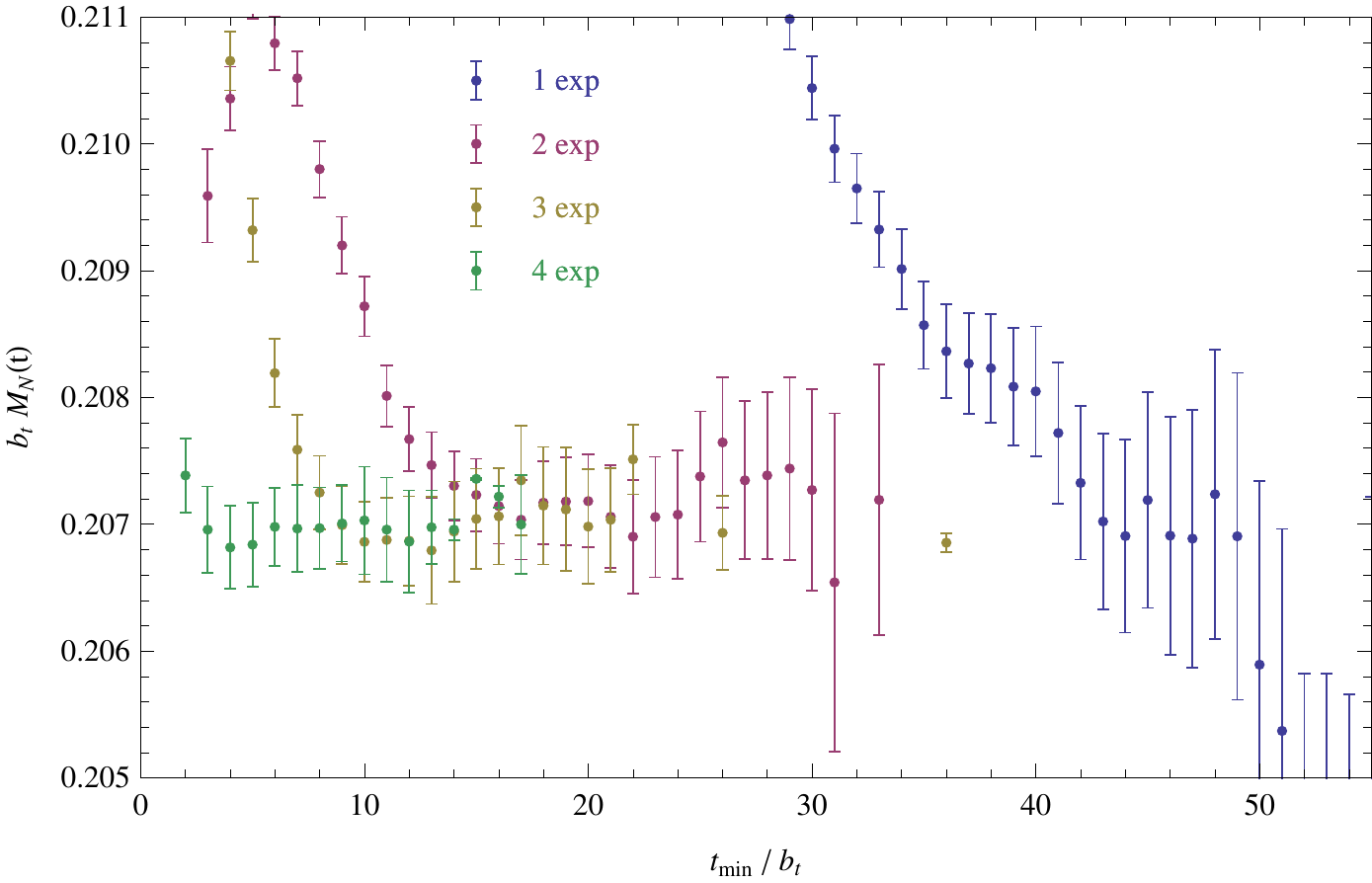}
  \end{tabular}
  \caption{\label{fig:nExpFits} The mass of the $\Xi$ (upper panel)
    and N (lower panel) extracted from multi-exponential fits as a
    function of the $t_{min}$ used in the fit.}
\end{figure}

The extracted mass of the $\Xi$ as a function of the number of
exponentials in the fit form is detailed in Table~\ref{tab:nExpFits}.
With the high statistics in this study, fitting a single exponential
yields a statistical uncertainty of less than $0.2\%$, with a slightly
larger fitting systematic uncertainty, however, 50 time-slices must be
discarded because of excited state contamination.  For our
multi-exponential fits, the fitting systematic uncertainty is defined
to be the standard deviation of all successful fits in a given
minimization.  To define a successful fit, we take a fixed length in
time, $R\equiv t_{\rm max} - t_{\rm min}$, and a fixed set of initial
parameters, and keep all fits with an integrated probability
distribution $Q > 0.1$ while varying $t_{min}$.\footnote{The quality
  of fit value, Q, is defined as the integrated probability
  distribution of $\chi^2$ with $d$ degrees of freedom, $Q \equiv
  \int_{\chi^2_{\rm min}}^\infty d \chi^2 \mathcal{P}(\chi^2, d)$,
  where ${\cal P}(x^2,d) = N (x^2)^{d/2-1} \exp(-x^2/2)$, with N the
  normalization constant.  The lower limit of the integration,
  $\chi^2_{\rm min}$, is the $\chi^2$ of the the fit under
  consideration.} One observes that the statistical and systematic
uncertainties are not further reduced by including more than three
exponentials in the fit.  The resulting ground state mass of the $\Xi$
as a function of the $t_{min}$ used in the fit is shown in upper panel
of fig.~\ref{fig:nExpFits} (in a style similar to an effective mass
plot) with the color and symbol shapes indicating the number of
exponentials in the fit.  The extraction of the nucleon mass is also
shown in the lower panel.  Increasing the number of exponentials in
the fit, $N_{\rm exp}$, allows the $t_{ min}$-interval over which the
ground-state energy is seen to plateau to be brought closer to the
source where statistical uncertainties are much reduced.

%%%%%%%%%%%%%%%
%% FIGURE MN MXi as function of Nexp
%%%%%%%%%%%%%%%

The full set of measurements have been used to generate the fits
presented in Table~\ref{tab:nExpFits}, and the correlation functions
from configurations nearby in Monte-Carlo time have not been
blocked.  Blocking is known to be important, since for correlated
configurations, unblocked correlation functions can lead to
underestimates of the true uncertainty.  For hadronic quantities, we
expect that the ensemble we have used has an auto-correlation time of
about $40$ Monte-Carlo time-steps~\cite{Lin:2008pr}.  Our calculations
have been performed on configurations separated by only $\tau = 10$.
Several different fits were performed to determine the effects of
blocking on our multi-exponential fits, the results of which are
collected in Table~\ref{tab:blocking}.  To normalize these fits, a
fitting interval, with a range of $R=30$, was determined from the
$N_{\rm exp}=3$ exponential fits.  For these fits, the blocking time
$\tau_{\rm block}$ has no detectable impact on either the statistical
or systematic uncertainties ($\tau_{\rm block} = 1$ corresponds to no
blocking, while $\tau_{\rm block} = 10$ corresponds to blocking every
10 configurations).
%%%%%%%%%%%%%%%%
%% Table BLOCKING
\begin{table}[!ht]
  \caption{\label{tab:blocking}{Effects of blocking on the determination
      of the ground state $\Xi$  mass.}}
  \begin{ruledtabular}
    \begin{tabular}{lccccccccc}
      $b_t M$ & $N_{\rm exp}$& R: $t_{\rm min},..,t_{\rm max}$& $N_{\rm suc}$& $\tau_{\rm block}$& $\xi_{\rm eff}$& $N_{\rm src}$ & $N_{\rm cfg}$& $\chi^2/{\rm dof}$& Q  \\
      \hline
      0.24113(25)(36)& 3&30: 14--44& 17& 1& 3.5& 245 & 1194& 0.84& 0.79\\
      0.24130(30)(09)& 3&30: 14--44& 15& 2& 3.5& 245 & 1194 & 0.83& 0.81\\
      0.24123(33)(31)& 3&30: 14--44& 21& 5& 3.5& 245 & 1194 & 0.91& 0.67\\
      0.24139(34)(34)& 3&30: 14--44& 11& 8& 3.5& 245 & 1194 & 1.1& 0.35\\
      0.24063(25)(36)& 3&30: 30--60& 2& 10& 3.5& 245 & 1194 & 0.93& 0.62\\
    \end{tabular}
  \end{ruledtabular}
\end{table}
%%%%%%%%%%%%%%%%%%

The range of time used in the fits also plays an important role in
minimizing the uncertainty.  The resulting fits for short and long
time-ranges used in three and four exponential fits are shown in
Table~\ref{tab:nExpRange}.  While the range does not have a
significant impact on the statistical uncertainty, it does
significantly reduce the systematic uncertainty in the fit.  To have
such long ranges of statistically useful time-slices, the anisotropy
$\xi = b_s / b_t$, which is 3.5 for this ensemble, is crucial.  We
have not performed calculations with a different anisotropy (including
isotropy), but this can be qualitatively studied by constructing
correlation functions using only every second or every third time
slice, with an effective anisotropy of $\xi_{\rm eff} =$ 1.75 and
1.17, respectively.  In the lower section of
Table~\ref{tab:nExpRange}, we display fits of 1, 2 and 3 exponentials
to these reduced sets of measurements.  This reduced anisotropy has a
significant impact on the resulting uncertainties, particularly for
$\xi_{\rm eff} = 1.17$.  We were unable to find successful four
exponential fits, and the number of successful fits with 1, 2 and 3
exponentials has been reduced.
Furthermore, the smaller number of time-slices in the same physical
extent, reduces our ability to control the systematics of the fits.
%%%%%%%%%%%%%%%%
%% Table RANGE
\begin{table}[!t]
  \caption{\label{tab:nExpRange}{Effects of fit range, R and anisotropy,
      $\xi_{\rm eff}$ on the determination of ground state $\Xi$ mass.}}
  \begin{ruledtabular}
    \begin{tabular}{lccccccccc}
      $b_t M$ & $N_{\rm exp}$& R: $t_{\rm min},..,t_{\rm max}$& $N_{\rm suc}$&
      $\tau_{\rm block}$
      & $\xi_{\rm eff}$& $N_{\rm src}$ & $N_{\rm cfg}$& $\chi^2/{\rm dof}$& Q  \\
      \hline
      0.24138(33)(44)& 1&20: 47--67 & 16& 1& 3.5 & 245 & 1194& 0.50& 1.00\\
      0.24108(28)(10)& 2&45: 24--69& 12& 1& 3.5& 245 & 1194& 0.70& 0.99\\
      0.24102(26)(25)& 3& 24: 20--44& 20& 1& 3.5& 245 & 1194 & 0.81& 0.80\\
      0.24112(26)(15)& 4& 25: 17--42& 15& 1& 3.5& 245 & 1194 & 0.94& 0.58\\
      0.24113(25)(07)& 3& 55: 14--69& 10& 1& 3.5& 245 & 1194 & 0.74& 0.98\\
      0.24106(24)(08)& 4& 60: 9--69& 8& 1& 3.5& 245 & 1194 & 0.79& 0.95\\
      \hline
      0.24062(49)(29)& 1& 24: 52,54,...,74 & 4& 1& 1.75& 245 & 1194 & 0.74& 0.78 \\
      0.24106(51)(34)& 2& 24: 34,36,...,58 & 11& 1& 1.75& 245 & 1194 & 0.55& 0.95 \\
      0.24111(25)(30)& 3& 28: 14,16,...,42 & 10& 1& 1.75& 245 & 1194 & 0.96& 0.51 \\
      \hline
      0.24119(37)(30)& 1& 15: 48,51,...,63 & 5& 1& 1.17& 245 & 1194 & 0.46& 0.9 \\
      0.24115(46)(14)& 2& 24: 33,36,...,57 & 2& 1& 1.17& 245 & 1194 & 0.72& 0.74 \\
      0.24110(21)(24)& 3& 27: 21,24,...,48 & 4& 1& 1.17& 245 & 1194 & 0.99& 0.45\\
    \end{tabular}
  \end{ruledtabular}
\end{table}
%%%%%%%%%%%%%%%%

Finally, the impact of the number of sources, $N_{\rm src}$ and number
of configurations $N_{\rm cfg}$, on the uncertainties in the extracted
mass of the $\Xi$ has been explored, the results of which are
collected in Table~\ref{tab:nExpNsrcNcfg}.  With $N_{\rm src} = 50$ or
$N_{\rm cfg} = 597$, the statistical uncertainties with $N_{\rm
  exp}=3$ exponential fits are the same as with the full set of
measurements.\footnote{The set of measurements with varying numbers of
  sources has been constructed by including all configurations which
  have at least $N_{\rm src} = 100$.}  However, in both cases the
systematic uncertainty is larger than that of the full set of
measurements. The corresponding dependence of more complicated
multi-particle observables on the number of configurations and sources
are under investigation.
%%%%%%%%%%%%%%%%
%% Table Nsrc NCFG
\begin{table}[!b]
  \caption{\label{tab:nExpNsrcNcfg}{Effects of the number of
      sources, $N_{\rm src}$ and number of configurations, $N_{\rm cfg}$ on the
      determination of the ground state $\Xi$ mass.}}
  \begin{ruledtabular}
    \begin{tabular}{lccccccccc}
      $b_t M$ & $N_{\rm exp}$& R: $t_{\rm min},..,t_{\rm max}$& $N_{\rm suc}$& $\tau_{\rm block}$& $\xi_{\rm eff}$& $N_{\rm src}$ & $N_{\rm cfg}$& $\chi^2/{\rm dof}$& Q  \\
      \hline
      0.24304(101)(54)& 1&20: 48--68& 18& 1& 3.5& 245 & 120 & 0.95& 0.55\\
      % --& 2&30,40& 0& 1& 3.5& 245 & 120 &\\
      % --& 3&30,40& 0& 1& 3.5& 245 & 120 &\\
      0.24248(64)(24)& 1&20: 45--65& 22& 1& 3.5& 245 & 239 & 0.65& 0.96\\
      0.24178(36)(34)& 1&20: 43--63& 17& 1& 3.5& 245 & 597 &0.66& 0.95\\
      0.24110(37)(13)& 3&40: 16--46& 8& 1& 3.5& 245 & 239 & 0.99& 0.50\\
      0.24134(24)(17)& 3&50: 22--72& 16& 1& 3.5& 245 & 597 &0.72& 0.98\\
      \hline
      0.24309(162)(137)& 1&20: 43--63& 24& 1& 3.5& 1& 1025 & 1.07& 0.35\\
      0.24075(129)(44)& 1&20: 52--72& 24& 1& 3.5& 10& 1025 & 0.72& 0.91\\
      0.24063(45)(56)& 1& 20:  46--56& 20& 1& 3.5& 50& 1025 & 0.72& 0.91\\
      0.24088(39)(48)& 1& 20:  46--56& 17& 1& 3.5& 100& 1025 & 1.00& 0.47\\
      0.24174(72)(60)& 3&50: 11--61& 13& 1& 3.5& 1& 1025 & 1.10& 0.24\\
      0.24116(38)(22)& 3&50: 15--65& 13& 1& 3.5& 10& 1025 & 0.87& 0.81\\
      0.24108(28)(15)& 3& 50:  15--65& 8& 1& 3.5& 50& 1025 & 0.94& 0.64\\
      0.24115(30)(04)& 3& 50:  20--70& 3& 1& 3.5& 100& 1025 & 1.15& 0.15\\
      \hline
      0.24138(33)(44)& 1&20: 47--67 & 16& 1& 3.5 & 245 & 1194& 0.50& 1.00\\
      0.24115(24)(07)& 3&50: 14--64 & 10& 1& 3.5& 245 & 1194& 0.77& 0.95\\
    \end{tabular}
  \end{ruledtabular}
\end{table}
%%%%%%%%%%%%%%%%

For multi-exponential fits, it appears that the most important feature
in controlling the uncertainty the ground state is the number of
exponentials with which a successful minimization can be performed.
Neither the statistical nor systematic uncertainties improve beyond
the inclusion of three exponentials in the fits.  To have confidence
in the $N_{\rm exp}=3$ exponential fits, the anisotropy is found to be
essential.  A quantitative exploration of the effects of the
anisotropy on the stability of multi-exponential fits is desirable,
but this would be a very costly numerical endeavor.  With three or
more exponentials, the fitting range, number of sources and number of
configurations have essentially the expected effect on the statistical
(and systematic) uncertainties.

%%%%%%%%%%%%%%%%%%%%%%%%%
\subsection{Generalized Effective Mass Plots}
\label{sec:GEMP}
\noindent
Correlation functions on an ensemble of configurations of infinite
extent in the time-direction become dominated by a single exponential
at large times with an argument that is the energy of the ground state
of the system,
\begin{eqnarray}
  C(t) & = & 
  \sum _{n=0}^\infty\ Z_n\ e^{- E_n t}
  \ \rightarrow\ Z_0\ e^{- E_0 t}
  \ \ \ .
  \label{eq:correlator}
\end{eqnarray}
It is conventional to define the effective mass (EM) from the
logarithm of the ratio of the correlation function on adjacent
time-slices. It is also possible \footnote{This was suggested by
  K.~Juge in a talk at Lattice 2008, see Ref.~\cite{JugeAllHands2008},
  but may have been used earlier.} to form a more general EM from
time-slices separated by $t_J>1$
\begin{eqnarray}
  M_{\rm eff,t_J}(t) & = & 
  {1\over t_J}\log\left( { C(t)\over C(t+t_J) } \right)
  \ \rightarrow\ E_0
  \ \ \ .
  \label{eq:Juge}
\end{eqnarray}
For exponentially decreasing signals with time-independent noise, this
will naturally reduce the statistical uncertainty in the EM and
improve the extraction of energy-eigenvalues as it increases the
``lever-arm'' of the exponential.  In such a case, the uncertainty in
$M_{\rm eff}(t_J)$ in Eq.~(\ref{eq:Juge}) will decrease as $1/t_J$.
Simple correlation functions involving $\pi$'s have time-independent
uncertainties, but this is not the case for baryonic correlation
functions, whose relative uncertainties grow exponentially with time.
We explore the improvements to baryon EMs, and ultimately the
extraction of baryon masses and the energy-eigenvalues in the volume,
that result from $t_J>1$. In fitting an energy to an EM (and other
generalizations), either the Bootstrap or Jackknife procedures are
used to generate the covariance matrix associated with the time-slices
in the range of the fit.\footnote{In the Bootstrap method, $N_{\rm
    boot}=N_{\rm cfg}$ randomly generated bootstrap samples are used
  after blocking over sets of five configurations, while the Jackknife
  ensembles are constructed by single omission after blocking over 10
  configurations. We have found consistent results using both methods
  and by using different blockings and values of $N_{\rm boot}$.}
This covariance matrix is then used to form the $\chi^2/{\rm dof}$,
which is minimized to determine the energy, and then explored to
determine the uncertainty in this energy. The statistical uncertainty
is obtained by finding values of the fit parameters where the $\chi^2$
function attains a value of $\chi^2_{\rm min}+1$.

To demonstrate the impact of $t_J>1$ for baryon EMs, we examine the
smeared-smeared correlation function of the $\Xi$-baryon.
%%%%%%%%%%%%%%%%%%%%%%%%%%%%%%%%%%%%%%%%%%%%%%%%%%%
%
% FIGURE: Juge on Xi Correlator sh_sh
%
%%%%%%%%%%%%%%%%%%%%%%%%%%%%%%%%%%%%%%%%%%%%%%%%%%%
\begin{figure}[!ht]
  \vskip0.5in \center
  \begin{tabular}{c}
    \includegraphics[width=0.99\textwidth]{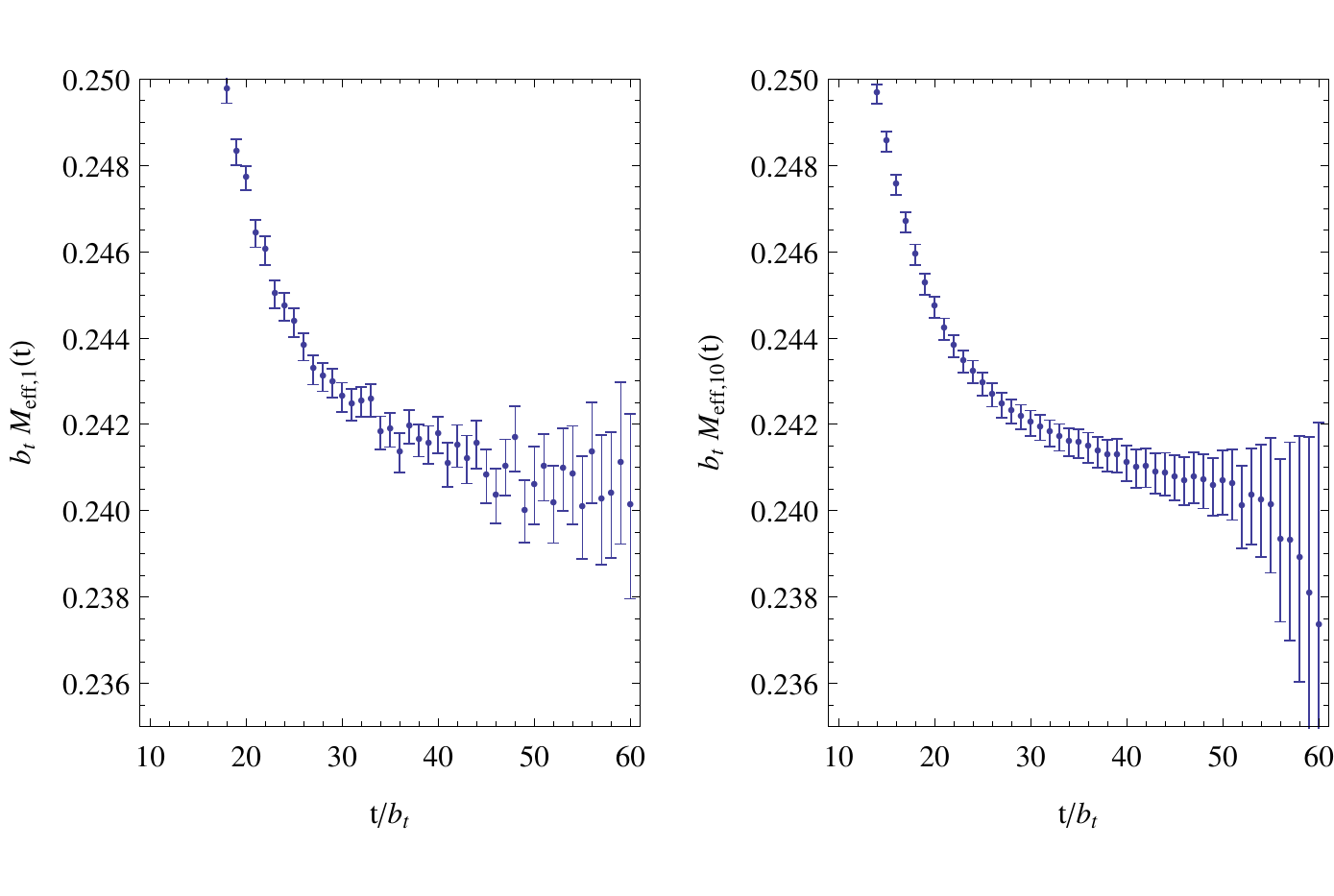}
  \end{tabular}
  \caption{\label{fig:JugeXi} The left panel shows the conventional EM
    ($t_J=1$) from the smeared-smeared $\Xi$ correlation function, and
    the right panel shows the EM for the same correlation function
    with $t_J=10$.  }
\end{figure}
Figure~\ref{fig:JugeXi} shows the EMs obtained with $t_J=1$ (left
panel) and $t_J=10$ (right panel).  The scatter of the effective mass
from time-slice to time-slice is significantly reduced with $t_J=10$
compared with $t_J=1$, allowing for a clear identification of the time
range over which it is reasonable to extract the (ground-state) mass
of the $\Xi$.  Therefore, the systematic uncertainty associated with
the fitting range in the EM is reduced.  The statistical uncertainty
in the mass of the $\Xi$ extracted from the EMs with the two different
values of $t_J$, when fit over the time time-slice interval, are
however very similar, as can be seen in the resulting fits to
time-slices $t=48$ to $t=58$,
\begin{eqnarray}
  M_\Xi^{t_J=1} & = & 0.24087\pm 0.00057\pm 0.00080
  \ \ ,\ \ 
  \chi^2/{\rm dof} \ =\ 0.68\,,
  \nonumber\\
  M_\Xi^{t_J=10} & = & 0.24060\pm 0.00061\pm 0.00060
  \ \ ,\ \ 
  \chi^2/{\rm dof} \ =\ 0.44\,.
  \label{eq:JugeFits}
\end{eqnarray}
The first uncertainty corresponds to the statistical uncertainty in
the mass determined from the $\chi^2/{\rm dof}$ minimization, while
the second corresponds to systematic uncertainty associated with the
fitting interval.  The systematic uncertainty of this fit is
determined by varying the fitting interval at each end by $0, \pm 1,
\pm 2$ time-slices, performing a $\chi^2/{\rm dof}$ minimization over
each interval and taking half of the spread of the extracted
masses. Alternative procedures such as using fits to rolling windows
of time-slices within the fitting interval return similar
uncertainties.

It is interesting to explore how different values of $t_J$ modify the
form of the covariance matrix that is input into the $\chi^2/{\rm
  dof}$ minimization.  The covariance matrices associated with the
time-interval $t=48$ to $t=51$ from these two EMs are shown in
Eq.~(\ref{eq:cov}).  They are quite different, with the distant
off-diagonal elements becoming more significant for increasing $t_J$.
\begin{eqnarray}
  \sigma_{t_J=1}^2 & = & 10^{-7}\left(
    \begin{array}{cccc}
      4.82 & 1.97 & 2.71 & 2.70\\
      1.97 & 6.21 & 3.03 & 3.15\\
      2.71 & 3.03 & 7.83 & 3.35\\
      2.70 & 3.15 & 3.35 & 7.06\\
    \end{array}
  \right)
  \ , \  
  \sigma_{t_J=10}^2 \ = \ 10^{-7}\left(
    \begin{array}{cccc}
      4.75 & 5.09 & 5.29 & 5.26 \\
      5.09 & 5.65 & 5.92 & 5.96\\
      5.29 & 5.91 & 6.41 & 6.48\\
      5.26 & 5.96 & 6.48 & 6.90
    \end{array}
  \right)
  \ .
  \label{eq:cov}
\end{eqnarray}
In this comparison, it is important to note that the two extractions
make use of different parts of the correlation function. The $t_J=1$
fit uses five time-slices, while the $t_J=10$ fit uses eight
well-separated time-slices.

%%%%%%%%%%%%%%%%%%%%%%%%%%%%%%%%%%%%%%%%%%%%%%%%%%%
%
% FIGURE: Juge on Xi Correlator sh_sh
%
%%%%%%%%%%%%%%%%%%%%%%%%%%%%%%%%%%%%%%%%%%%%%%%%%%%
\begin{figure}[!ht]
  \vskip0.5in \center
  \begin{tabular}{c}
    \includegraphics[width=0.99\textwidth]{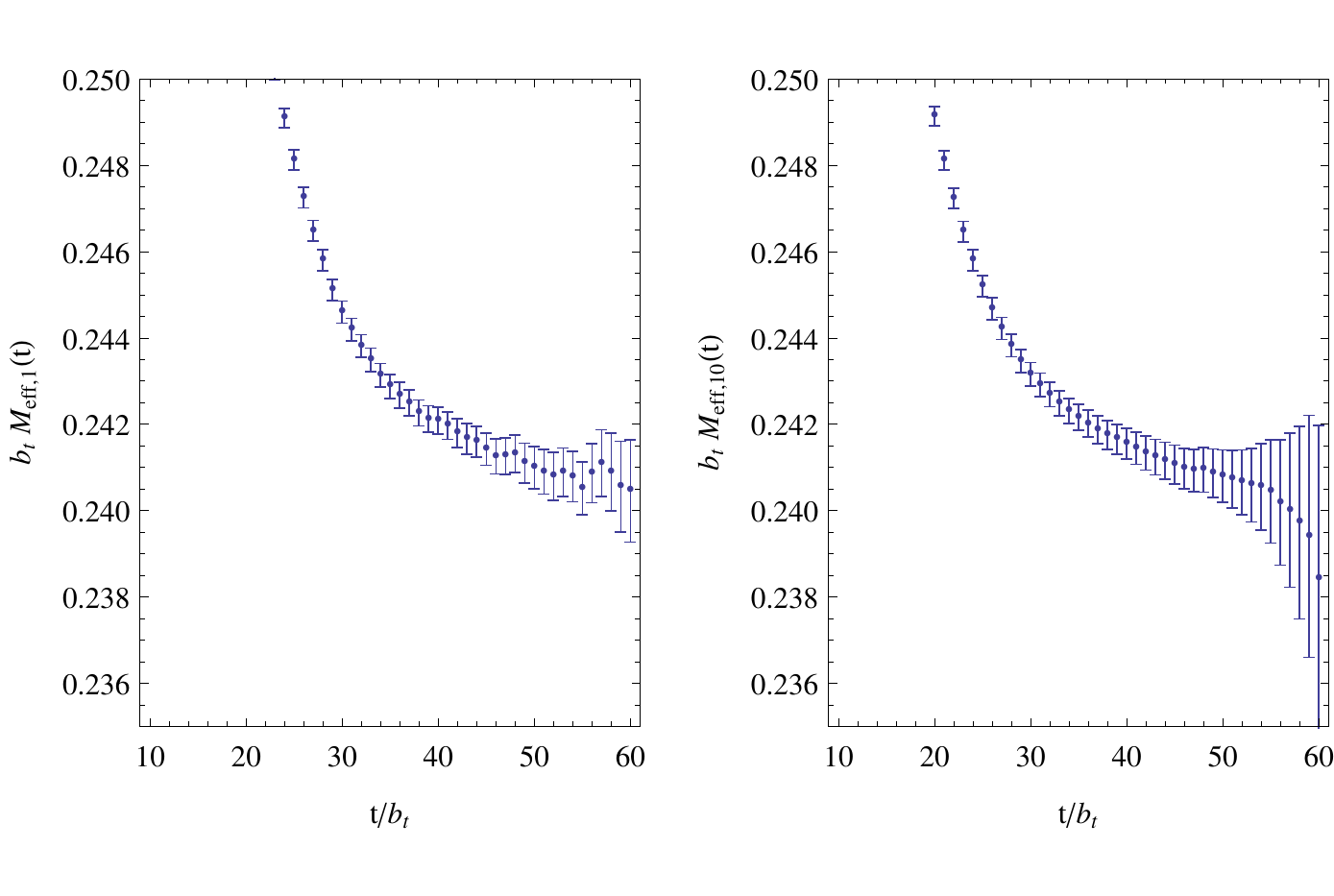}
  \end{tabular}
  \caption{\label{fig:JugeXishpt} The left panel shows the
    conventional EM ($t_J=1$) from the smeared-point $\Xi$ correlation
    function, and the right panel shows the EM for the same
    correlation function with $t_J=10$.  }
\end{figure}
The EMs from the smeared-point $\Xi$ correlation functions with
$t_J=1$ and $t_J=10$ are shown in fig.~\ref{fig:JugeXishpt}. The
scatter in the EM for $t_J=1$ is substantially less than for the
smeared-smeared correlation function, as the overlap of the
interpolating operator onto the ground-state is larger.  However, the
overlap onto excited states is even larger, and the ground state
component of the correlation function does not become dominant until
later times, increasing the fitting systematic uncertainty in the
extraction of the ground-state mass.

%%%%%%%%%%%%%%%%%%%%%%%%%
\subsection{Prony's Method/Linear Prediction}

\noindent
As the signal-to-noise ratio of baryon correlation functions degrades
exponentially with time, it is important to extract the ground-state
signal (or excited state signal if that is the state of interest) from
a range of time-slices starting at the earliest possible time.
Significant effort has been placed into determining interpolating
operators that maximize the overlap onto the ground-states of the
baryons in order to facilitate this.  Further, there has been
significant effort put into using the variational method
\cite{Michael:1985ne,Luscher:1990ck}, for which the correlation
functions resulting from a number of hadronic interpolating operators
are diagonalized on each time-slice to give the eigen-energies with
the appropriate quantum numbers.  A few years ago, Fleming suggested
that generalizing the EM method to two or more exponential functions
might be useful in LQCD analysis based on findings of NMR
spectroscopists~\cite{Fleming:2004hs}~\footnote{ The method is more
  generally referred to as Prony's method~\cite{Hamming} after Gaspard
  Riche de Prony who first constructed it in 1795~\cite{Prony}. These
  techniques and other related methods are known as linear prediction
  theory in the signal analysis community.}.  At that time, we
explored this technique with sets of correlation functions that were
available to us at that time, and found the method was quite unstable
to the statistical fluctuations in those measurements. More recently,
Lin and Cohen \cite{Lin:2007iq} compared this method favorably to the
variational approach.  Given the small statistical uncertainties in
the correlation functions we are presently considering, and the
reduction in the systematic uncertainties achieved with $t_J>1$, we
return to explore this technique.

In LQCD, two-point correlation functions have the form
\begin{eqnarray}
  \label{eq:333}
  G(t) & = & A_0 e^{-\alpha_0 t}\ +\  A_1 e^{-\alpha_1 t}\ +\ ...\ +\  A_{k-1}
  e^{-\alpha_{k-1} t}+\ldots
  \ \ \ ,
\end{eqnarray}
where $t$ denotes the time-slice (time-slices are implicitly taken to
be evenly-spaced).  It follows from Eq.~(\ref{eq:333}) that
\begin{eqnarray}
  G(t+ n k)\  +\  C_{k-1}\ G(t+n(k-1)) \  +\  C_{k-2}\ G(t+n(k-2))  + .. +  C_0\ G(t) & = & 0
  \ ,
  \label{eq:TheCs}
\end{eqnarray}
where the integer $n$ is the generalization of $t_J$ to the case of a
multi-exponential function.  In order to determine the $k$
coefficients $C_i$, $k$ equations are required to be formed from the
measured correlation function.  Given the $C_i$, the roots of
\begin{eqnarray}
  \left(e^{-n\alpha}\right)^k\ +\  C_{k-1}\left(e^{-n\alpha}\right)^{k-1}\ +\  
  C_{k-2}\left(e^{-n\alpha}\right)^{k-2}
  \ +\ ..\ + \ C_0 & = & 0
  \ \ \ ,
  \label{eq:solve}
\end{eqnarray}
and in particular the $\alpha$'s, provide the energies of the states
contributing to the correlation function.

%%%%%%%%%%%%%%%%
\subsubsection{One Exponential : The Standard Effective Mass}

In the case of $k=1$, where the correlation function is assumed to be
a single exponential, and taking $n=1$,
\begin{eqnarray}
  G(t+ 1)\  +\  C_0\ G(t) & = & 0
  \ \ ,\ \ 
  \left(e^{-\alpha}\right)\ +\  C_0 \ = \ 0
  \ ,
  \label{eq:TheCsONE}
\end{eqnarray}
and the usual expression for the EM follows trivially.

%%%%%%%%%%%%%%%%
\subsubsection{Two Exponentials}

In the case of two exponentials in the correlation function, the most
general pair of equations that can be used to extract the two
effective masses is
\begin{eqnarray}
  G(t+ 2 n)\  +\  C_{1}\ G(t+n) \ +\ C_0\ G(t) & = & 0
  \nonumber\\
  G(t+ 2 n + q_1)\  +\  C_{1}\ G(t + n + q_1) \ +\ C_0\ G(t+q_1) & = & 0
  \ \ \ ,
\end{eqnarray}
where $q_1$ is an arbitrary integer off-set between the two equations.
Inserting the values of the calculated correlation function allows for
an extraction of $C_{0,1}$ on each time-slice, $j$.  These
coefficients are then inserted into
\begin{eqnarray}
  \left(e^{-n\alpha}\right)^2\ +\  C_{1}\left(e^{-n\alpha}\right)\ +\  C_{0} & = & 0
  \ \ \ ,
  \label{eq:solve2}
\end{eqnarray}
to recover a numerical value for $e^{-n\alpha}$.  By choosing
$n=m=1$, the expressions of Fleming~\cite{Fleming:2004hs} are
recovered.  In order to optimize the two-exponential extraction, a
search over values of the pair $(n,q_1)$ must be performed. A further
systematic uncertainty can be assigned from this choice.

\begin{figure}[!ht]
  \vskip0.5in \center
  \begin{tabular}{c}
    \includegraphics[width=0.99\textwidth]{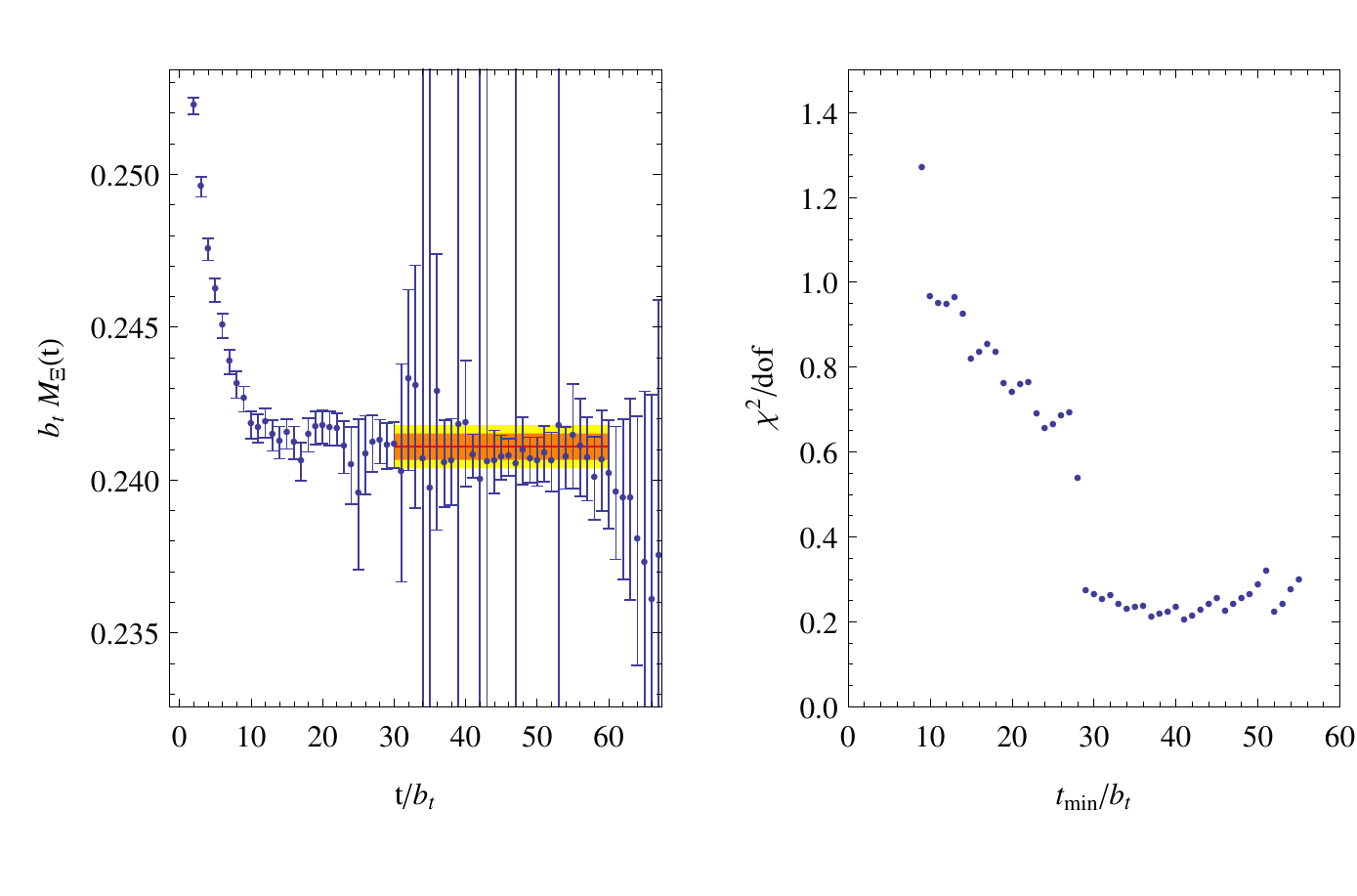}
  \end{tabular}
  \caption{\label{fig:2expXi} The left panel shows the ground-state
    extracted from the smeared-smeared $\Xi$ correlation function with
    a 2-exponential Prony determination with $n=q_1=5$, and the
    correlated fit to the time-slices between $t=30$ and $t=60$. The
    inner (darker) region corresponds to the statistical uncertainty,
    while the outer (lighter) region corresponds to the statistical
    and fitting systematic uncertainties combined in quadrature.  The
    right panel is the $\chi^2/{\rm dof}$ for fits between time-slices
    $t=t_{\rm min}$ and $t=60$.}
\end{figure}

The ground-state extracted from the smeared-smeared $\Xi$ correlation
function with $n=q_1=5$ is shown in fig.~\ref{fig:2expXi}.
%%%%%%%%%%%%%%%%%%%%%%%%%%%%%%%%%%%%%%%%%%%%%%%%%%%
%
% FIGURE: k=2 Prony on Xi Correlator
%
%%%%%%%%%%%%%%%%%%%%%%%%%%%%%%%%%%%%%%%%%%%%%%%%%%%
It is clear that the ground-state signal can be isolated from the
correlation function for a large number of time-slices, many more than
using the single exponential EM (fig.~\ref{fig:JugeXi}) alone.  We
have shown the fit to the ground-state result between time slices
$t=30$ and $t=60$.  The lower-limit of the time interval was chosen to
be within an interval for which $\chi^2/{\rm dof} < 1$.  Extending the
fit interval to lower time-slices gradually increases the $\chi^2/{\rm
  dof}$, as shown in the right panel of fig.~\ref{fig:2expXi},
indicating contamination from higher energy states.  The upper-limit
of the fitting interval was chosen to be in the region for which
backward propagating states (due to the anti-periodic BC's in the
time-direction) were not visible in the EM (or in the $\chi^2/{\rm
  dof}$).  The ground-state $\Xi$ mass we extract from this
2-exponential analysis is
\begin{eqnarray}
  M_\Xi & = & 0.24109\pm 0.00043\pm 0.00057
  \ \ ,\ \ 
  \chi^2/{\rm dof} \ =\ 0.38
  \ \ \ ,
  \label{eq:2expFits}
\end{eqnarray}
where the first uncertainty is statistical and the second is the
fitting systematic (as defined previously).  The statistical
uncertainty in the 2-exponential extraction is significantly smaller
than that obtained from the one-exponential analysis
(Eq.~(\ref{eq:JugeFits})).  This is due to the substantially increased
number of time-slices in the ground-state plateau in the generalized
EM.

One aspect of this method that is less appealing is the ambiguity in
the association of the two roots that result from
Eq.~(\ref{eq:solve2}) to the two states on different time-slices and
on different jackknife/bootstrap ensembles. This (mis-)identification
issue is the cause of the anomalously large uncertainties at
time-slices 31,\ldots,36 in fig.~\ref{fig:2expXi} - this should not be
interpreted as variance of the signal for the ground
state. Additionally, on different time-slices and jackknife/bootstrap
ensembles, this method can, and likely will, select different terms in
Eq.~(\ref{eq:333}) particularly for the sub-dominant excited state,
adding additional artificial variance to the signals for particular
energy eigenstates. Consequently, the extracted second state is not
physically meaningful.

%%%%%%%%%%%%%%%%
\subsubsection{Three and More Exponentials}

The generalization of the method to arbitrary numbers of exponential
functions is straightforward.  In the case of three exponentials,
inserting the values of the calculated correlation functions,
\begin{eqnarray}
  G({t+ 3 n})\  +\  C_{2}\ G({t+ 2 n})\  +\  C_{1}\ G({t+n}) \ +\ C_0\ G({t}) 
  & = & 0
  \nonumber\\
  G({t+ 3 n + q_1})\  +\  C_{2}\ G({t+ 2 n + q_1})\  +\  C_{1}\ G({t+n + q_1}) 
  \ +\ C_0\ G({t + q_1}) & = & 0
  \nonumber\\
  G({t+ 3 n + q_2})\  +\  C_{2}\ G({t+ 2 n + q_2})\  +\  C_{1}\ G({t+n + q_2}) 
  \ +\ C_0\ G({t + q_2}) & = & 0
  \ \ \ ,
  \label{eq:3expsys}
\end{eqnarray}
with $q_1\ne q_2\ne 0$ allows for an extraction of $C_{0,1,2}$ on each
time-slice, $t$.  Again, these coefficients can be extracted uniquely
in terms of the $G(t)$ due to the fact that the system is linear.
These coefficients $C_{0,1,2}$ are then inserted into
\begin{eqnarray}
  \left(e^{-n\alpha}\right)^3\ +\  C_{2}\ \left(e^{-n\alpha}\right)^2\ +\  
  C_{1}\left(e^{-n\alpha}\right)\ +\  C_{0} & = & 0
  \ \ \ ,
  \label{eq:solve3}
\end{eqnarray}
to recover a numerical value of $e^{-n\alpha}$.  Analysis of a given
correlation function involves searching for the values of the triplet
$(n,q_1,q_2)$ that optimizes the extraction (in each equality in
Eq.~(\ref{eq:3expsys}), different $n$ can be used).  In this case,
statistical fluctuations occasionally result in complex roots of
Eq.~(\ref{eq:solve3}) on a particular Jackknife or Bootstrap
ensemble. At present, we simply omit these contributions in our
analysis.  Such complex roots correspond to an oscillatory solution
and arise from short distance noise in the correlation function (or
nearly degenerate states in the spectrum), and are a well-known issue
with the simple Prony method.  More advanced
methods~\cite{osbornesmyth} can mitigate this issue, but do not result
in improved extractions of the ground-state so we do not discuss them
in detail.

\begin{figure}[!ht]
  \vskip0.5in \center
  \begin{tabular}{c}
    \includegraphics[width=0.99\textwidth]{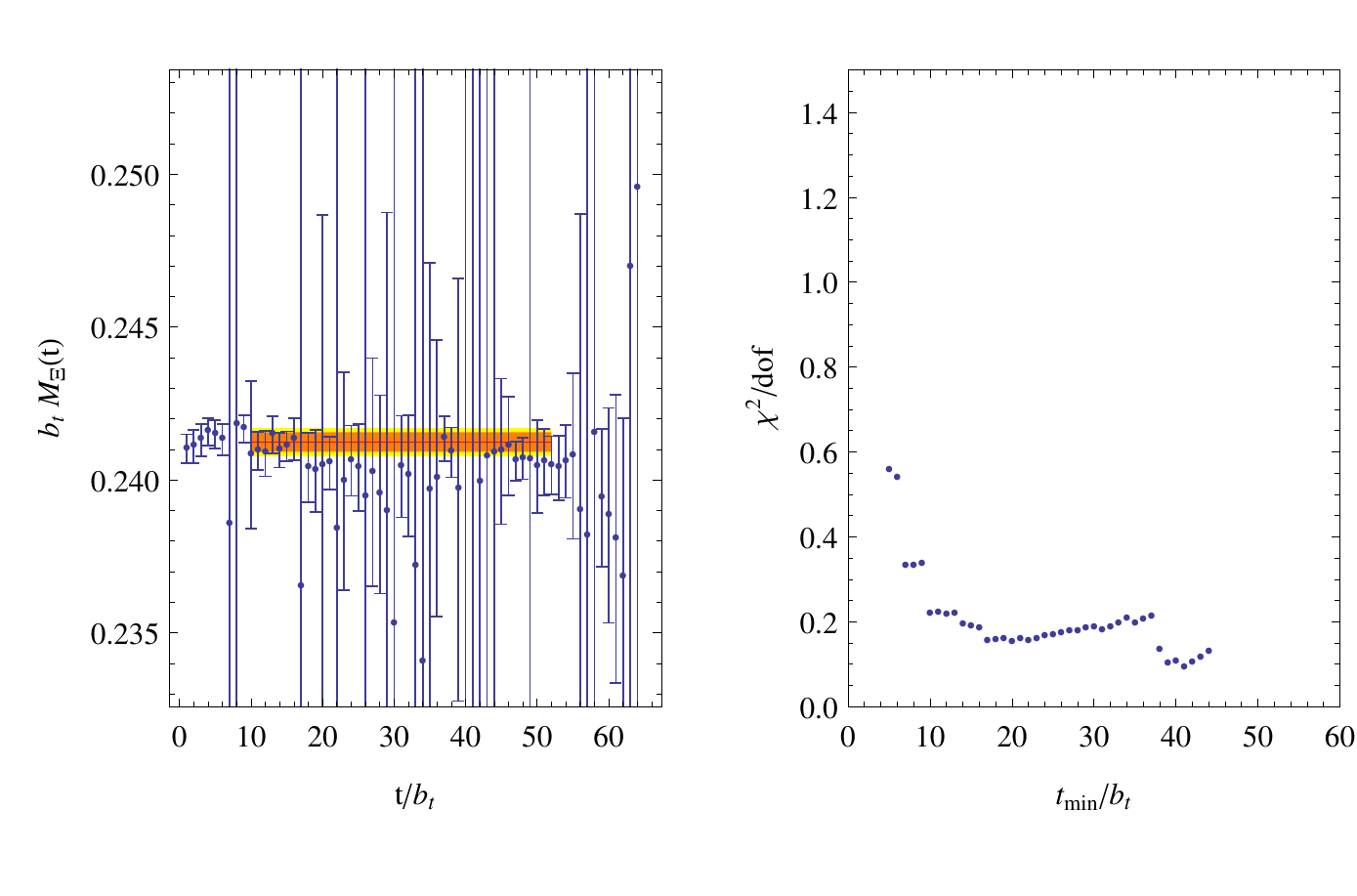}
  \end{tabular}
  \caption{\label{fig:3expXi} The left panel shows the ground-state
    extracted from the smeared-smeared $\Xi$ correlation function with
    a 3-exponential Prony determination with $n=10, q_1=3, q_2=6$, and
    the correlated fit to the time-slices between $t=10$ and
    $t=52$. The inner (yellow) region corresponds to the statistical
    uncertainty, while the outer (red) region corresponds to the
    statistical and fitting systematic uncertainties combined in
    quadrature.  The right panel is the $\chi^2/{\rm dof}$ for fits
    between the time-slices $t=t_{\rm min}$ (the horizontal axis) and
    $t=52$.}
\end{figure}

The ground-state energy extracted from the smeared-smeared $\Xi$
correlation function with $n=10, q_1=3, q_2=6$ is shown in
fig.~\ref{fig:3expXi}.
%%%%%%%%%%%%%%%%%%%%%%%%%%%%%%%%%%%%%%%%%%%%%%%%%%%
%
% FIGURE: k=3 Prony on Xi Correlator 4,2,6
%
%%%%%%%%%%%%%%%%%%%%%%%%%%%%%%%%%%%%%%%%%%%%%%%%%%%
It is clear that the ground-state signal is extractable from
time-slices even closer to the source than with two-exponential
analysis. In fig.~\ref{fig:3expXi}, the fit to the ground-state
between time slices $t=10$ and $t=52$ is shown.  The extracted mass is
\begin{eqnarray}
  M_\Xi & = & 0.24124\pm 0.00032\pm 0.00034
  \ \ ,\ \ 
  \chi^2/{\rm dof} \ =\ 0.22
  \ \ \ ,
  \label{eq:3expFits}
\end{eqnarray}
with the statistical uncertainty being slightly less than in the
two-exponential analysis.  It is important to realize that this level
of precision corresponds to a statistical uncertainty of $\sim 2~{\rm
  MeV}$ in the $\Xi$ mass.

We have successfully applied the four- and five- state Prony method to
our data but no improvement is seen beyond the three-exponential
extractions.

%%%%%%%%%%%%%%%%%%%%%%%%%%%%%%%%%%%%%%%
\subsection{Multi-Correlation Function Prony Method}
\label{sec:modif-prony-meth}

There are a number of extensions of the Prony method that exist in the
literature (see for example \cite{osbornesmyth}), some of which we
have investigated in detail.  For the correlation functions we have in
hand, these extensions do not significantly improve on the standard
Prony method.  Typically, these methods are applied in cases where
only a single set of measurements is available.  However, we have two
sets of correlation functions (smeared-smeared and smeared-point)
whose energy spectra are identical in the limit of a large number of
configurations.  It is straightforward to generalize Prony's method
to include both correlation functions--the matrix-Prony method.  This
form leads to a further reduction in the uncertainty of the extraction
of the energy eigenvalues. A similar approach, has been briefly
discussed in~Ref.~\cite{Fleming:2006zz}.

Assume we have $N$ ($N=2$ in our case) correlation functions from
which we want to extract the energy levels. If these correlation
functions are a sum of exponentials they satisfy the following
recursion relation,
\begin{equation}
  M y(\tau+t_J) - V y(\tau) = 0
  \ \ \ ,
  \label{eq:recursion}
\end{equation}
where $M$ and $V$ are $N\times N$ matrices and and $y(t)$ is a column
vector of $N$ components corresponding to the $N$ correlation
functions.  Eq.~(\ref{eq:recursion}) implies then the correlation
functions are
\begin{equation}
  y(t) = \sum_{n=1}^N  C_n q_n \lambda_n^{t}
  \ \ \ ,
  \label{eq:signal}
\end{equation}
where $q_n$ and $\lambda_n=\exp(m_n t_J)$ the eigenvectors and
eigenvalues of the following generalized eigenvalue problem
\begin{equation}
  M q = \lambda V q\ \ \ .
  \label{eq:gev}
\end{equation}
Given the $N$ sets of correlation functions, the masses can be found
by determining the matrices $M$ and $V$ that are needed in order for
the signal to satisfy Eq.~(\ref{eq:recursion}). Solving
Eq.~(\ref{eq:gev}), then leads to the eigenvalues $\lambda_n=\exp(m_n
t_J)$ and the eigenvectors $q_n$ needed to reconstruct the amplitudes
with which each exponential enters the correlation functions.  A
simple solution can be constructed as follows. First note that
\begin{equation}
  M \sum_{\tau=t}^{t+t_W} y(\tau+t_J)y(\tau)^T - V  \sum_{\tau=t}^{t+t_W}
  y(\tau)y(\tau)^T = 0
  \ \ \ .
  \label{eq:recursion3}
\end{equation}
Clearly, a solution for $M$ and $V$ is
\begin{eqnarray}
  M &=& \left[\
    \sum_{\tau=t}^{t+t_W} y(\tau+t_J)y(\tau)^T\
  \right]^{-1}
  \ \ \ \ ,\
  \ \ \
  V \ =\  \left[\ \sum_{\tau=t}^{t+t_W} y(\tau)y(\tau)^T \ \right]^{-1}
  \ \ \ ,
  \label{eq:solution}
\end{eqnarray}
where these inverses exist provided that the range, $t_W$, is large
enough to make the matrices in the brackets full rank ($t_W\ge N -
1$).  In our case with two exponentials the range has to be two for
achieving full rank.  Once the eigenvalues, $\lambda_n$ and
eigenvectors $q_n$ are determined, the amplitudes, $C_n$, can be
reconstructed using $t$ as a normalization point.  The shift parameter
$t_J$ can be used it improve stability.  The above solution is equivalent to determining $M$ and
$V$ by requiring that
\begin{eqnarray}
  \Psi^2 &=& \sum_{\tau=t}^{t+t_W}\left[ M y(\tau+t_J) - V
    y(\tau)\right]^T  \left[ M y(\tau+t_J) - V y(\tau)\right] 
  \label{eq:chi2}
\end{eqnarray}
is minimized.

To go beyond extracting two states, one can construct and solve a
second order recursion relation. The minimization condition of
Eq.~(\ref{eq:chi2}), augmented to contain the second order terms in the
recursion, can be used to determine the unknown matrices.  The
resulting eigenvalue problem is a second order nonlinear generalized
eigenvalue problem which is straightforward to solve. However, to
isolate the ground-state, which is our present focus, the two state
model is sufficient and we do not pursue this further.

To demonstrate how this method works, we return to the $\Xi$ correlation
function discussed above.
%%%%%%%%%%%%%%%%%%%%%%%%%%%%%%%%%%%%%%%%%%%%%%%%%%%
%
% FIGURE: Matrix prony tJ=tW=10
%
%%%%%%%%%%%%%%%%%%%%%%%%%%%%%%%%%%%%%%%%%%%%%%%%%%%
\begin{figure}[!ht]
  \centering
  \includegraphics[width=0.99\textwidth]{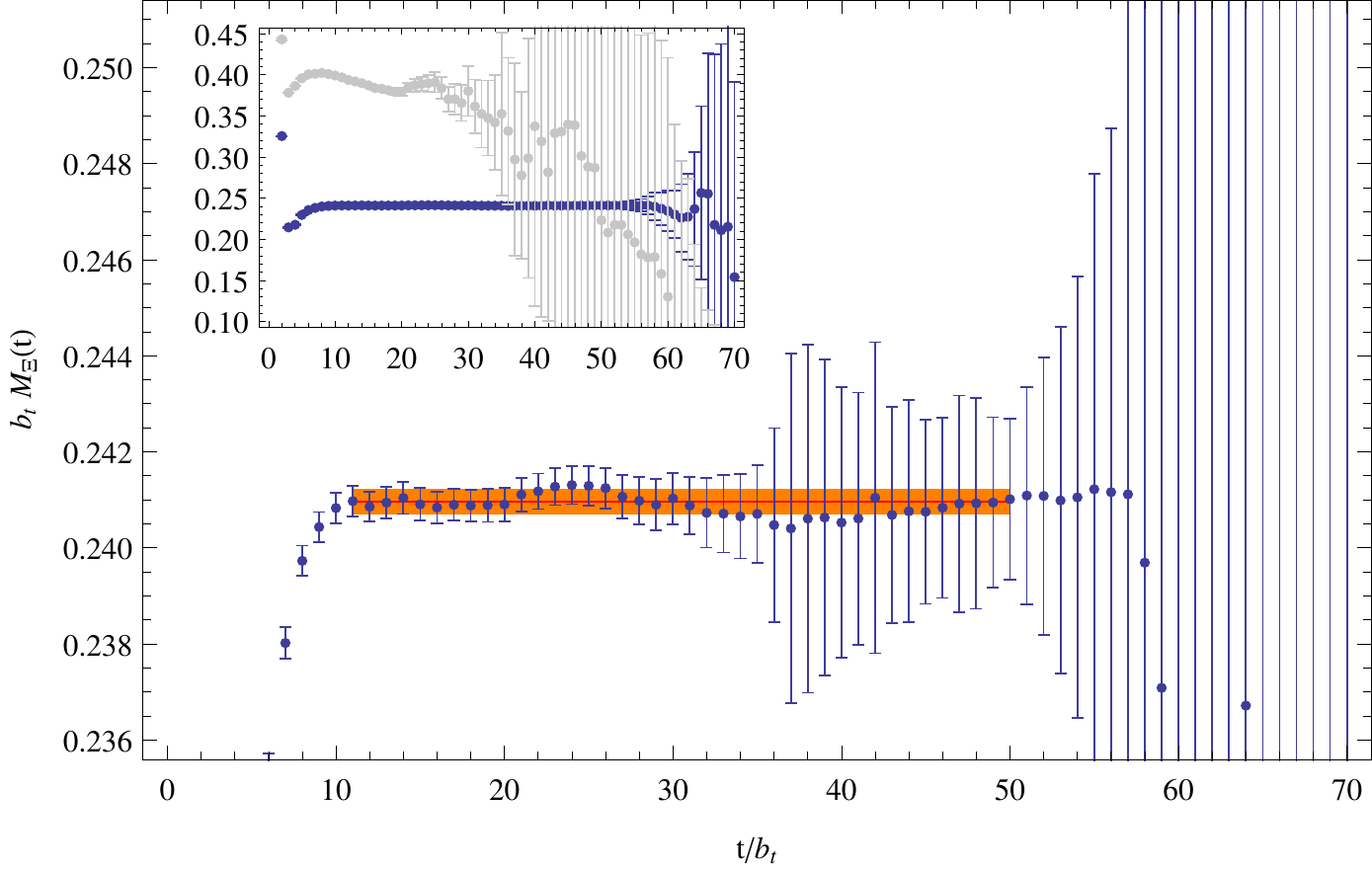}\
  \caption{\label{fig:2x2MatrixexpXi} The generalized EMP for the mass
    of the $\Xi$ using a Matrix-Prony analysis with $t_J=7$ and
    $t_W=11$, and the correlated fit to the time-slices between $t=11$
    and $t=50$. The inner (darker) region corresponds to the
    statistical uncertainty, while the outer (lighter) region
    corresponds to the statistical and fitting systematic
    uncertainties combined in quadrature. The inset shows both states
    extracted with the matrix-Prony method.}
\end{figure}
Figure~\ref{fig:2x2MatrixexpXi} shows the generalized EMP for the
$\Xi$ mass as a function of time determined with a $N=2$ matrix-Prony
extraction, using both the smeared-smeared and smeared-point
correlation functions.  The inset shows the second extracted state in
addition to the ground state. The extracted value of the $\Xi$ mass,
determined by fitting in the time interval $t=11$ to $t=50$, is
\begin{eqnarray}
  M_\Xi & = & 0.24097\pm 0.00025\pm 0.00003
  \ \ ,\ \
  \chi^2/{\rm dof} \ =\ 0.81
  \ \ \ .
  \label{eq:Neq2expFits}
\end{eqnarray}
The EM of the dominant state in fig.~\ref{fig:2x2MatrixexpXi} plateaus
around time-slice $t=10$, and is well-defined over a large interval.
In addition to being somewhat more visually appealing than the
previous Prony analyses of single correlation functions, this method
provides the smallest uncertainties, particularly for the fitting
systematic.

In our final extractions of baryon masses, our EM analysis will use
the matrix-Prony method. This method yields ground state energies that
are in complete agreement with those from the other methods discussed.
The generalized EMs from the matrix-Prony method are consistently
clean, and the quality of fits are uniformly good for the ground
state. Since they involve only one fit parameter, one can easily
assess the quality of the fits.  The procedure for fitting parameters
and determining their statistical uncertainty has been described in
Section~\ref{sec:GEMP}.  Systematic uncertainties are calculated by
performing fits over rolling windows of time-slices within the quoted
overall range and looking at the standard deviation of the central
values of those fits. This is combined in quadrature with a further
systematic uncertainty that is generated by sampling a large range of
possible values of $t_J$ and $t_W$ and taking the standard deviation
of the central values of the resulting fits.
%%%%%%%%%%%%%%%%%%%%%%%%%%%%%%%%%%%%%%%%%%%%%%%%%%%
%
% FIGURE: Matrix prony varying tJ=tW
%
%%%%%%%%%%%%%%%%%%%%%%%%%%%%%%%%%%%%%%%%%%%%%%%%%%%
\begin{figure}[!ht]
  \centering\hspace*{-1cm}
  \includegraphics[width=1.1\textwidth]{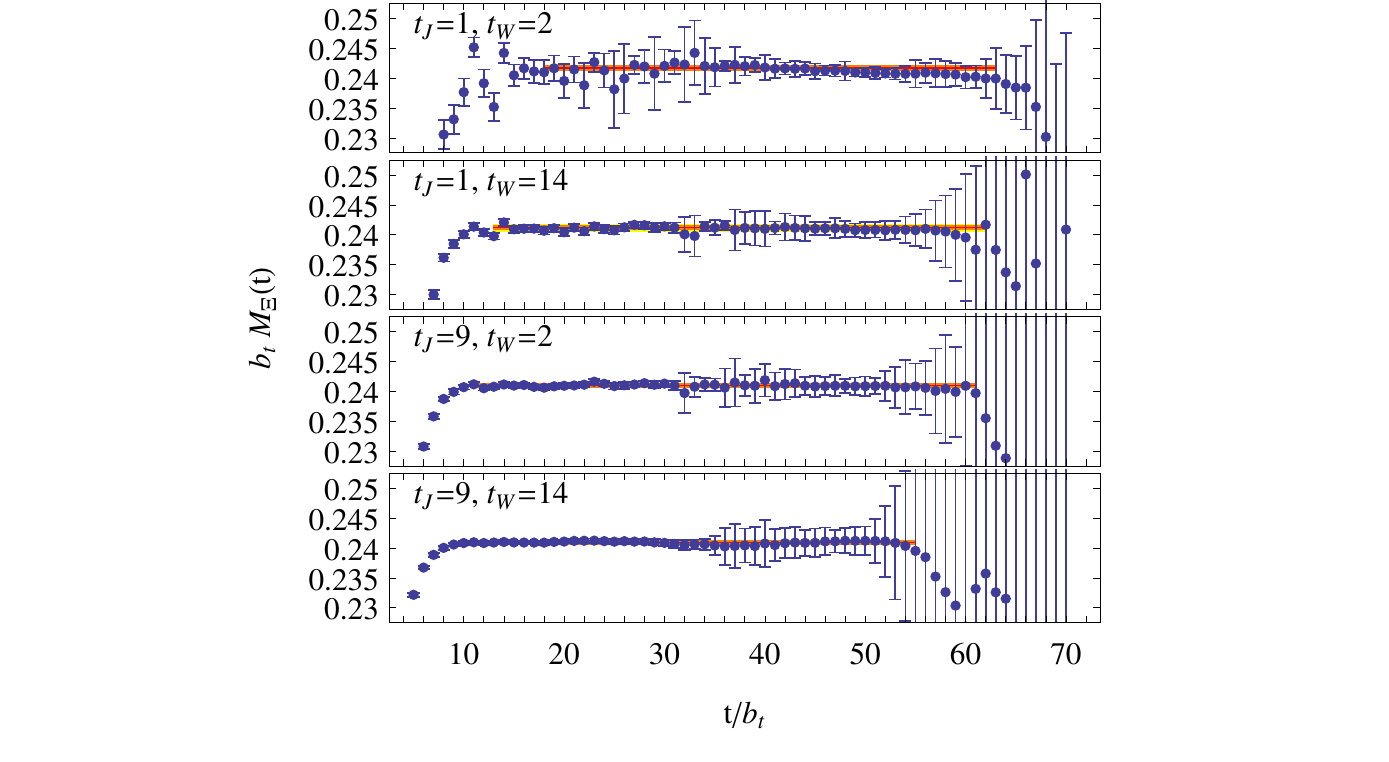}\
  \caption{\label{fig:2x2MatrixexpXiMULTI} The generalized EMP for the
    mass of the $\Xi$ using a Matrix-Prony analysis for a variety of
    values for $t_J$ and $t_W$.}
\end{figure}
The generalized EMP for the $\Xi$ extracted with the matrix-Prony
method for a variety of values of $t_W$ and $t_J$ can be seen in
fig.~\ref{fig:2x2MatrixexpXiMULTI}.

%%%%%%%%%%%%%%%%
\subsection{Prony-Histograms}
\label{sec:prony-histograms}

\noindent
In extracting the energies of the states through the Prony procedure,
a set of roots are produced on each time-slice for each member of the
Bootstrap or Jackknife ensemble.  In general these roots are real and
there is an ambiguity in associating the roots with energy-levels in
the finite volume (only the single particle masses are approximately
known).  In order to aid identification of energy-levels it is useful
to form histograms of the complete set of roots generated through the
Bootstrap procedure.  The simplest histogram is formed by accumulating
all of the roots obtained on a sub-set of, or all of, the time-slices
over all bootstrap/jackknife ensembles.  The dominant components of
the correlation function will appear as well-defined peaks in the
histogram.
%%%%%%%%%%%%%%%%%%%%%%%%%%%%%%%%%%%%%%%%%%%%%%%%%%%
%
% FIGURE: Xi Hists
%                     
%
%%%%%%%%%%%%%%%%%%%%%%%%%%%%%%%%%%%%%%%%%%%%%%%%%%%
\begin{figure}[!ht]
  \vskip0.5in \center
  \begin{tabular}{c}
    \includegraphics[width=0.99\textwidth]{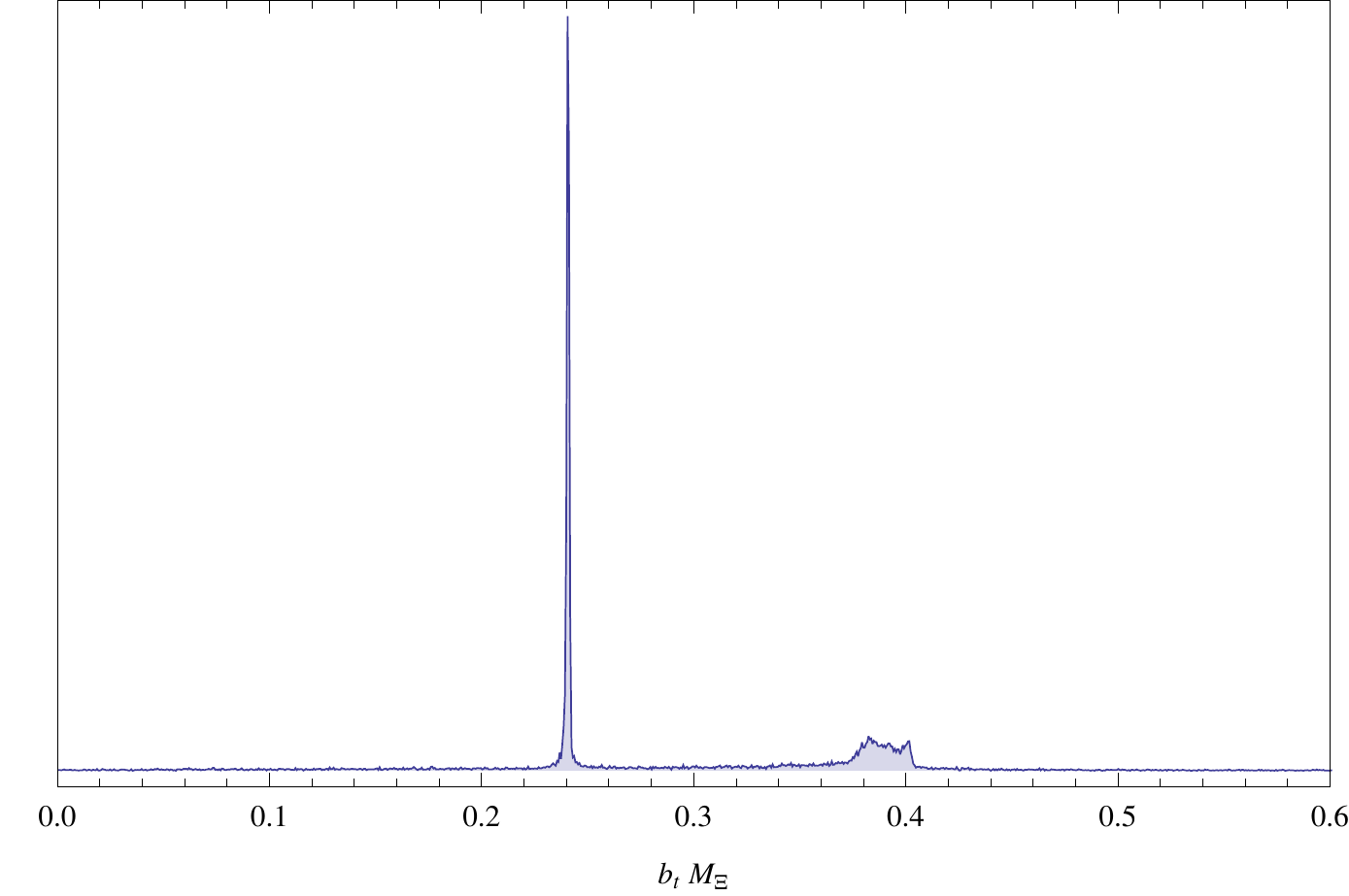}
  \end{tabular}
  \caption{\label{fig:XipsssHIST} The histogram of the positive roots
    extracted from time-slices $t=11$ to $t=50$ from $N=2$
    matrix-Prony analysis of the $\Xi$ correlation functions, with
    $t_J=7$ and $t_W=11$.}
\end{figure}

In most cases, this histogramming procedure produces very similar
results when either two, three or four exponential Prony, or
matrix-Prony analyses are used. Only atypically do the higher
exponential analyses reveal a clean state that is not present in the
two-exponential analysis. Additionally, since our baryon correlation
functions are asymmetric in time because of the parity projectors used
in Eq.~(\ref{eq:2}), noise is reduced in these histograms by
separately accumulating the roots over the two half configuration.  As
expected, the excited states have a larger presence in the
smeared-point correlation function.  An example histogram is shown in
fig. \ref{fig:XipsssHIST}, corresponding to the $\Xi$ correlation
function analyzed in fig. \ref{fig:2x2MatrixexpXi}. There is one clear
peak in the histogram, corresponding to the $\Xi$ ground-state and one
broad structure at higher mass, which the histogram suggests is likely
to be a collection of closely spaced states that currently are not
resolvable. This interpretation of the excited state is consistent
with expectations for the $\Xi$ spectrum and with the instability of
the extractions of the first excited state in the exponential fits
discussed earlier.

%%%%%%%%%%%%%%%%%%%%%%%%
\section{Meson Spectrum \label{sec:mesons}}

%%%%%%%%%%%%%%%%%%%%%%%%%%%%%%%%%%%%%%%%%%%%%%%%%%%
%
% FIGURE: Pion correlator and histogram
%
%%%%%%%%%%%%%%%%%%%%%%%%%%%%%%%%%%%%%%%%%%%%%%%%%%%
\begin{figure}[!ht]
  \vskip0.5in \center
  \begin{tabular}{c}
    \includegraphics[width=0.9\textwidth]{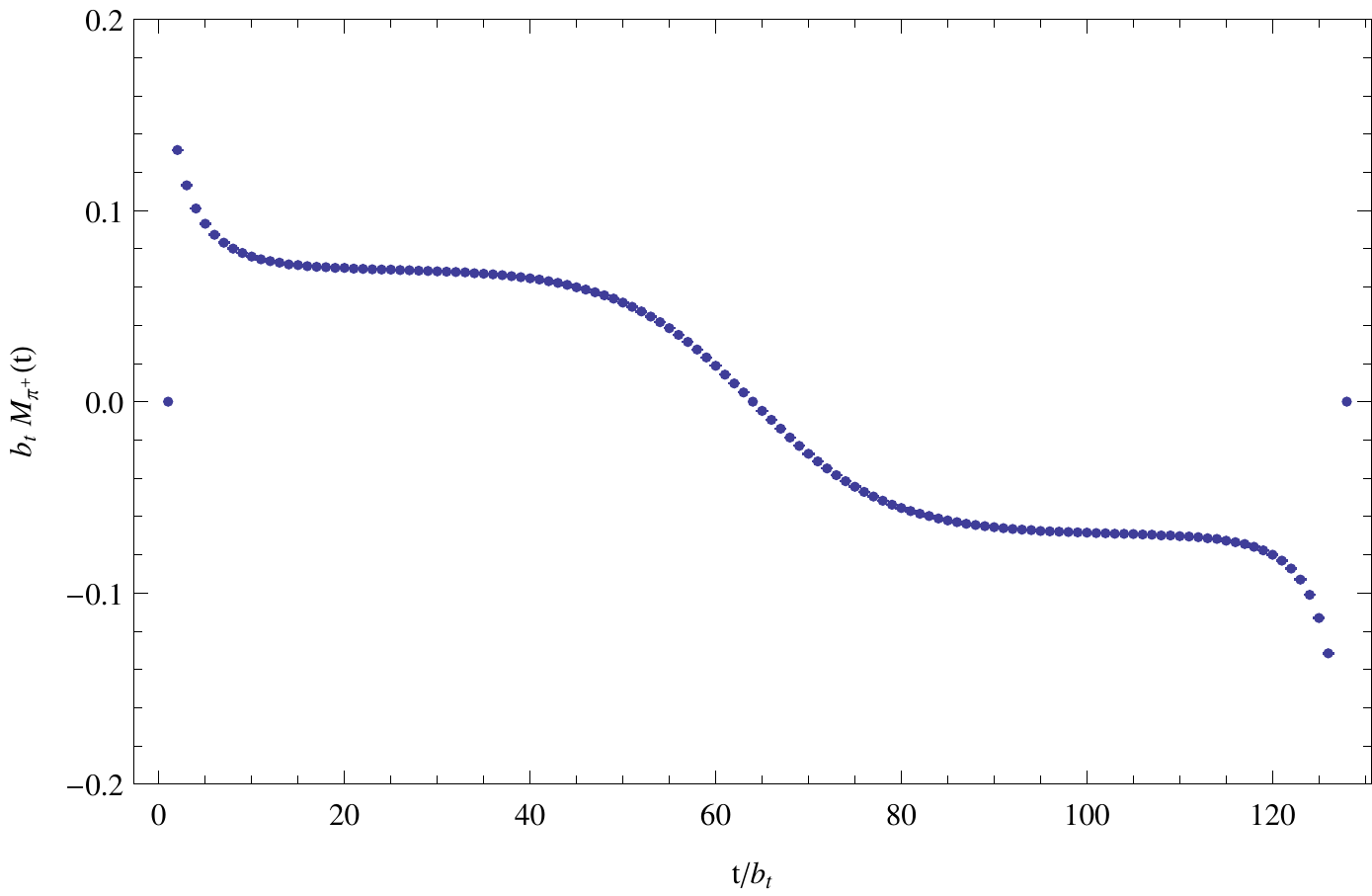}
    \\
    \includegraphics[width=0.9\textwidth]{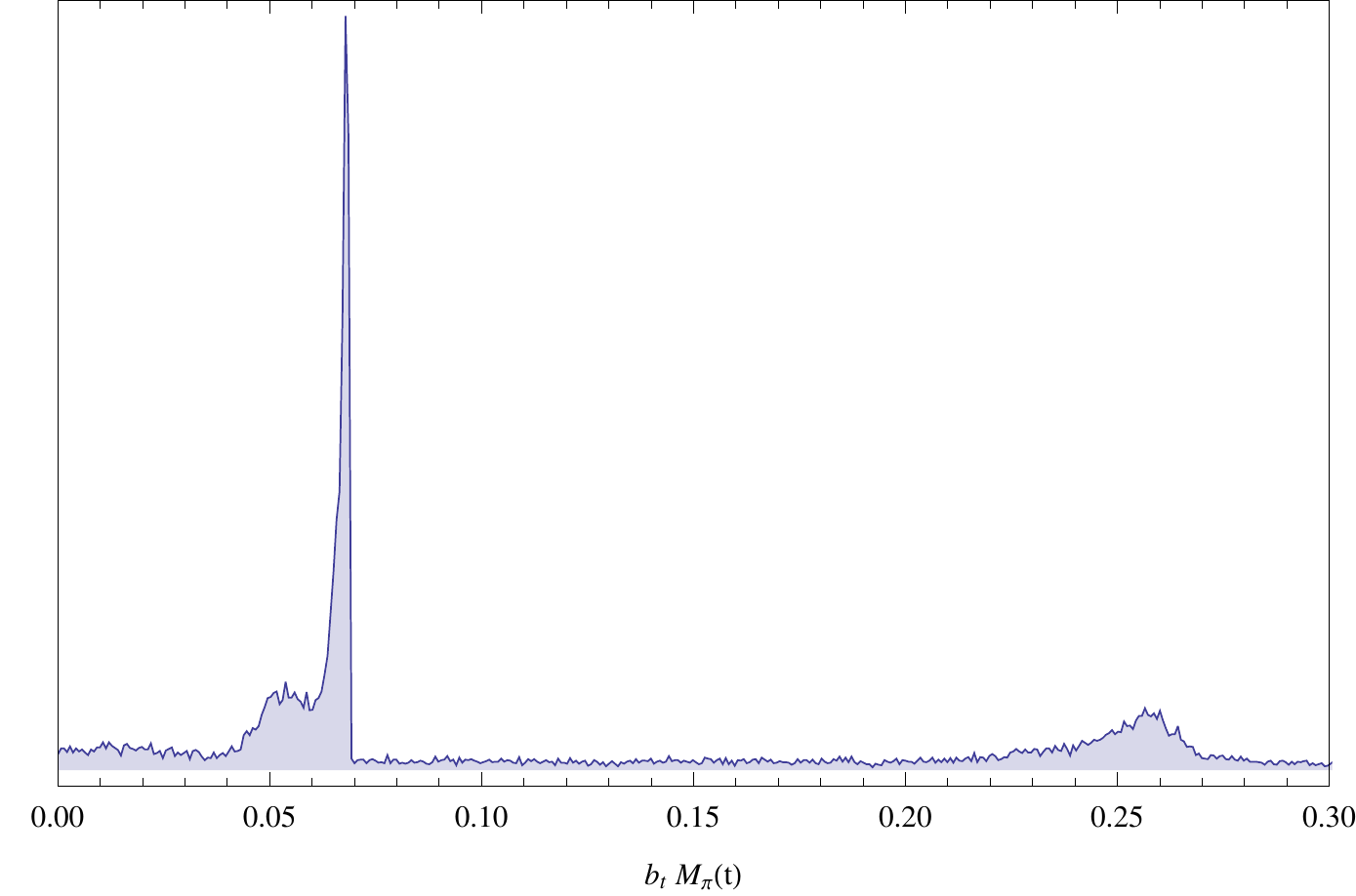} 
  \end{tabular}
  \caption{\label{fig:pionEM} The upper panel shows the EM for the
    smeared-point $\pi^+$ correlation function, while the lower panel
    shows the associated matrix-Prony histogram.  }
\end{figure}
\noindent
The $\pi^+$ correlation functions do not suffer from exponential
signal-to-noise degradation for configurations of infinite temporal
extent (in Section \ref{sec:ston}, we will find that this is not true
for a finite time-direction even for the pion).  As a result, they can
be calculated with small statistical uncertainty on each time-slice,
as shown in fig.~\ref{fig:pionEM}.  As the EM for the $\pi^+$ does not
exhibit a plateau, the $\pi^+$ mass is determined by fitting $\cosh
\left(\mpi (t-{T\over 2})\right)$ to a (large) number of time-slices
of the correlation function.  Performing a double cosh fit to
time-slices $t=21$ to $t=41$ yields
\begin{eqnarray}
  \mpi & = & 0.06936\pm 0.00012\pm 0.00005
  \ \ ,\ \ 
  \chi^2/{\rm dof} \ =\ 0.73
  \ \ \ ,
  \label{eq:pionmass}
\end{eqnarray}
where the first uncertainty is statistical and the second is fitting
systematic.  The statistical uncertainty in the mass is determined
with the Jackknife procedure, and the fitting systematic is determined
by varying the fitting interval over a reasonable range.

The second set of peaks that are visible in the histogram are at an
energy consistent with the $I=1$ $K K \pi$ state (with a threshold at
$\mpi + 2 \mK \sim 0.2636$) that can couple to the source that
produces a single $\pi^+$.  With even greater statistics, the energy
of this state could be calculated with enough precision to extract the
$I={1\over 2}$ $K\pi$ and $I=0$ $KK$ scattering lengths and the $I=1$
$KK\pi$ three-body interaction.  An expression for the energy-levels
of this system in a finite volume in terms of the $KK$ and $K\pi$
scattering amplitudes and various three-body interactions has recently
been derived \cite{Smigielski:2008pa} and would be useful in analyzing
this state.  There is no clear peak that can be associated with $I=1$
$\pi\pi\pi$, which one would naively thought would have been present.
It must be the case that the source does not couple with any
appreciable strength to this state.

\begin{figure}[!ht]
  \vskip0.5in \center
  \begin{tabular}{c}
    \includegraphics[width=0.9\textwidth]{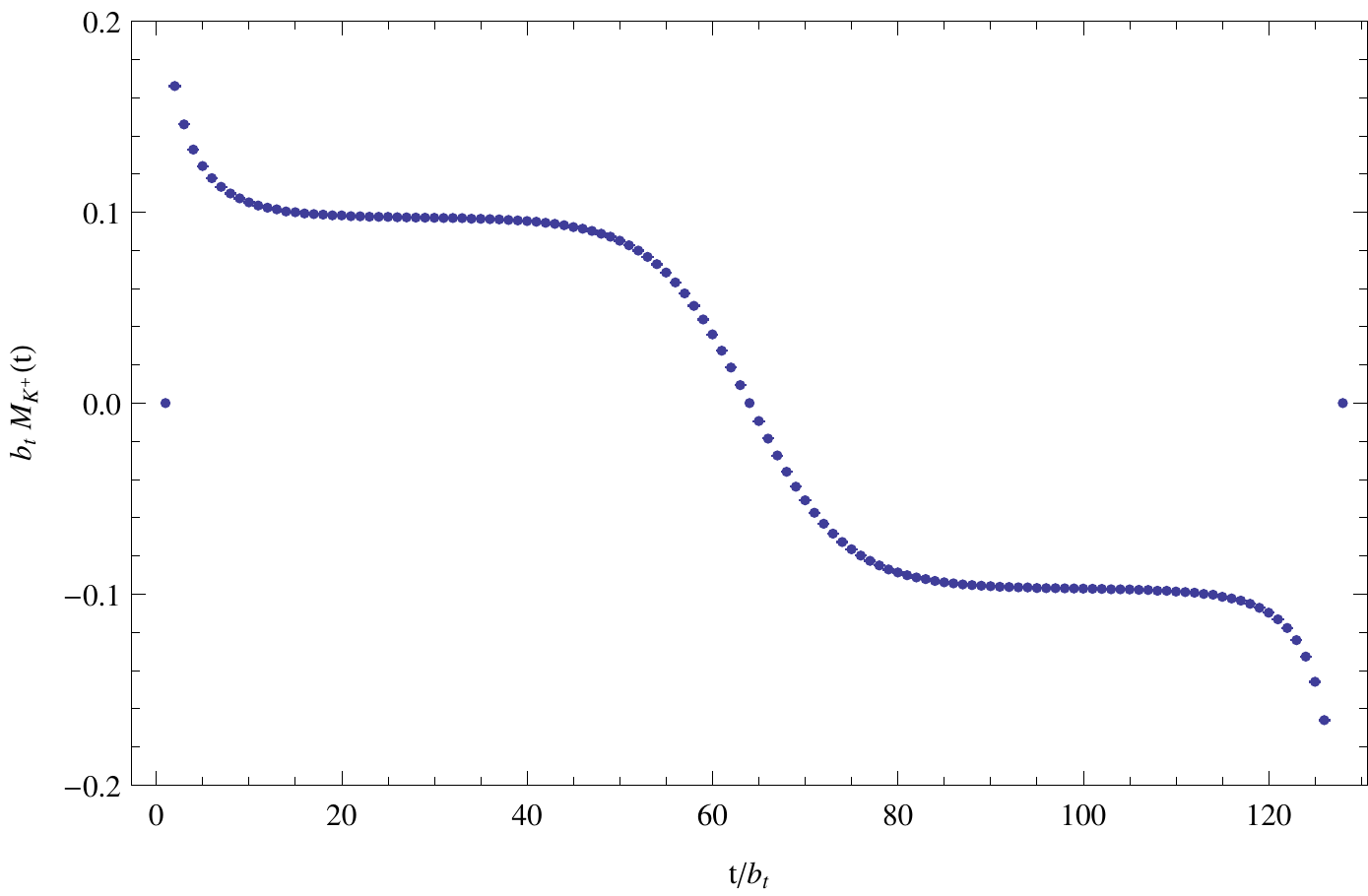}\\
    \includegraphics[width=0.9\textwidth]{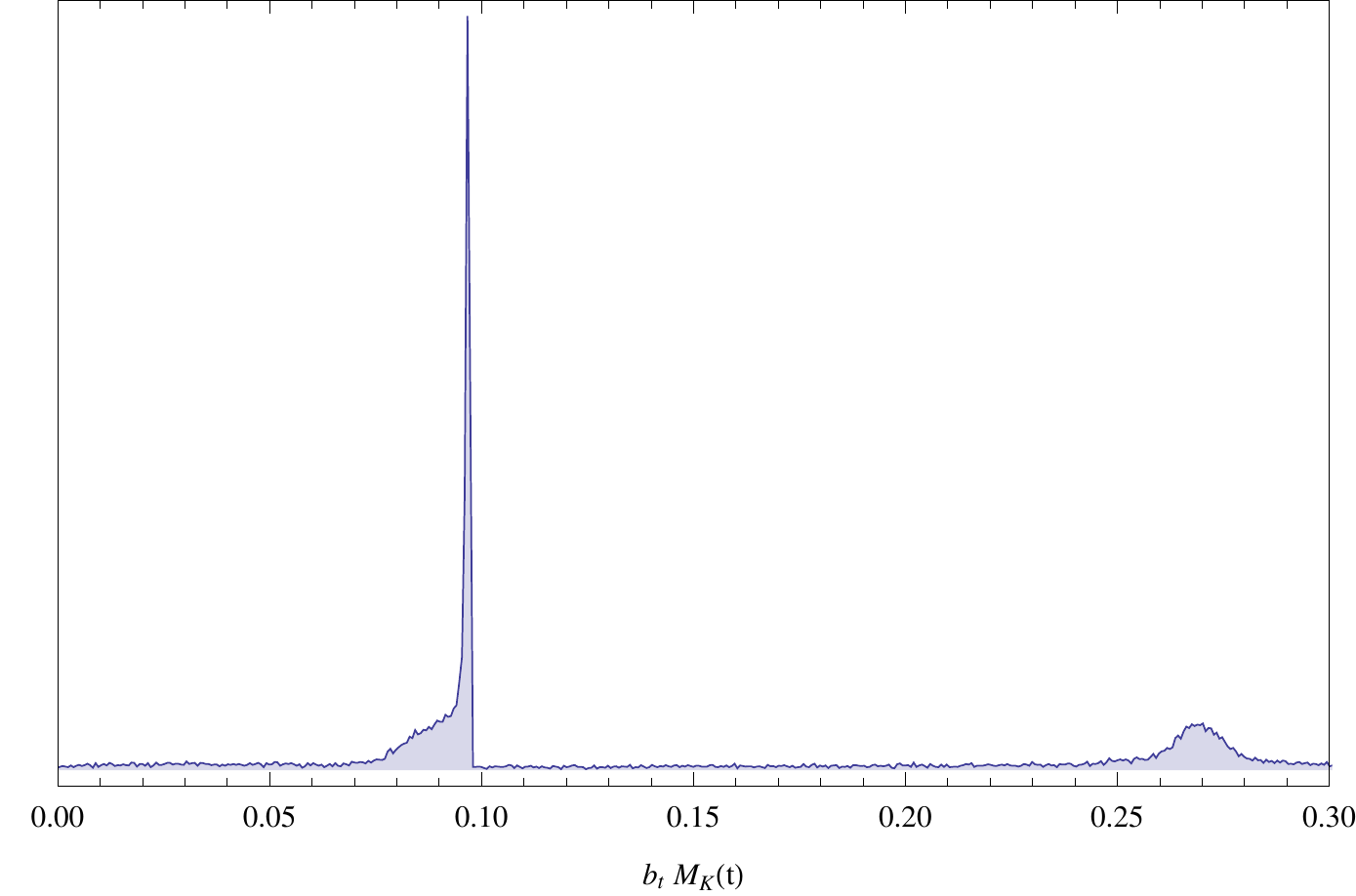} 
  \end{tabular}
  \caption{\label{fig:kaonEM} The upper panel shows the EM for the
    smeared-point $K^+$ correlation function, while the lower panel
    shows the associated Prony histogram.  }
\end{figure}
\noindent
The EM associated with the smeared-point kaon correlation function is
shown in fig.~\ref{fig:kaonEM}, along with the bootstrap-Prony
histogram.  Despite the appearance of the EM, no plateau is found in
the EM, and the kaon mass is extracted by fitting $\cosh \left(\mK
  (t-{T\over 2})\right)$ to a number of time-slices of the correlation
function.  Performing a double cosh fit over the time-slices between
$t=29$ and $t=49$, yields a $K^+$ mass of
\begin{eqnarray}
  \mK\ =\ 0.097016\pm 0.000099\pm 0.000033
  \ \ ,\ \ 
  \chi^2/{\rm dof} \ =\ 1.01  \ \ \ .
  \label{eq:kaonmass}
\end{eqnarray}

The excited state(s) that are seen in the histogram in
fig.~\ref{fig:kaonEM} are consistent with the $I={1\over 2}$ $KKK$.  A
better measurement of this state, in analogy with the pion correlation
function, would allow for a determination of the $I=0$ $KK$ scattering
amplitude.

%%%%%%%%%%%%%%%%%%%%%%%%
\section{Ground-State Baryon Spectrum \label{sec:baryons}}
\noindent
With the methodology we have presented in Section \ref{sec:methods},
we are in a position to extract the masses of the lowest-lying octet
baryons.  The $\Xi$ correlation functions have been used extensively
to demonstrate the strengths and weaknesses of the various methods,
with the resulting mass extraction given in
Eq.~(\ref{eq:Neq2expFits}), and we do not repeat that discussion here.

\begin{figure}[!ht]
  \vskip0.5in \center
  \begin{tabular}{c}
    \includegraphics[width=0.9\textwidth]{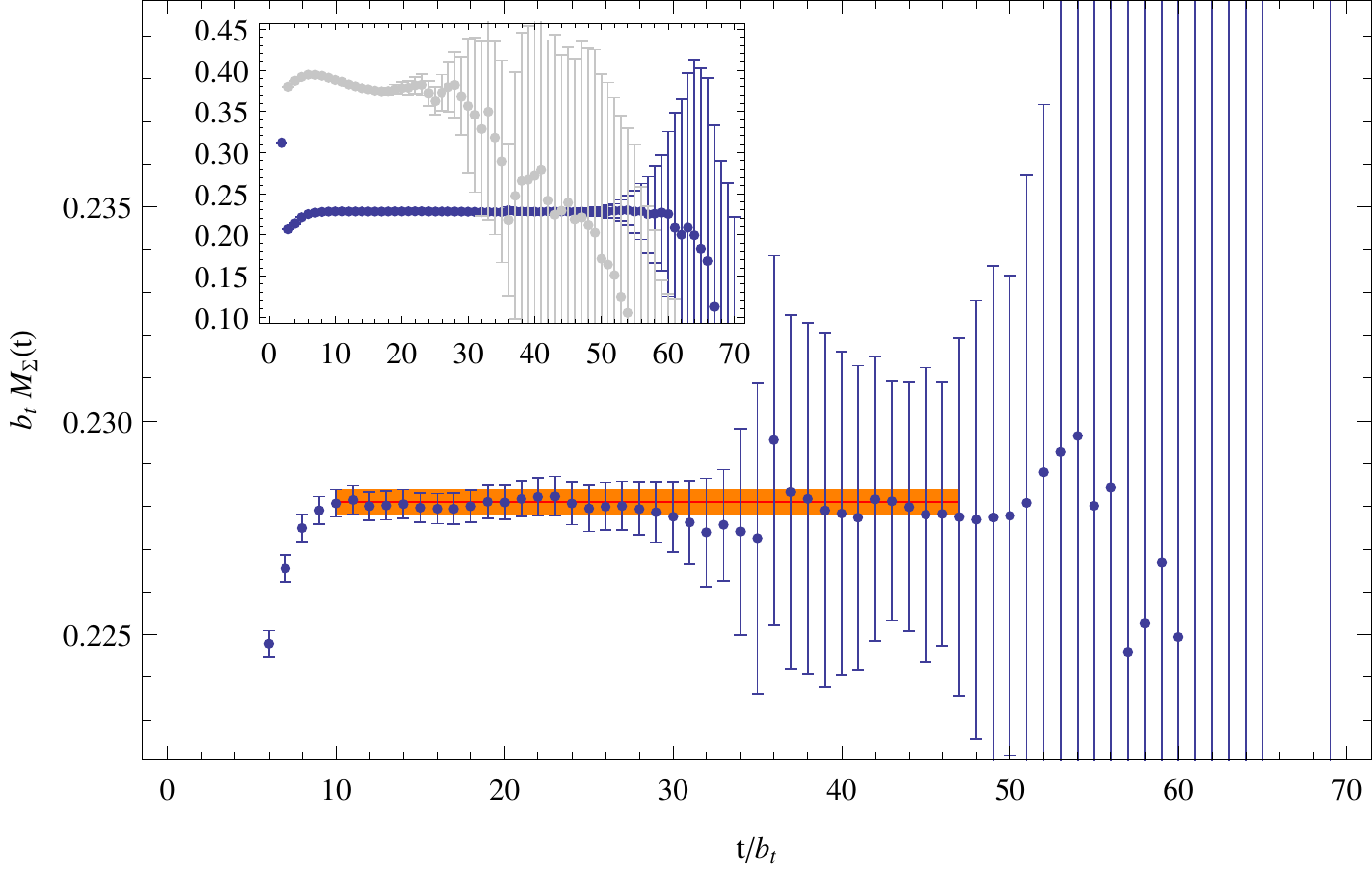}\\
    \includegraphics[width=0.9\textwidth]{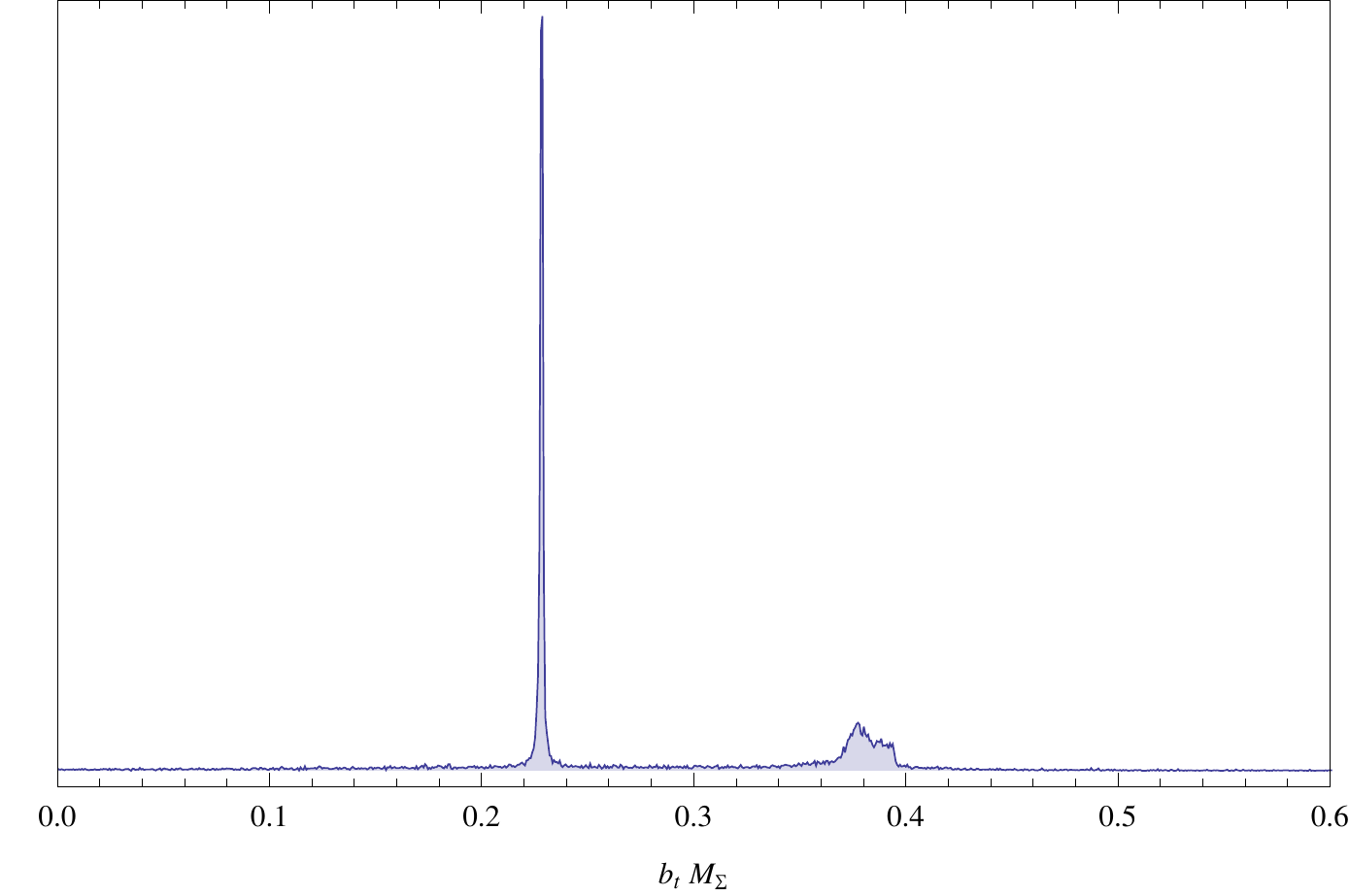}
  \end{tabular}
  \caption{\label{fig:2x2MatrixexpSigma} The upper panel shows the
    generalized EM for the mass of the $\Sigma$ using a Matrix-Prony
    analysis with $t_J=9$ and $t_W=17$, and the correlated fit to the
    time-slices between $t=12$ and $t=47$. The inner (darker) region
    corresponds to the statistical uncertainty, while the outer
    (lighter) region corresponds to the statistical and fitting
    systematic uncertainties combined in quadrature.  The inset show
    the same ground-state EM plot along with that of the excited state
    (light points).  The lower panel shows the associated Prony
    histogram of the positive roots for the time-slices $t=12$ to
    $t=47$.  }
\end{figure}

\begin{figure}[!ht]
  \vskip0.5in \center
  \begin{tabular}{c}
    \includegraphics[width=0.9\textwidth]{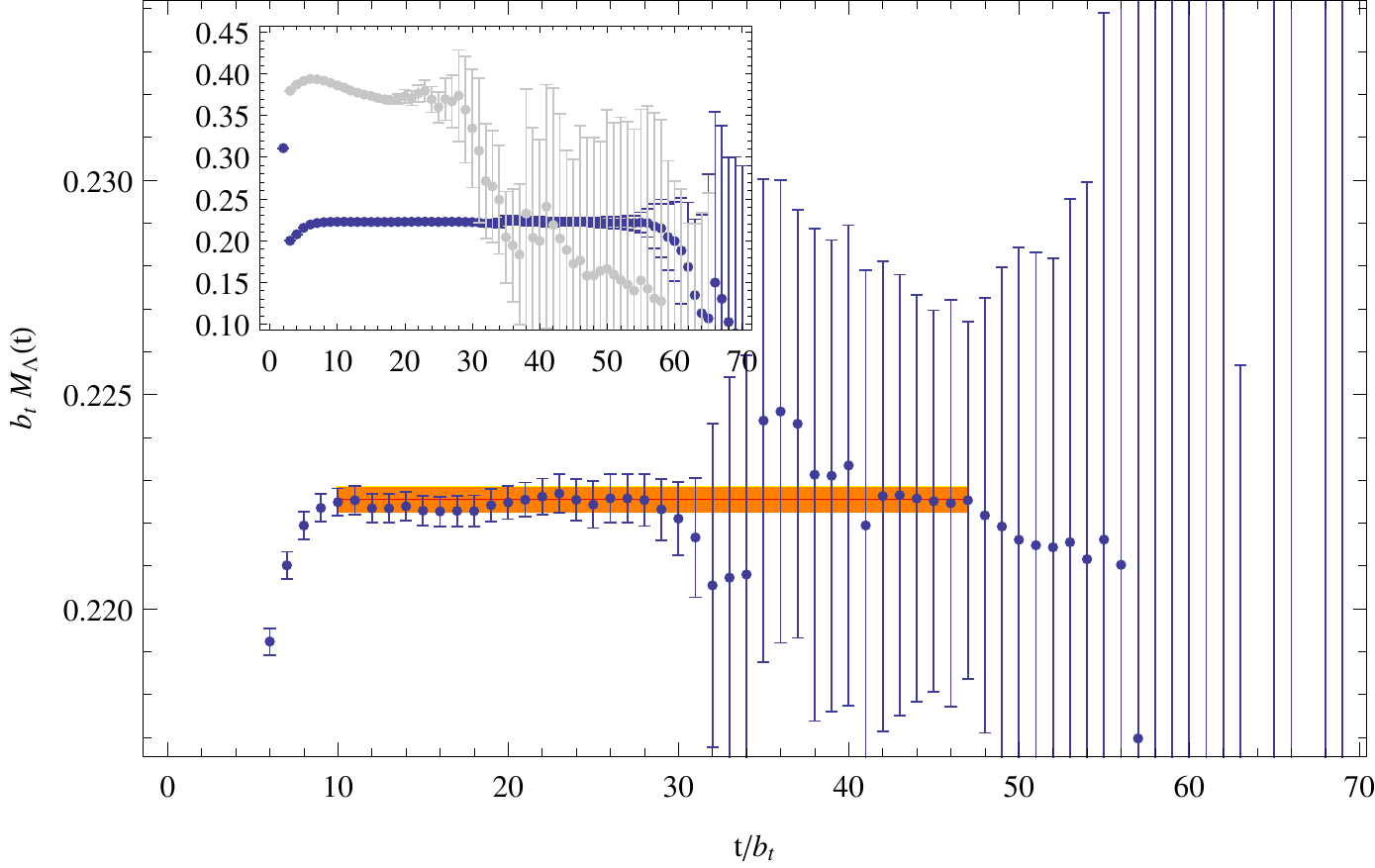}
    \\
    \includegraphics[width=0.9\textwidth]{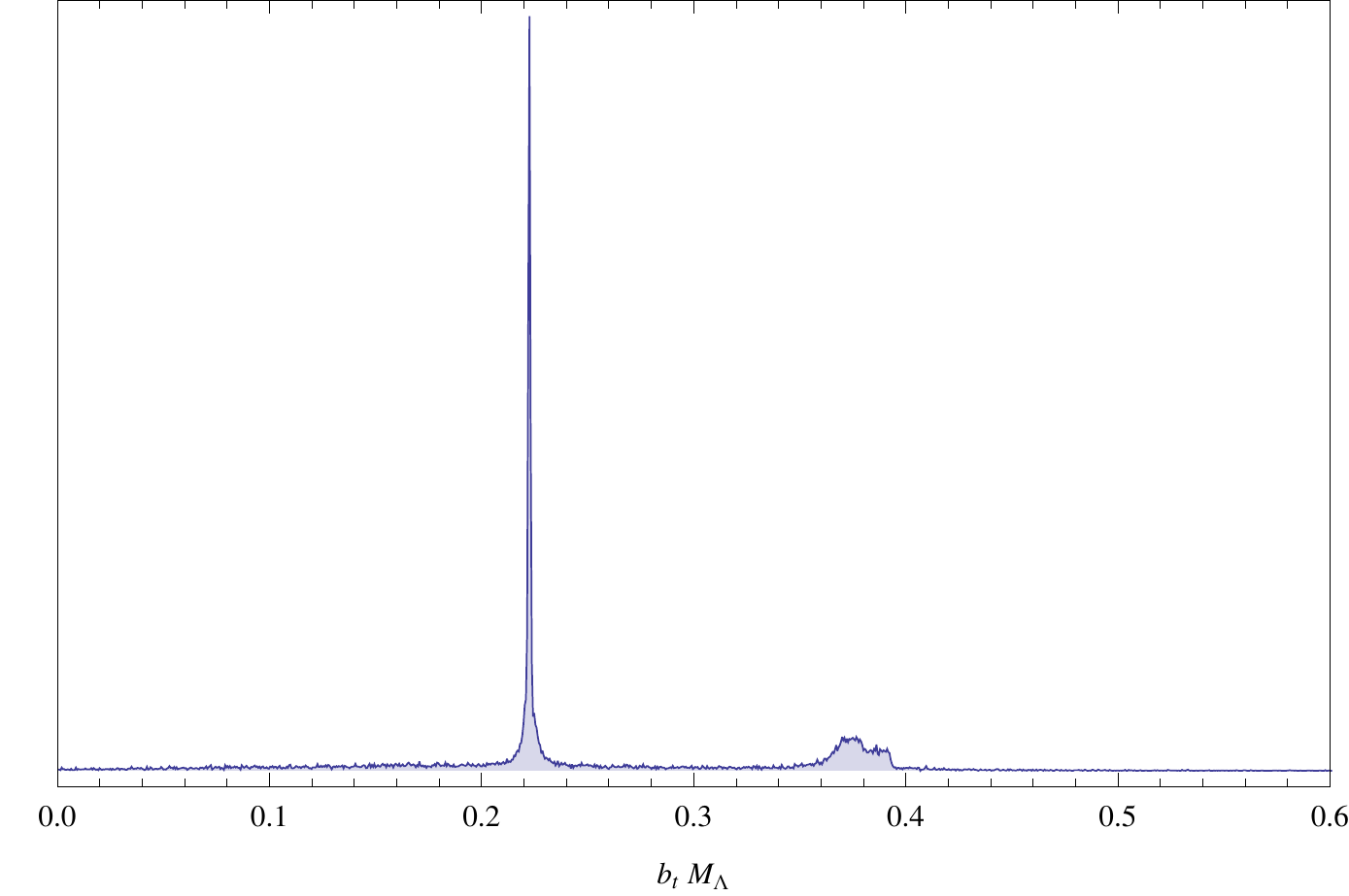}
  \end{tabular}
  \caption{\label{fig:2x2MatrixexpLambda} The upper panel shows the
    generalized EM for the mass of the $\Lambda$ using a Matrix-Prony
    analysis with $t_J=9$ and $t_W=11$, and the correlated fit to the
    time-slices between $t=10$ and $t=47$. The inner (darker) region
    corresponds to the statistical uncertainty, while the outer
    (lighter) region corresponds to the statistical and fitting
    systematic uncertainties combined in quadrature.  The inset show
    the same ground-state EM plot along with that of the excited state
    (light points).  The lower panel shows the associated Prony
    histogram of the positive roots for the time-slices $t=10$ to
    $t=47$.  }
\end{figure}

\begin{figure}[!ht]
  \vskip0.5in \center
  \begin{tabular}{c}
    \includegraphics[width=0.9\textwidth]{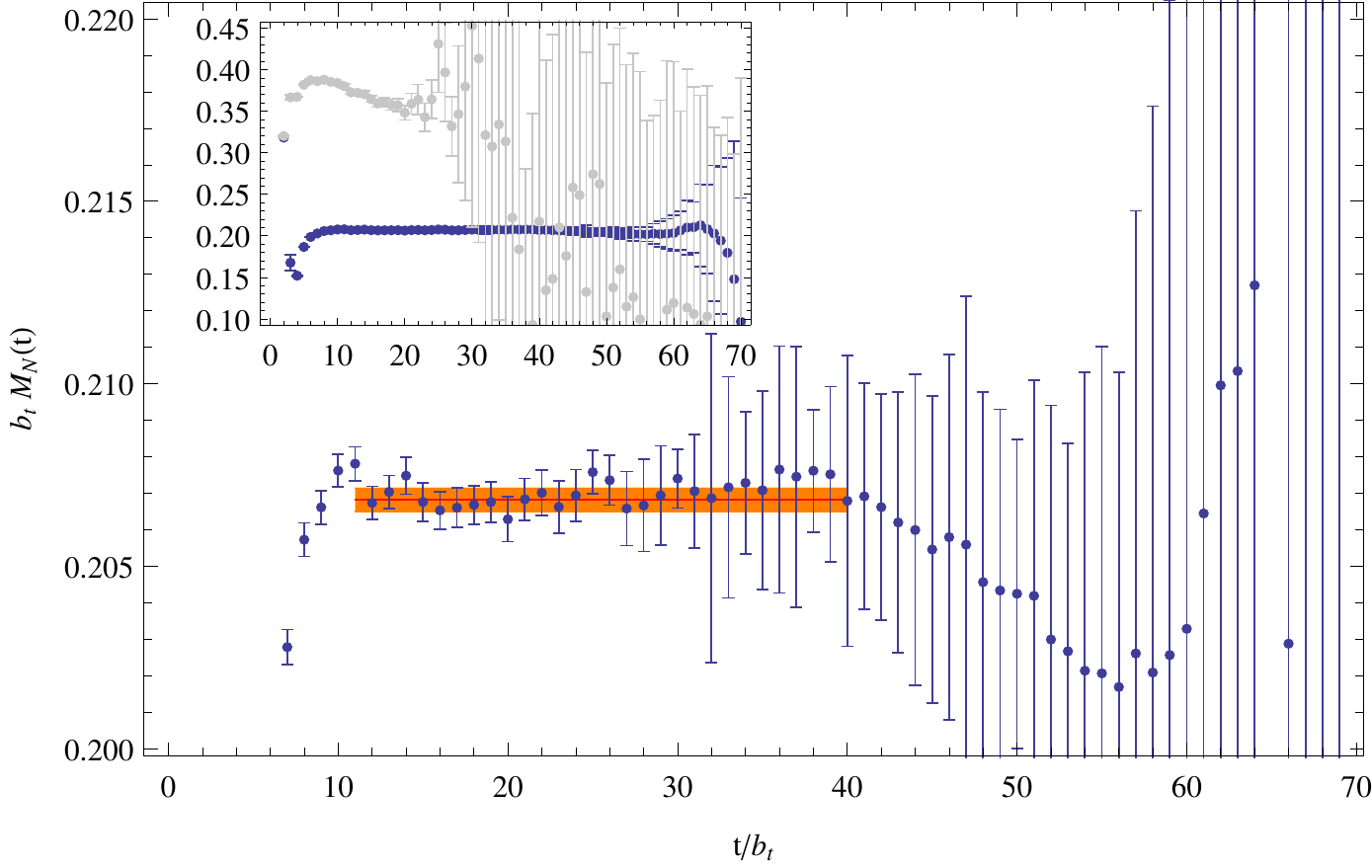}\\
    \includegraphics[width=0.9\textwidth]{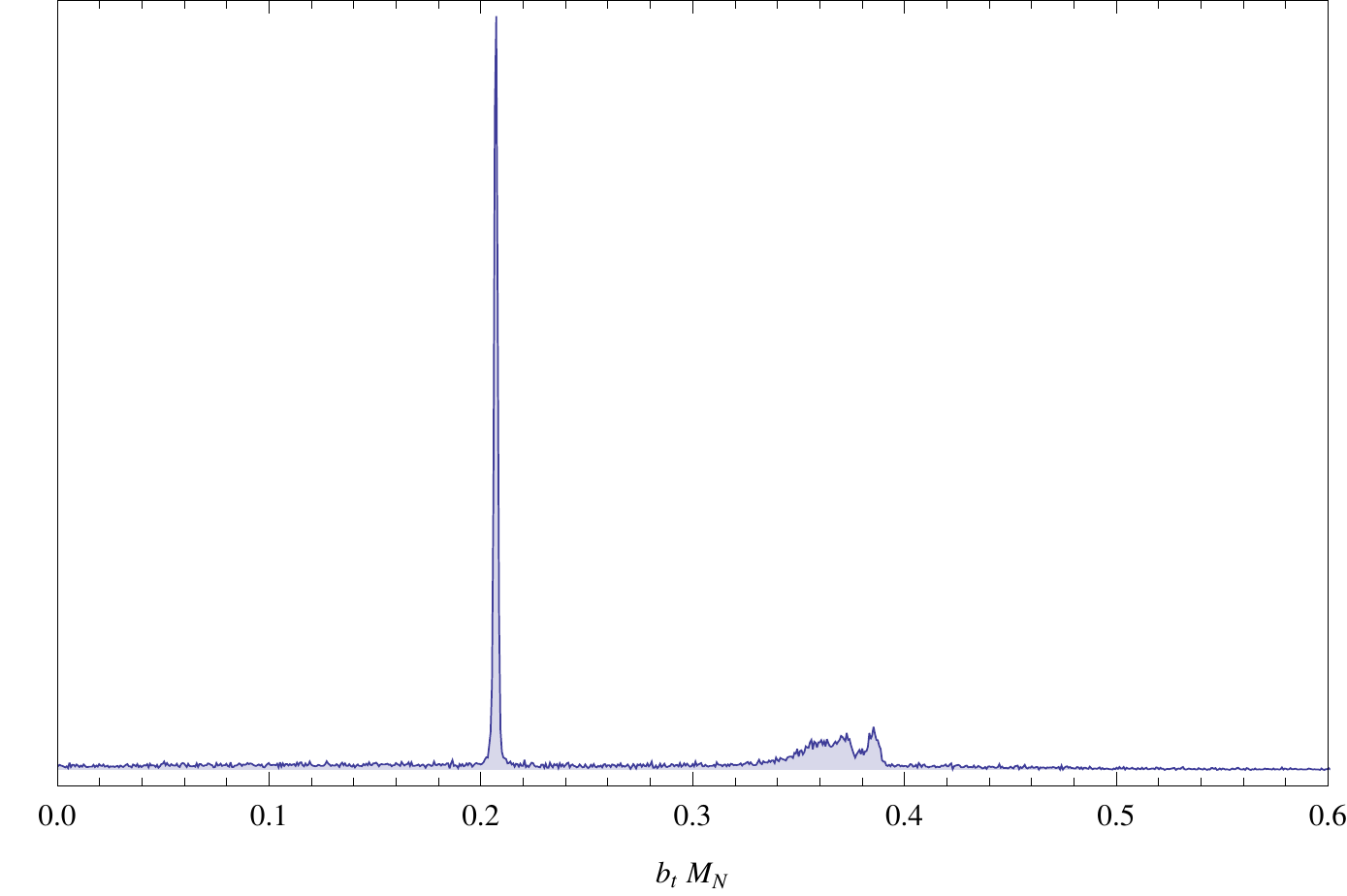}
  \end{tabular}
  \caption{\label{fig:2x2MatrixexpNucleon} The upper panel shows the
    generalized EM for the mass of the N using a Matrix-Prony analysis
    with $t_J=7$ and $t_W=2$, and the correlated fit to the
    time-slices between $t=11$ and $t=40$. The inner (darker) region
    corresponds to the statistical uncertainty, while the outer
    (lighter) region corresponds to the statistical and fitting
    systematic uncertainties combined in quadrature.  The inset show
    the same ground-state EM plot along with that of the excited state
    (light points).  The lower panel shows the associated Prony
    histogram of the positive roots for the time-slices $t=11$ to
    $t=40$.  }
\end{figure}

The matrix-Prony method applied to the smeared-smeared and
smeared-point correlation functions associated with the $\Sigma$,
$\Lambda$ and $N$ produces the Prony-histograms and generalized EMs
shown in fig.~\ref{fig:2x2MatrixexpSigma},
fig.~\ref{fig:2x2MatrixexpLambda}, and
fig.~\ref{fig:2x2MatrixexpNucleon}.
%%%%%%%%%%%%%%%%%%%%%%%%%%%%%%%%%%%%%%%%%%%%%%%%%%%
%
% FIGURE: Sigma Matrix prony tJ=tW=10
%
%%%%%%%%%%%%%%%%%%%%%%%%%%%%%%%%%%%%%%%%%%%%%%%%%%%
Fitting the $\Sigma$ EM between time-slices $t=12$ to $t=47$ yields
$\Sigma$ mass,
\begin{eqnarray}
  M_\Sigma & = & 0.22811\pm 0.00028\pm 0.00018
  \ \ ,\ \ 
  \chi^2/{\rm dof} \ =\ 0.77
  \ \ \ .
  \label{eq:SigGS}
\end{eqnarray}
%
%%%%%%%%%%%%%%%%%%%%%%%%%%%%%%%%%%%%%%%%%%%%%%%%%%%
%
% FIGURE: Lambda Matrix prony tJ=tW=10
%
%%%%%%%%%%%%%%%%%%%%%%%%%%%%%%%%%%%%%%%%%%%%%%%%%%%
Fitting the $\Lambda$ EM between time-slices $t=12$ to $t=52$ yields
$\Lambda$ mass,
\begin{eqnarray}
  M_\Lambda & = & 0.22255\pm 0.00028\pm 0.00005
  \ \ ,\ \ 
  \chi^2/{\rm dof} \ =\ 1.21
  \ \ \ .
  \label{eq:LamGS}
\end{eqnarray}
%
%
%%%%%%%%%%%%%%%%%%%%%%%%%%%%%%%%%%%%%%%%%%%%%%%%%%%
%
% FIGURE: Nucleon Matrix prony tJ=tW=10
%
%%%%%%%%%%%%%%%%%%%%%%%%%%%%%%%%%%%%%%%%%%%%%%%%%%%
Finally, fitting N the EM between time-slices $t=11$ to $t=40$ yields
N mass,
\begin{eqnarray}
  M_N & = & 0.20682\pm 0.00032\pm 0.00010
  \ \ ,\ \ 
  \chi^2/{\rm dof} \ =\ 1.5
  \ \ \ .
  \label{eq:NucGS}
\end{eqnarray}

The results of the best extractions of the ground-state baryon masses
using multi-exponential fitting and the matrix-Prony method, which
give consistent results for each species of baryon, are collected in
Table~\ref{tab:nExpResults}. These results are completely consistent
within their uncertainties, giving us confidence that our extractions
are correct.
%%%%%%%%%%%%%%%%
%% TABLE FINAL EXP MASSES
\begin{table}[!ht]
  \caption{\label{tab:nExpResults}{
      The ground-state masses of the $J^\pi={1\over 2}^+$ baryons extracted by
      fitting four exponentials and by the matrix-Prony
      method. The first uncertainty is statistical 
      while the second is the fitting systematic.}} 
  \begin{ruledtabular}
    \begin{tabular}{c|cccc|ccc}
      & & Exponential Fitting & & & & Matrix-Prony &\\
      \hline
      state & $b_t M$ &  range & $\chi^2/{\rm dof}$ & Q &  $b_t M$ & range  & $\chi^2/{\rm dof}$\\
      \hline
      N & 0.20693(33)(07) &  7--64 & 0.72 & 0.99 & 0.20682(32)(10)& 11--40 & 1.50\\
      $\Lambda$ & 0.22265(25)(16) &  9--64 & 0.89 & 0.78  & 0.22255(28)(5) &
      10--47 & 1.21 \\
      $\Sigma$ & 0.22819(25)(07) &  8--64 & 0.85 & 0.86  & 0.22811(28)(18) & 12--47 &0.77\\
      $\Xi$ & 0.24112(21)(06) &  7--64 & 0.84 & 0.87  & 0.24097(25)(3) & 11--50 & 0.81
    \end{tabular}
  \end{ruledtabular}
\end{table}
%%%%%%%%%%%%%%%%

%%%%%%%%%%%%%%%%%%%%%%%%
\section{Negative-Parity Excited Baryon States \label{sec:odds}}

\noindent
The interpolating operators that produce even-parity baryons moving
forward in time also produce negative-parity partners moving backwards
in time.  As the interpolating operators couple to continuum states
such as $N\pi$, it is possible that, by using L\"uscher's method (and
ideally, multiple spatial volumes), the phase-shifts for meson-baryon
scattering can be extracted in channels with contributions from
disconnected diagrams.

In addition to excited single baryon states, and the continuum states
that carry zero units of momentum in the volume, there are also
continuum states where each hadron carries one or more units of
momentum in the volume, while having vanishing total momentum.  The
lowest energy state containing hadrons $A$ and $B$ with back-to-back
momenta $\pm{\bf p}=\pm\frac{2\pi}{L}{\bf n}$ (where ${\bf n}$ is an
integer triplet) occurs at
\begin{eqnarray}
  E_{AB}^{|\bf n|}=\sqrt{M_A^2 + \left({2\pi |{\bf n}|\over
        L\xi_t}\right)^2}\ +\  
  \sqrt{M_B^2 + \left({2\pi |{\bf n}|\over L\xi_t}\right)^2}
  \ \ \ .
\end{eqnarray}
In attempting to unravel the spectrum of states contributing to the
correlation functions, we must also consider such continuum states.

%
%%%%%%%%%%%%%%%%%%%%%%%%%%%%%%%%%%%%%%%%%%%%%%%%%%%
%
% FIGURE: Odd Parity Nucleon Matrix prony tJ=tW=3
%
%%%%%%%%%%%%%%%%%%%%%%%%%%%%%%%%%%%%%%%%%%%%%%%%%%%
\begin{figure}[!ht]
  \vskip0.5in \center
  \begin{tabular}{c}
    \includegraphics[width=0.9\textwidth]{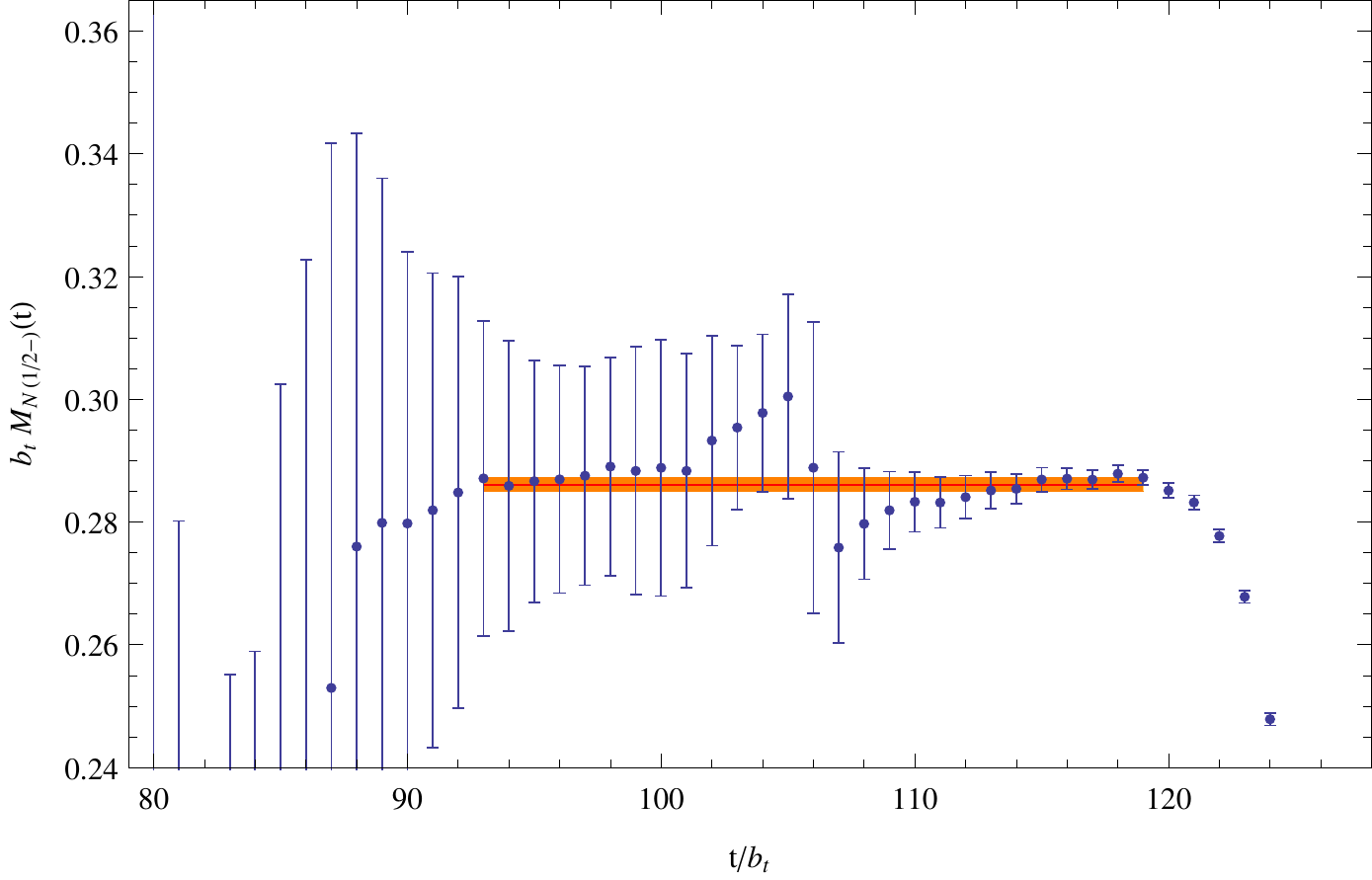}\\
    \includegraphics[width=0.9\textwidth]{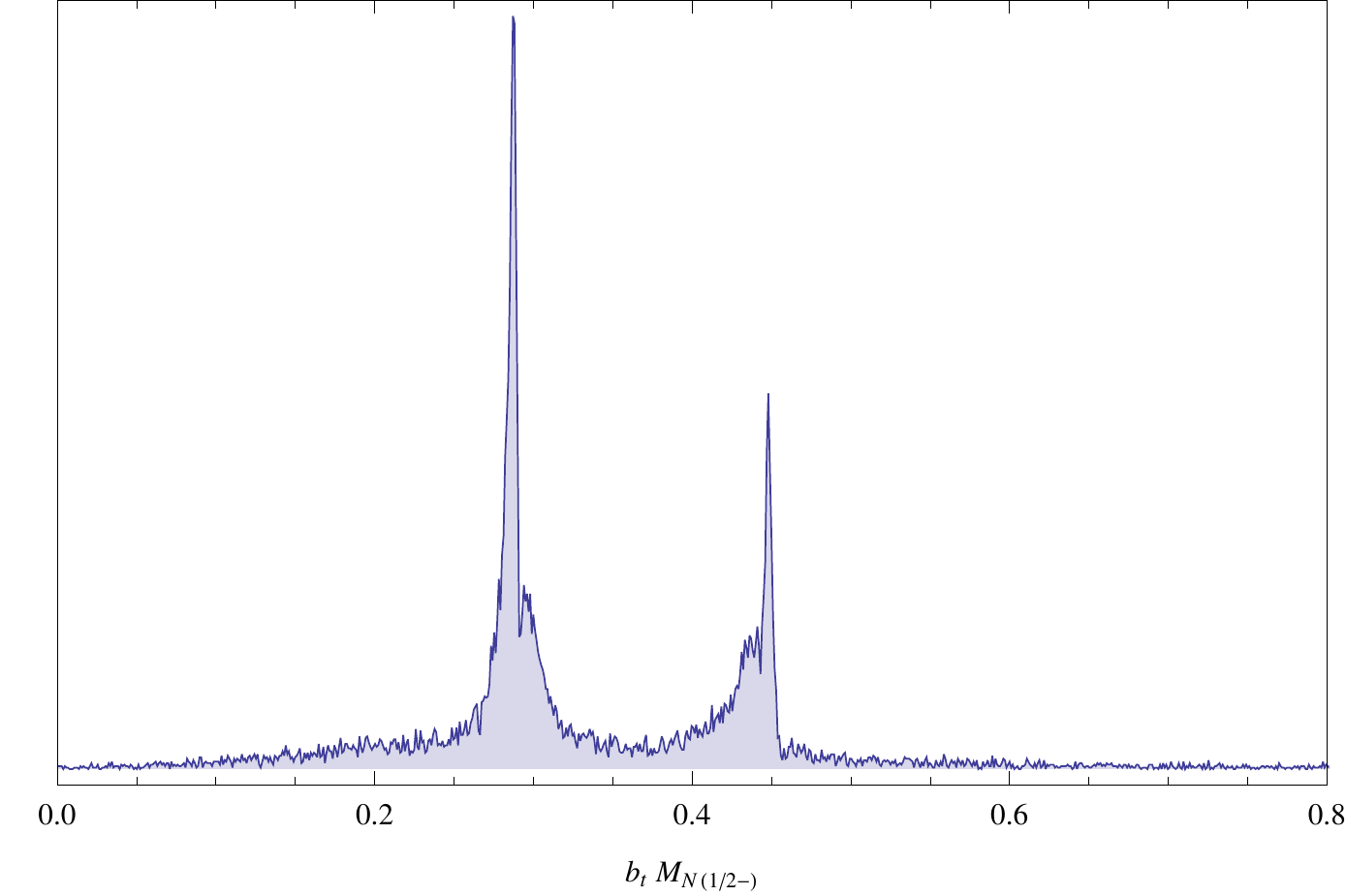}
  \end{tabular}
  \caption{\label{fig:2x2MatrixexpOddNucleon} The upper panel shows
    the generalized EM for the lowest-lying negative-parity state
    coupling to the N-source using a Matrix-Prony analysis with
    $t_J=5$ and $t_W=8$, and the correlated fit to the time-slices
    between $t=93$ and $t=119$. The inner (darker) region corresponds
    to the statistical uncertainty, while the outer (lighter) region
    corresponds to the statistical and fitting systematic
    uncertainties combined in quadrature.  The lower panel shows the
    associated Prony histogram of the positive roots for the
    time-slices $t=93$ to $t=119$.  }
\end{figure}
The lowest-lying negative parity state that is expected to couple to
the interpolating operator for the single nucleon is the $s$-wave
$N\pi$ state (more precisely, we refer to the $A_1^+$ representation
of the hyper-cubic group), which has a threshold, neglecting
interactions, of $\mpi+M_N = 0.27618\pm 0.00034\pm 0.00011$.  Fitting
the EM shown in fig.~\ref{fig:2x2MatrixexpOddNucleon} between
time-slices $t=93$ to $t=119$ yields,
\begin{eqnarray}
  E_{N\pi} & = & 0.2861\pm 0.0011\pm 0.0020
  \ \ ,\ \ 
  \chi^2/{\rm dof} \ =\ 0.91
  \ \ \ ,
  \label{eq:ODDNuc}
\end{eqnarray}
significantly above threshold.
Therefore, we conclude that this state is an $s$-wave $\pi N$
scattering state with isospin $I={1\over 2}$, as it has energy
considerably below that of the first momentum excitation in the volume
at $E_{N\pi}^{|{\bf n}|=1} = 0.33889\pm 0.00042\pm 0.00014$, and the
$\pi\pi N$ state.  Given that there are no other channels that are
energetically allowed for this state to mix with, the $s$-wave $\pi N$
phase-shift\footnote{Here we ignore possible contributions from
  $L=4,6,\ldots$ partial waves that also contribute in the $A_1^+$
  representation of the cubic group H(3).} in this channel can
be extracted at an energy of $\delta E_{\pi N} = E_{N ({1\over 2}^-)}
- M_N - \mpi\ =\ 0.0095\pm 0.0011\pm 0.0020$ ($\delta E_{\pi N} =
15.3\pm 1.8\pm 3.2~{\rm MeV}$) where the uncertainties are dominated
by the uncertainty in $E_{N ({1\over 2}^-)}$. It is important to note
that this channel receives contributions from disconnected diagrams,
and in the calculation we are doing, these contributions are
completely accounted for in the gauge-configurations.  Using the
standard L\"uscher procedure, a phase-shift of $\delta_{\pi N}=-26\pm
7 \pm 6$ degrees is found at this energy.  The Prony histogram in
fig.~\ref{fig:2x2MatrixexpOddNucleon} shows significant structure in
this channel, and one could argue that there is a single level at
$E\sim 0.45$, but this would require further exploration.

%
%%%%%%%%%%%%%%%%%%%%%%%%%%%%%%%%%%%%%%%%%%%%%%%%%%%
%
% FIGURE: Odd Parity Lambda Matrix prony tJ=tW=3
%
%%%%%%%%%%%%%%%%%%%%%%%%%%%%%%%%%%%%%%%%%%%%%%%%%%%
\begin{figure}[!ht]
  \vskip0.5in \center
  \begin{tabular}{c}
    \includegraphics[width=0.9\textwidth]{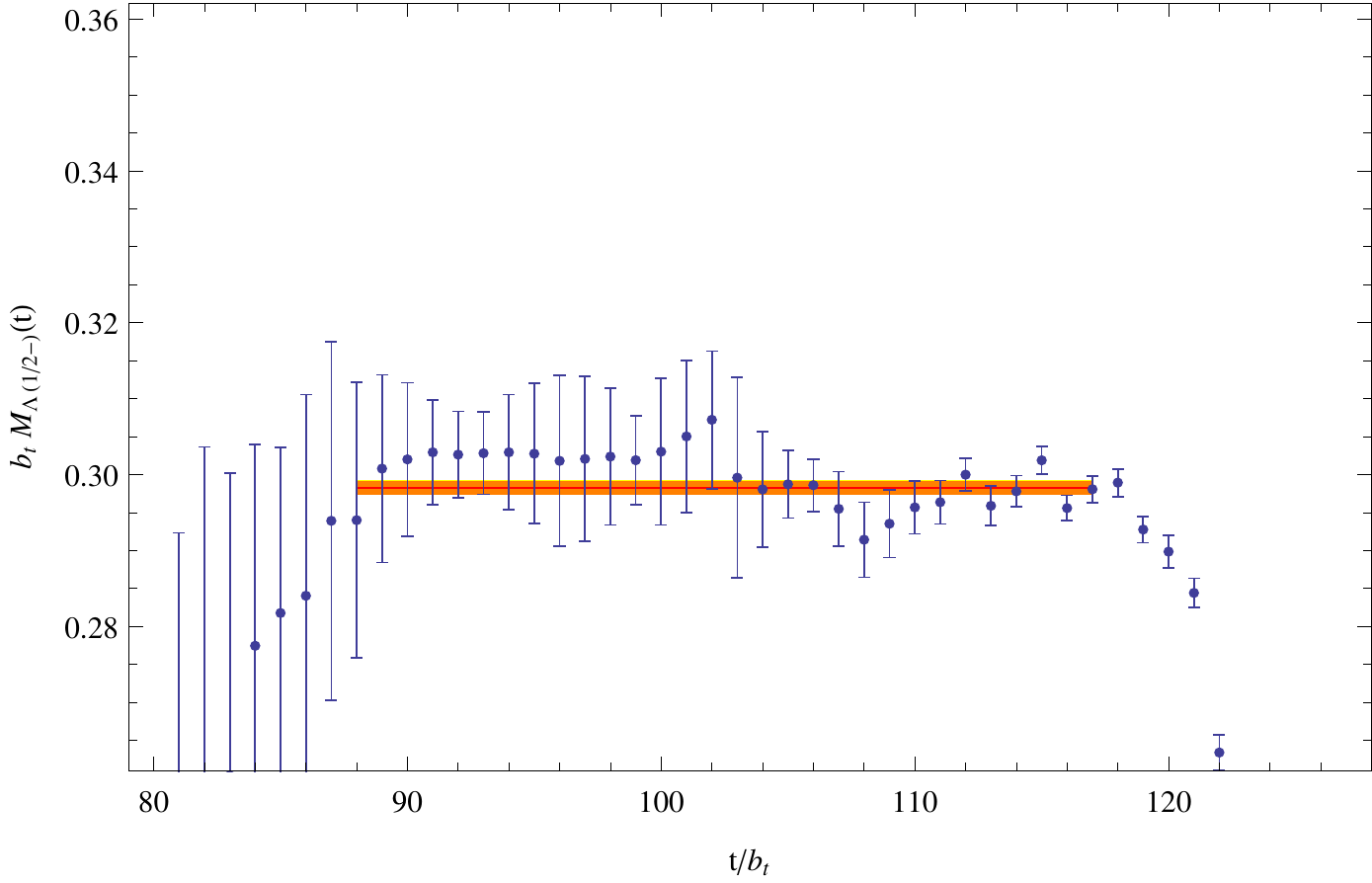}\\
    \includegraphics[width=0.9\textwidth]{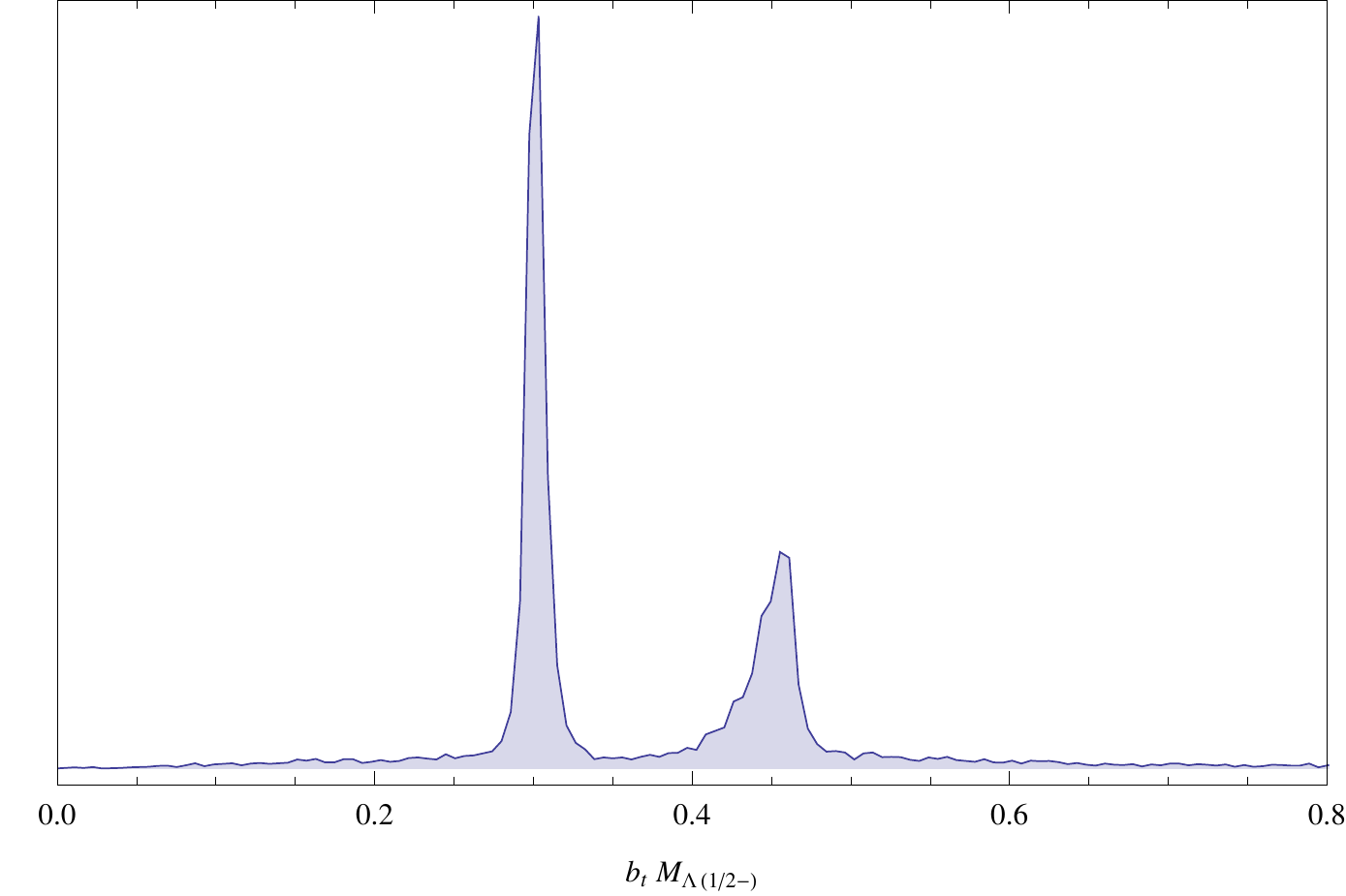}
  \end{tabular}
  \caption{\label{fig:2x2MatrixexpOddLambda} The upper panel shows the
    generalized EM for the lowest-lying negative-parity state coupling
    to the $\Lambda$-source using a Matrix-Prony analysis with $t_J=1$
    and $t_W=5$, and the correlated fit to the time-slices between
    $t=88$ and $t=117$. The inner (darker) region corresponds to the
    statistical uncertainty, while the outer (lighter) region
    corresponds to the statistical and fitting systematic
    uncertainties combined in quadrature.  The lower panel shows the
    associated Prony histogram of the positive roots for the
    time-slices $t=88$ to $t=117$.  }
\end{figure}
For the negative-parity state that couples to the interpolating
operators for the $\Lambda$, the situation is not so clean.  The
thresholds for the lowest-lying continuum states, $\Sigma\pi$ and
$NK$, are located at $M_\Sigma+\mpi = 0.29747\pm 0.00030 \pm 0.00019$
and $M_N+M_K=0.30384\pm 0.00033 \pm 0.00011$, respectively, in the
absence of interactions.  The corresponding lowest-lying states with
one unit of momentum occur at $E_{\Sigma\pi}^{|{\bf n}|=1} =
0.35857\pm 0.00037\pm 0.00023$, and $E_{NK}^{|{\bf n}|=1} = 0.35763\pm
0.00039\pm 0.00012$.  Fitting the EM shown in
fig.~\ref{fig:2x2MatrixexpOddLambda} between time-slices $t=88$ to
$t=117$ yields,
\begin{eqnarray}
  E_{\Lambda ({1\over 2}^-)} & = & 0.2983\pm 0.0008\pm 0.0004
  \ \ ,\ \ 
  \chi^2/{\rm dof} \ =\ 1.02
  \ \ \ .
  \label{eq:ODDLam}
\end{eqnarray}
This is, within uncertainties, at the threshold for $\Sigma\pi$ or
$NK$. The eigenstates will be a combination of these two systems and
it is likely that we have not resolved the two nearby-states in the
EM, and the result in Eq.~(\ref{eq:ODDLam}) is actually an average of
two closely-spaced energies.

%
%%%%%%%%%%%%%%%%%%%%%%%%%%%%%%%%%%%%%%%%%%%%%%%%%%%
%
% FIGURE: Odd Parity Sigma Matrix prony tJ=tW=3
%
%%%%%%%%%%%%%%%%%%%%%%%%%%%%%%%%%%%%%%%%%%%%%%%%%%%
\begin{figure}[!ht]
  \vskip0.5in \center
  \begin{tabular}{c}
    \includegraphics[width=0.9\textwidth]{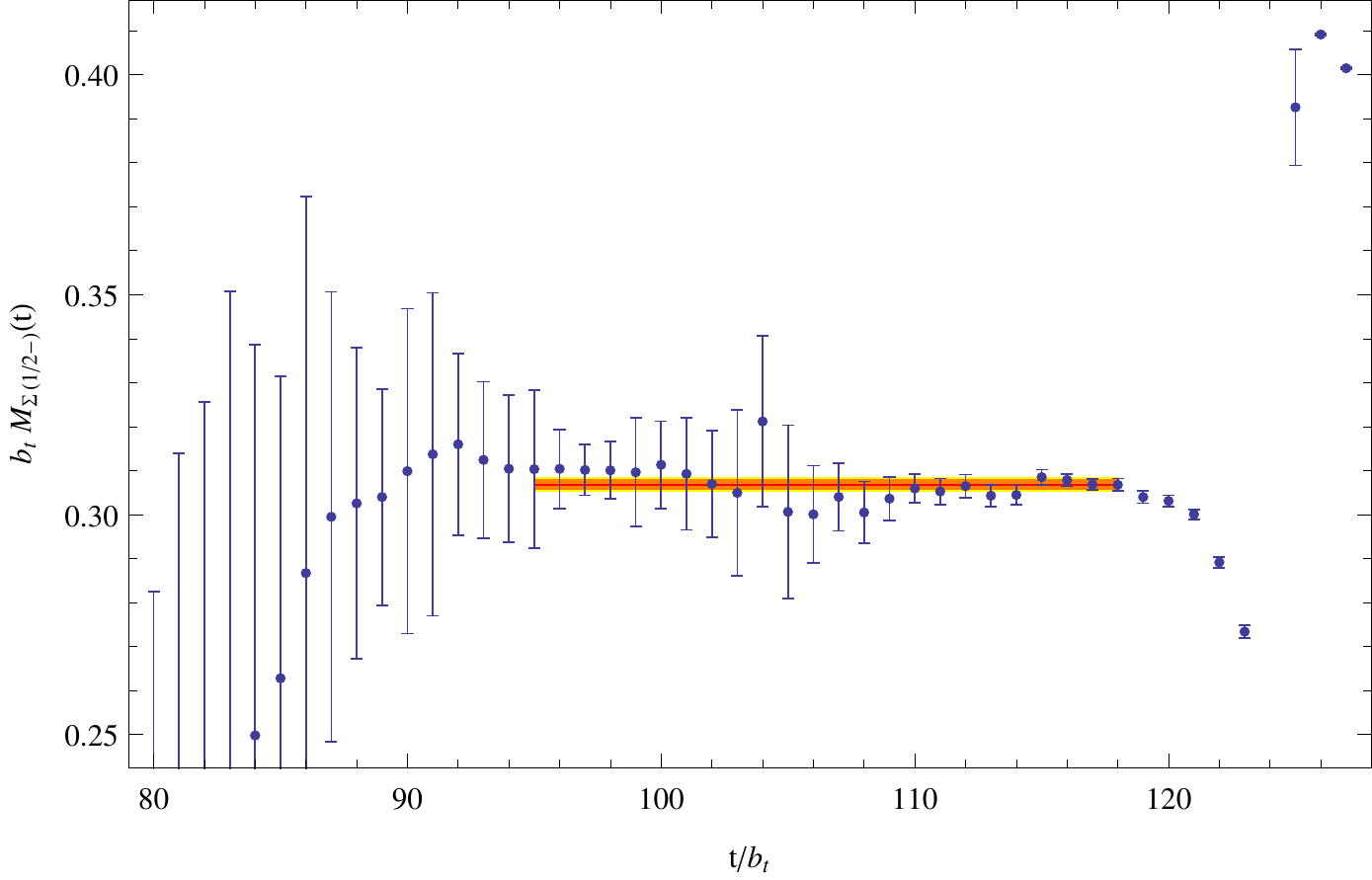}\\
    \includegraphics[width=0.9\textwidth]{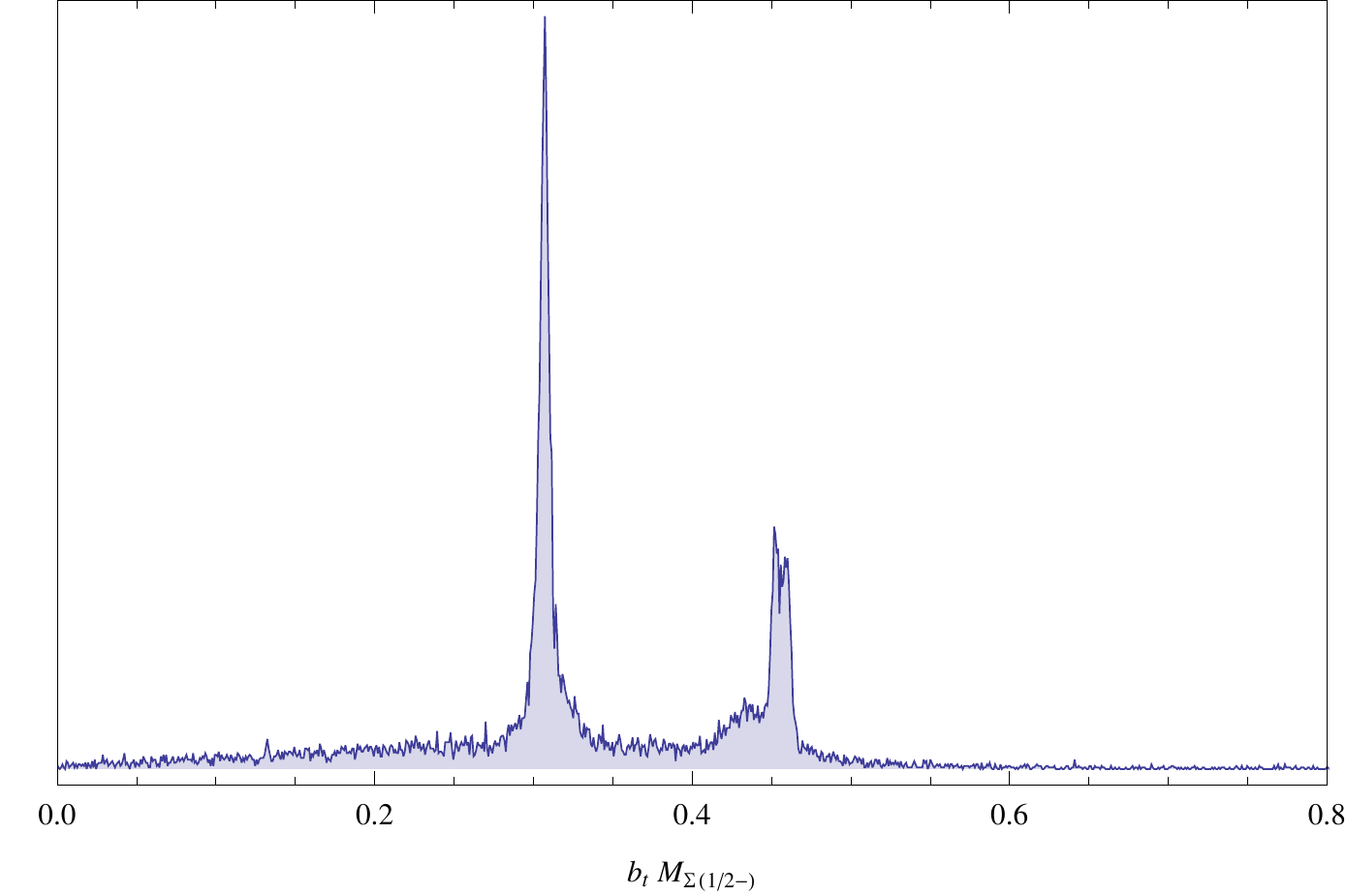}
  \end{tabular}
  \caption{\label{fig:2x2MatrixexpOddSigma} The upper panel shows the
    generalized EM for the lowest-lying negative-parity state coupling
    to the $\Sigma$-source using a Matrix-Prony analysis with $t_J=7$
    and $t_W=2$, and the correlated fit to the time-slices between
    $t=95$ and $t=118$. The inner (darker) region corresponds to the
    statistical uncertainty, while the outer (lighter) region
    corresponds to the statistical and fitting systematic
    uncertainties combined in quadrature.  The lower panel shows the
    associated Prony histogram of the positive roots for the
    time-slices $t=95$ to $t=118$.  }
\end{figure}
For the lowest-lying negative-parity state(s) produced by the
interpolating operator for the $\Sigma$, the situation is even more
complicated.  The threshold of the non-interacting s-wave $\Sigma\pi$
state is at $M_\Sigma+\mpi = 0.29747\pm 0.00030 \pm 0.00019$, for the
$\Lambda\pi$ state is $M_\Lambda+\mpi = 0.29191\pm 0.00030 \pm
0.00007$, and for the $NK$ state is $M_N+M_K=0.30384\pm 0.00033 \pm
0.00011$. Therefore, in this large volume, we expect to observe three
eigenstates that are nearly degenerate.  The lowest-lying states with
one unit of momentum occur at $E_{\Sigma\pi}^{|{\bf n}|=1} =
0.35857\pm 0.00037\pm 0.00023$, $E_{\Lambda\pi}^{|{\bf n}|=1} =
0.35341\pm 0.00030\pm 0.00007$, and $E_{NK}^{|{\bf n}|=1} = 0.35763\pm
0.00039\pm 0.00012$ and are well separated from the ${\bf n}=0$
states.  Fitting the EM shown in fig.~\ref{fig:2x2MatrixexpOddSigma}
between time-slices $t=95$ to $t=118$ yields,
\begin{eqnarray}
  E_{\Sigma ({1\over 2}^-)} & = & 0.3068\pm 0.0011\pm 0.0011
  \ \ ,\ \ 
  \chi^2/{\rm dof} \ =\ 0.80
  \ \ \ ,
  \label{eq:ODDSig}
\end{eqnarray}
in the region where one expects to find three closely-spaced states,
corresponding to the eigenstates dominated by $\Sigma\pi$,
$\Lambda\pi$, and $NK$.  Given how closely spaced these states are
expected to be, the extraction in Eq.~(\ref{eq:ODDSig}) is likely a
complicated average of three energies.

%
%%%%%%%%%%%%%%%%%%%%%%%%%%%%%%%%%%%%%%%%%%%%%%%%%%%
%
% FIGURE: Odd Parity Xi Matrix prony tJ=tW=3
%
%%%%%%%%%%%%%%%%%%%%%%%%%%%%%%%%%%%%%%%%%%%%%%%%%%%
\begin{figure}[!ht]
  \vskip0.5in \center
  \begin{tabular}{c}
    \includegraphics[width=0.9\textwidth]{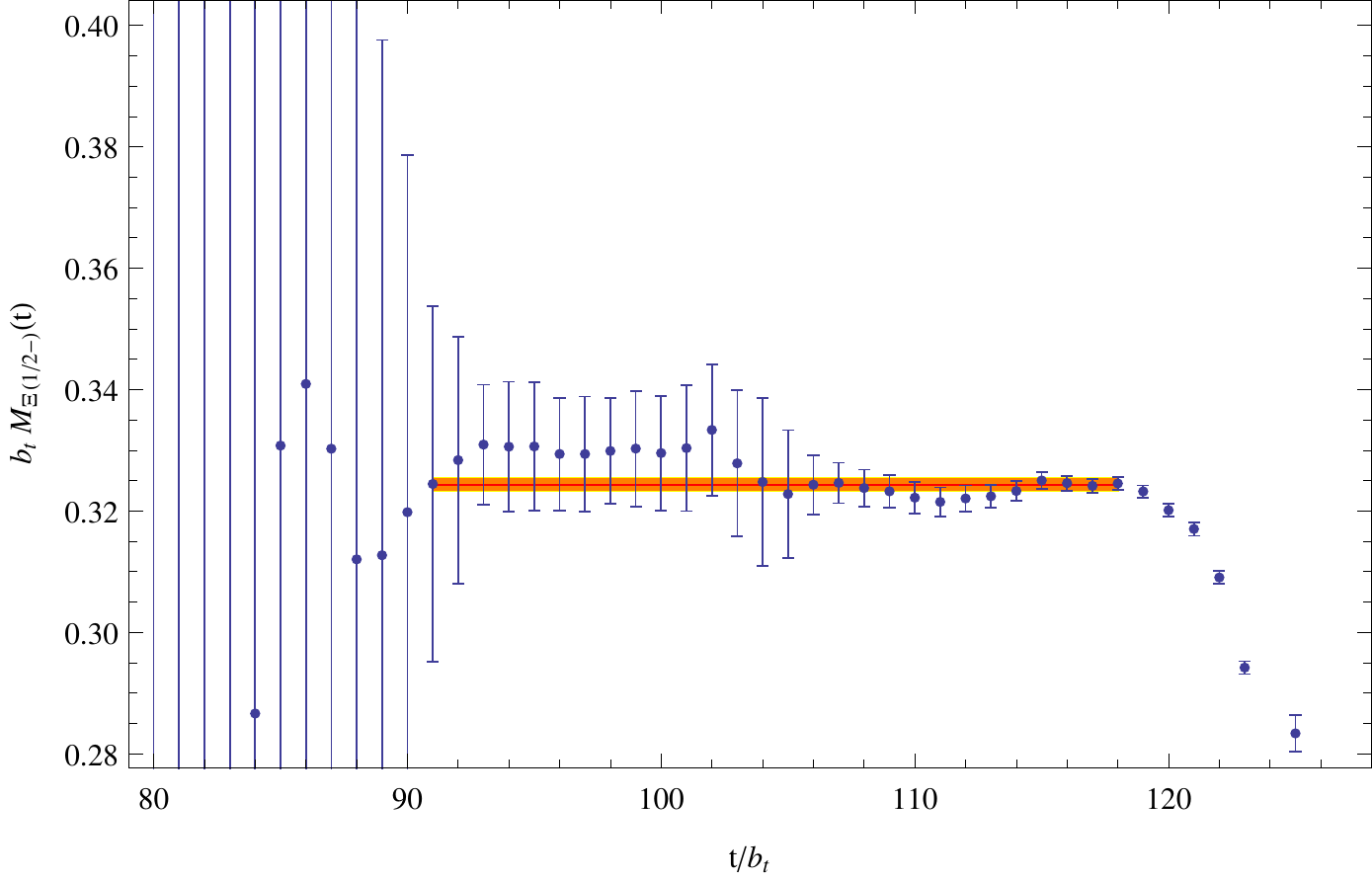}\\
    \includegraphics[width=0.9\textwidth]{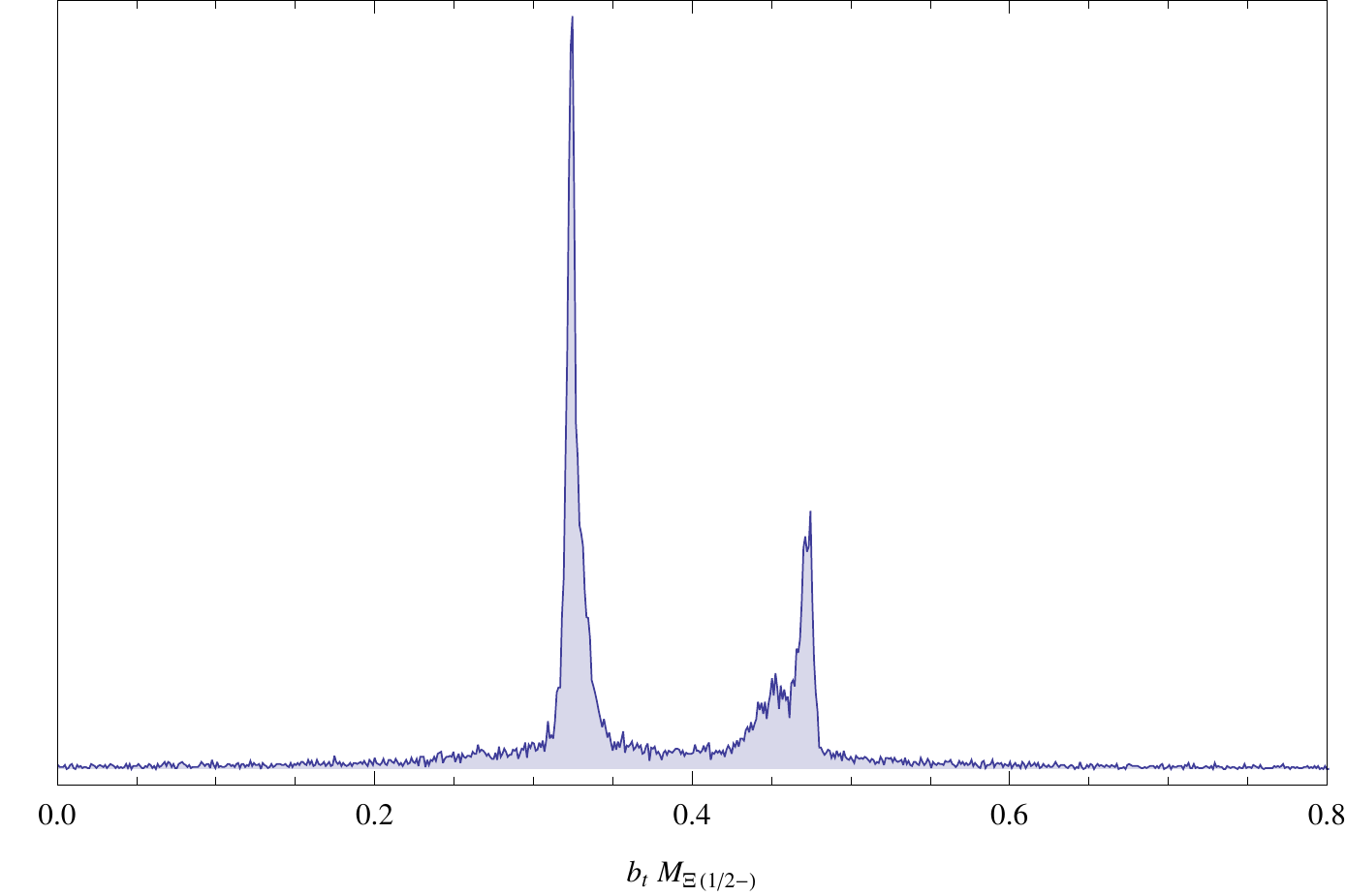}
  \end{tabular}
  \caption{\label{fig:2x2MatrixexpOddXi} The upper panel shows the
    generalized EM for the lowest-lying negative-parity state coupling
    to the $\Xi$-source using a Matrix-Prony analysis with $t_J=5$ and
    $t_W=11$, and the correlated fit to the time-slices between $t=91$
    and $t=118$. The inner (darker) region corresponds to the
    statistical uncertainty, while the outer (lighter) region
    corresponds to the statistical and fitting systematic
    uncertainties combined in quadrature.  The lower panel shows the
    associated Prony histogram of the positive roots for the
    time-slices $t=91$ to $t=118$.  }
\end{figure}
The situation is no better for the lowest-lying negative-parity states
that are expected to couple to the interpolating operator for the
$\Xi$.  The lowest-lying s-wave continuum states are $\pi\Xi$,
$\Lambda K$ and $\Sigma K$.  The threshold for these states, in the
absence of interactions, are $E_{\pi\Xi}^{|\bf n|=0} = 0.31033\pm
0.00028\pm 0.00006$, $E_{K\Lambda}^{|\bf n|=0} = 0.31957\pm 0.00030\pm
0.00006$, and $E_{K\Sigma}^{|\bf n|=0}=0.32513\pm 0.00030\pm 0.00018$,
respectively.  The corresponding states where both hadrons carry one
unit of momentum have thresholds state $E_{\Xi\pi}^{|{\bf n}|=1} =
0.37058\pm 0.00033\pm 0.00007$, $E_{\Lambda K}^{|{\bf n}|=1} =
0.37214\pm 0.00035\pm 0.00007$, $E_{\Sigma K}^{|{\bf n}|=1} =
0.37731\pm 0.00034\pm 0.00021$, respectively, Therefore, we expect to
observe two sets of three nearly degenerate eigenstates.  Fitting the
EM shown in fig.~\ref{fig:2x2MatrixexpOddXi} between time-slices
$t=91$ to $t=118$ yields,
\begin{eqnarray}
  E_{\Xi({1\over 2}^-)} & = & 0.3243\pm 0.0010\pm 0.0009
  \ \ ,\ \ 
  \chi^2/{\rm dof} \ =\ 0.72
  \ \ \ ,
  \label{eq:ODDXi}
\end{eqnarray}
in the region where one expects to find three closely-spaced states,
corresponding to the eigenstates dominated by ${\bf n}=0$ $\Xi\pi$,
$\Lambda K$, and $\Sigma K$.  Given how closely spaced these states
are expected to be, the extraction in Eq.~(\ref{eq:ODDXi}) is likely
an average of three unresolved energies.  There are hints of a couple
of other peaks in the Prony-histogram, but nothing conclusive.

The results of the best extractions of the ground-state baryon masses
using multi-exponential fitting and the matrix-Prony method, which
give consistent results for each species of baryon, are collected in
Table~\ref{tab:nExpResultsODD}.
%%%%%%%%%%%%%%%%
%% TABLE FINAL EXP MASSES
\begin{table}[!ht]
  \caption{\label{tab:nExpResultsODD}{
      The masses of the lowest-lying $J^\pi={1\over 2}^-$ states with
      unit baryon number  extracted by fitting three  exponentials and
      by the matrix-Prony method. The first
      uncertainty is statistical while the second is the fitting
      systematic. }}
  \begin{ruledtabular}
    \begin{tabular}{c|cccc|ccc}
      & & Exponential Fitting & & & & Matrix-Prony &\\
      \hline
      state & $b_t M$ &  range & $\chi^2/{\rm dof}$ & Q &  $b_t M$ & range  & $\chi^2/{\rm dof}$\\
      \hline
      $N$($\frac{1}{2}^-$)  & 0.2871(18)(10)&   90--117 & 1.11 & 0.28 & 0.2861(11)(20)
      & 93--119 & 0.91\\
      $\Lambda$($\frac{1}{2}^-$) & 0.2954(05)(15)&  90--113 & 0.89 & 0.64 &
      0.2983(8)(4) & 88--117 & 1.02\\
      $\Sigma$($\frac{1}{2}^-$) & 0.3074(15)(15)&   90--118 & 1.02 & 0.41 &
      0.3068(11)(11) & 95--118 & 0.80 \\
      $\Xi$($\frac{1}{2}^-$) &  0.3261(09)(15)&   89--115 & 1.07 & 0.36 &
      0.3243(10)(9) & 91--118 & 0.72\\
    \end{tabular}
  \end{ruledtabular}
\end{table}
%%%%%%%%%%%%%%%%

%%%%%%%%%%%%%%%%%%%%%%%%
\section{Signal-To-Noise Ratios \label{sec:ston}}
\noindent
Many observables of importance to particle physics that are currently
being calculated with LQCD, such as the pion decay constant and the
Gasser-Leutwyler coefficients, require the calculation of mesonic
correlation functions.  Statistical fluctuations on each time-slice of
these correlation function are well-behaved.  In contrast, as argued
by Lepage~\cite{Lepage:1989hd}, correlation functions involving one or
more baryons exhibit exponentially growing statistical noise.  In the
case of a single positive parity nucleon, the correlation function has
the form
\begin{eqnarray}
  \langle \theta_{N}(t)\rangle &=&
  \sum_{\bf x}\ \Gamma_+^{\beta\alpha}\langle N^\alpha({\bf x},t) \overline{N}^{\beta} ({\bf 0},0)\rangle
  \ \rightarrow\ Z_{0} \ e^{-M_N t}
  \ \ ,
  \label{eq:Gfunproton}
\end{eqnarray}
where $N$ is an interpolating field that has non-vanishing overlap
with the nucleon and the angle brackets indicate statistical averaging
over measurements on an ensemble of configurations.  The variance of
this correlation function is
\begin{eqnarray}
  N \sigma^2 & \sim & \langle \theta^{\dagger}_N(t)
  \theta_N(t)\rangle  - \langle \theta_{N}(t) \rangle^2 \nonumber \\
  & = & \sum_{\bf x,y} \Gamma_+^{\beta\alpha}\Gamma_+^{\gamma\delta}
  \langle N^\alpha({\bf x},t) \overline{N}^{\beta}({\bf y},t) N^\gamma({\bf 0},0)
  \overline{N}^{\delta}({\bf 0},0) \rangle\ \ -\ \langle \theta_{N}(t) \rangle^2
  \nonumber\\
  & \rightarrow & Z_{3\pi}\  e^{-3 \mpi t}- Z_{N}^2\  e^{-2M_N t}
  \ \ \rightarrow \ Z_{3\pi}\  e^{-3 \mpi t}
  \ \ ,
  \label{eq:GGdaggerfunproton}
\end{eqnarray}
and therefore, as Lepage~\cite{Lepage:1989hd} argued, the
noise-to-signal ratio behaves as
\begin{eqnarray} {\sigma\over\overline{x}} & = & {\sigma (t)\over
    \langle \theta(t) \rangle } \sim {1\over \sqrt{N}} \ e^{\left( M_N
      - {3\over 2} \mpi\right) t} \ \ .
  \label{eq:NtoSproton}
\end{eqnarray}
More generally, for a system of $A$ nucleons, the noise-to-signal
ratio behaves as
\begin{eqnarray} {\sigma\over\overline{x}} & & \sim {1\over \sqrt{N}}
  \ e^{A \left( M_N - {3\over 2} \mpi\right) t} \ \ .
  \label{eq:NtoSnucleus}
\end{eqnarray}
Therefore, in addition to the signal itself falling as $G\sim e^{-A
  M_N t}$, the noise-to-signal associated with the correlation
function grows exponentially, as in Eq.~(\ref{eq:NtoSnucleus}).

These arguments are constructed for a system with an infinite
time-direction and are modified in an important way for systems with a
finite time-direction with given BCs.  The calculations that are
presented in this work have employed anti-periodic BC's in the
time-direction.  With such BCs the positive parity nucleon correlation
function in Eq.~(\ref{eq:Gfunproton}) becomes
\begin{eqnarray}
  \langle \theta_{N}(t) \rangle &  \rightarrow & Z_{N} \ e^{-M_N t}
  \ +\ Z_{N\pi} \ e^{+E_{N\pi} (t-T)}
  \ \ ,
  \label{eq:GfunprotonABC}
\end{eqnarray}
where $E_{N\pi}$ is the energy of the lowest-lying negative-parity
state in the volume, which, for this ensemble of configurations, is a
continuum nucleon and pion at rest.  The arrow denotes the behavior of
the correlation function far from source (in both time-directions).
Further, the correlation function dictating the behavior of the
variance of the nucleon correlation function is modified similarly,
with Eq.~(\ref{eq:GGdaggerfunproton}) becoming
\begin{eqnarray}
  N \sigma^2 
  &\rightarrow &
  A_{3\pi} \ 
  e^{-{3\over 2} \mpi T}\ 
  \cosh\left(3 \mpi \left[t-{T\over 2}\right]\right)
  \ +\ 
  A_{\pi} 
  \ e^{-{3\over 2} \mpi T}\ 
  \cosh\left(\mpi \left[t-{T\over 2}\right]\right)
  \nonumber\\
  & &\ +\ 
  A_0\ e^{-M_N T}
  \ +\ ...
  \ \ .
  \label{eq:GGdaggerfunprotonABC}
\end{eqnarray}
The first term in Eq.~(\ref{eq:GGdaggerfunprotonABC}) arises from
$3\pi$'s propagating forward and $3\pi$'s propagating backwards, the
second term arises from $2\pi$'s propagating forward along with one
$\pi$ propagating backward and vice versa, the third
(time-independent) term arises from a nucleon propagating forward and
a nucleon propagating backward, and the ellipses denotes terms
involving larger masses.  As the negative-parity state is more massive
than the nucleon, the nucleon is the dominate component in the
correlation function, Eq.~(\ref{eq:GfunprotonABC}), for a number of
time-slices beyond the mid-point of the configuration.  From this
argument, one expects to see the signal-to-noise ratio degrade even
more rapidly than the expectation shown in Eq.~(\ref{eq:NtoSproton})
in the time-slices near the mid-point of the configuration where the
correlation function is still dominated by the nucleon.  One expects
to find regions of the correlation function, depending on the
structure of the source, which have the noise-to-signal scaling as $
e^{(m_p-{3\over 2} \mpi)t}$, $ e^{-{1\over 2} \mpi T} e^{(m_p-{1\over
    2} \mpi)t}$, $ e^{-\mpi T} e^{(m_p+{1\over 2} \mpi)t}$ $
e^{-{3\over 2} \mpi T} e^{(m_p+{3\over 2} \mpi)t}$, and $e^{m_p
  (t-T)}$, or combinations thereof.

The high statistics calculations we are presenting here enable a
detailed study of the behavior of the signal-to-noise ratio associated
with the correlation functions formed with quark propagators generated
with anti-periodic BCs.
%%%%%%%%%%%%%%%%%%%%%%%%%%%%%%%%%%%%%%%%%%%%%%%%%%%
%
% FIGURE: sh-pt signal-to-noise
%
%%%%%%%%%%%%%%%%%%%%%%%%%%%%%%%%%%%%%%%%%%%%%%%%%%%
\begin{figure}[!ht]
  \vskip0.5in \center
  \begin{tabular}{c}
    \includegraphics[width=0.99\textwidth]{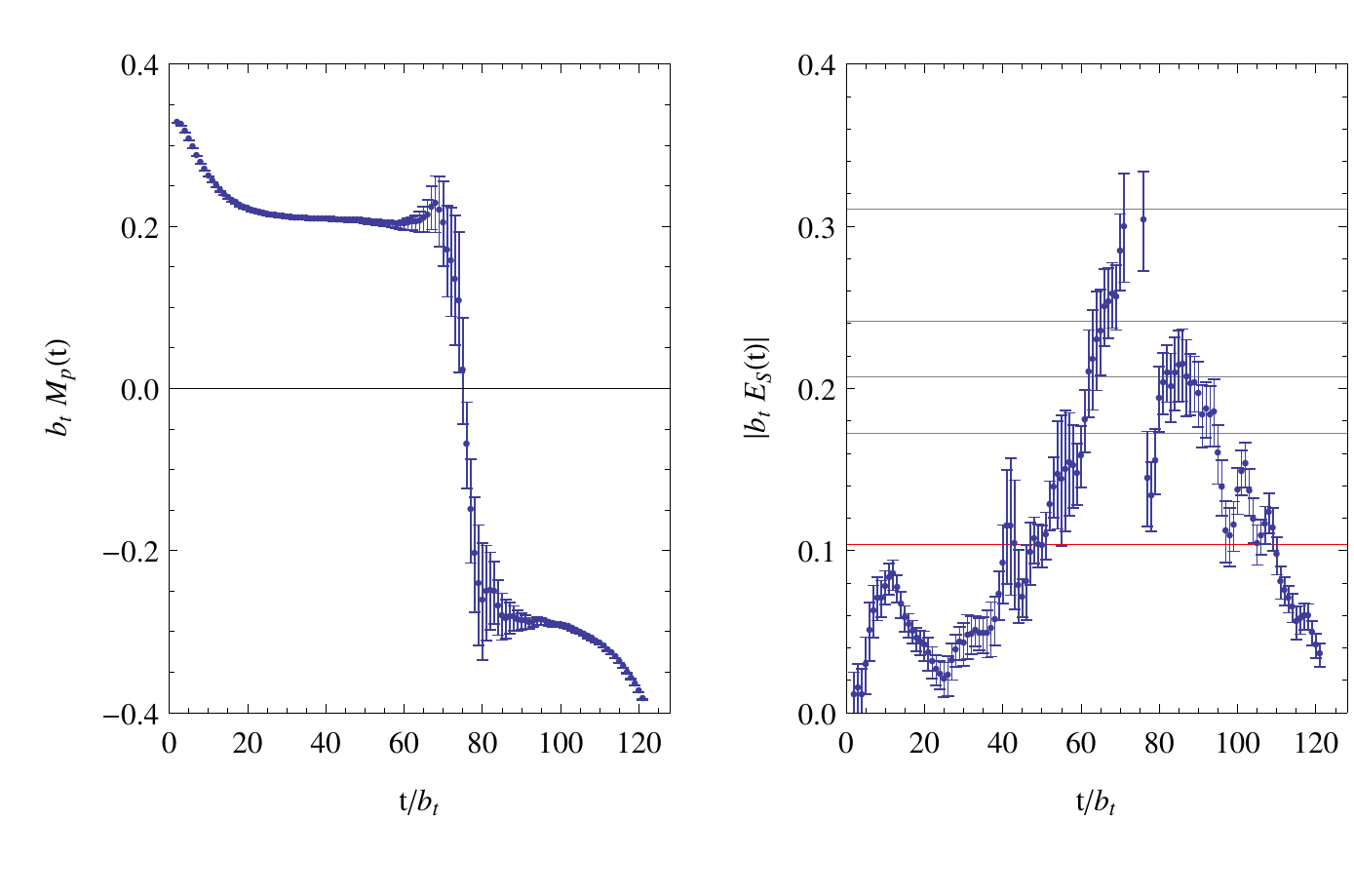}
  \end{tabular}
  \caption{\label{fig:ptshPROTStoN} The left panel shows the EM of the
    smeared-point N correlation function formed with with $t_J=3$.
    The right panel shows the energy-scale, $E_{\cal S}$, associated
    with the growth of the noise-to-signal ratio, as defined in
    Eq.~(\protect\ref{eq:Estondefn}).  The horizontal lines correspond
    to the energy scales $m_p-{3\over 2} \mpi$, $m_p-{1\over 2} \mpi$,
    $m_p$, $m_p+{1\over 2} \mpi$, and $m_p+{3\over 2} \mpi$ (from
    lowest energy to highest energy).  }
\end{figure}
%
%%%%%%%%%%%%%%%%%%%%%%%%%%%%%%%%%%%%%%%%%%%%%%%%%%%
%
% FIGURE: sh-sh signal-to-noise
%
%%%%%%%%%%%%%%%%%%%%%%%%%%%%%%%%%%%%%%%%%%%%%%%%%%%
\begin{figure}[!ht]
  \vskip0.5in \center
  \begin{tabular}{c}
    \includegraphics[width=0.99\textwidth]{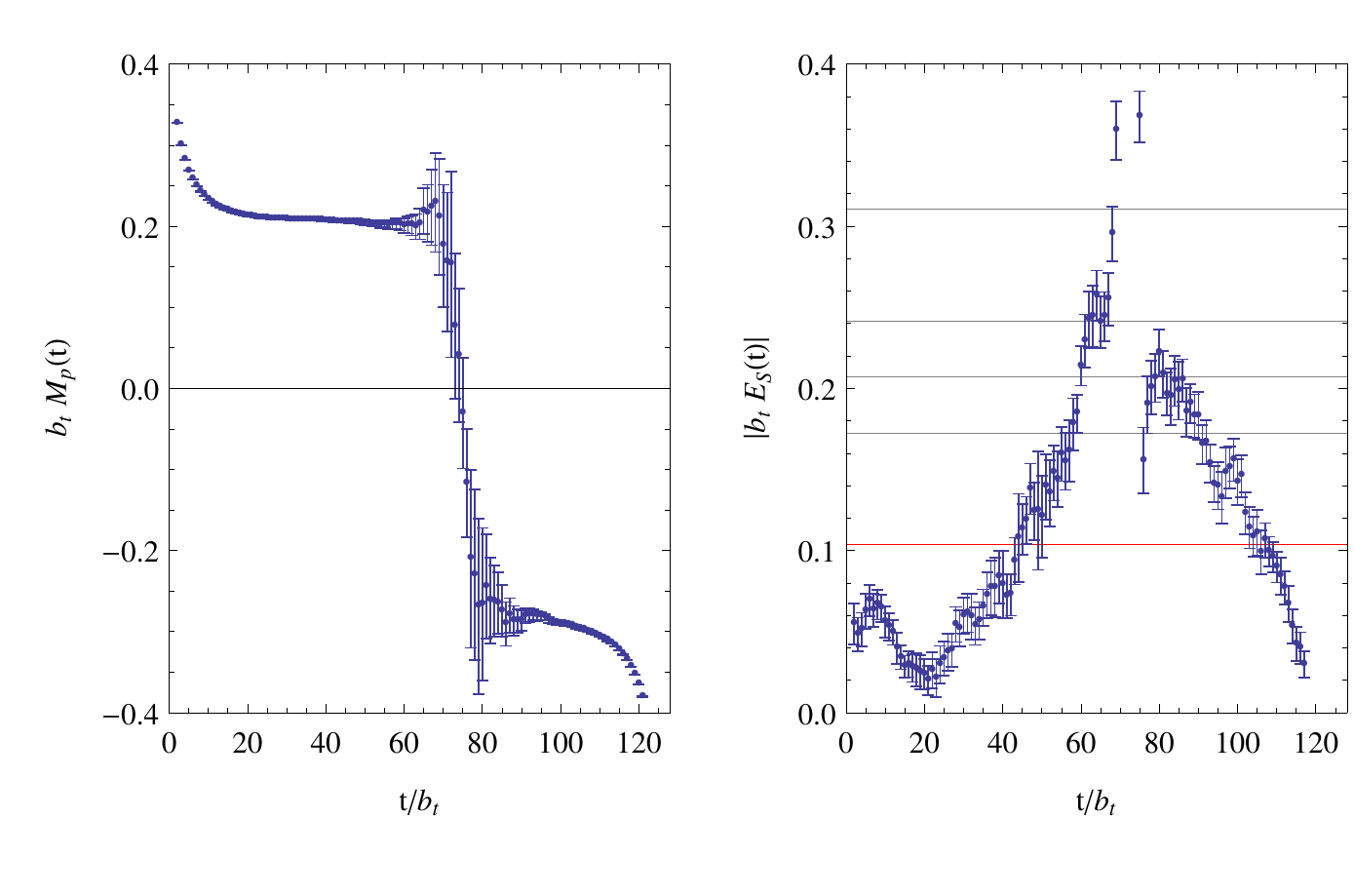}
  \end{tabular}
  \caption{\label{fig:shshPROTStoN} The left panel shows the EM of the
    smeared-smeared N correlation function formed with $t_J=5$.  The
    right panel shows the energy-scale, $E_{\cal S}$, associated with
    the growth of the noise-to-signal ratio, as defined in
    Eq.~(\protect\ref{eq:Estondefn}).  The horizontal lines correspond
    to the energy scales $m_p-{3\over 2} \mpi$, $m_p-{1\over 2} \mpi$,
    $m_p$, $m_p+{1\over 2} \mpi$, and $m_p+{3\over 2} \mpi$ (from
    lowest energy to highest energy).  }
\end{figure}
It is useful to form the effective noise-to-signal plot, in analogy
with the EMs.  On each time slice, the quantity
\begin{eqnarray} {\cal S}(t) & = & {\sigma (t)\over\overline{x}(t)} \
  \ \ ,
  \label{eq:stondefn}
\end{eqnarray}
is formed, from which the energy governing the exponential behavior
can be extracted via
\begin{eqnarray}
  E_{\cal S}(t;t_J) & = & 
  {1\over t_J}\ \log\left({ {\cal S}(t+t_J) \over {\cal S}(t)}\right)
  \ \ \ .
  \label{eq:Estondefn}
\end{eqnarray}
If the correlation function is dominated by a single state, and a
single energy-scale determines the behavior of the noise-to-signal
ratio, the quantity $E_{\cal S}(t;t_J)$ will be independent of both
$t$ and $t_J$.

In fig.~\ref{fig:ptshPROTStoN}, the full EM of the smeared-point
nucleon correlation function is shown (with $t_J=3$), and in
fig.~\ref{fig:shshPROTStoN}, the full EM of the smeared-smeared
nucleon correlation function is shown (with $t_J=5$).  Also shown are
the energy-scales associated with the growth of the noise-to-signal
ratio from Eq.~(\ref{eq:Estondefn}), with uncertainties generated
using the Jackknife procedure.  Considering the smeared-point
correlation function in fig.~\ref{fig:ptshPROTStoN}, after time-slice
$t=35$ or so, the correlation function is dominated by the
ground-state nucleon which persists until time-slice $t\sim 70$.
Beyond this time-slice the backward propagating negative-parity
$N\pi$-state becomes dominant.  Between time-slices $t\sim 40$ and
$t\sim 50$, the noise-to-signal ratio is determined by the expectation
of $m_p-{3\over 2} \mpi$.  However, after $t\sim 50$ the
signal-to-noise ratio degrades exponentially faster than this, and by
$t\sim 65$ the relevant energy-scale is $\sim m_p+{1\over 2} \mpi$ and
increasing with $t$.  Similar behavior is clear in the smeared-smeared
correlation function, for which the nucleon ground state dominates
from an earlier time-slice.

It is clear from this analysis of the noise-to-signal ratio, that the
length of the time-direction of these configurations and resulting
thermal states are limiting the precision of the ground-state nucleon
mass determination.  This will be even more true for the multiple
baryon correlation functions for which the signal-to-noise degrades
exponentially faster than in the single nucleon correlation functions.
Increasing the length of the time direction will lead to exponential
improvement of the correlation function at large times where the
nucleon component dominates the correlation function.  It is
interesting to note that the coefficients of the backward propagating
contributions to the noise-to-signal ratio are suppressed by powers of
$e^{-{1\over 2}\mpi T}$.  On the current configurations with $T=128$,
in order to reduce the contribution to the noise from the $m_p-{1\over
  2} \mpi$ component by an order of magnitude, the time extent would
need to be increased to $T\sim 192$.  This will reduce the
$m_p+{1\over 2} \mpi$ component by a factor of $\sim 84$ and the
$m_p+{3\over 2} \mpi$ component by $\sim 770$. Such an increase in the
temporal extent would significantly decrease the statistical
uncertainties with which ground-state signals are extracted.

The noise-to-signal analysis of the $\Xi$ correlation functions is
somewhat more complex, because there are a number of low-lying states
which can contribute to the variance.  For the $\Xi\bar\Xi$ noise
correlation function, the lightest intermediate states that can couple
to quark-content $ssu\overline{s}\overline{s}\overline{u}$ are
$KK\eta$, $\eta\eta\pi$ and $\eta\eta\eta$.
%%%%%%%%%%%%%%%%%%%%%%%%%%%%%%%%%%%%%%%%%%%%%%%%%%%
%
% FIGURE: Xi sh-pt signal-to-noise
%
%%%%%%%%%%%%%%%%%%%%%%%%%%%%%%%%%%%%%%%%%%%%%%%%%%%
\begin{figure}[!ht]
  \vskip0.5in \center
  \begin{tabular}{c}
    \includegraphics[width=0.99\textwidth]{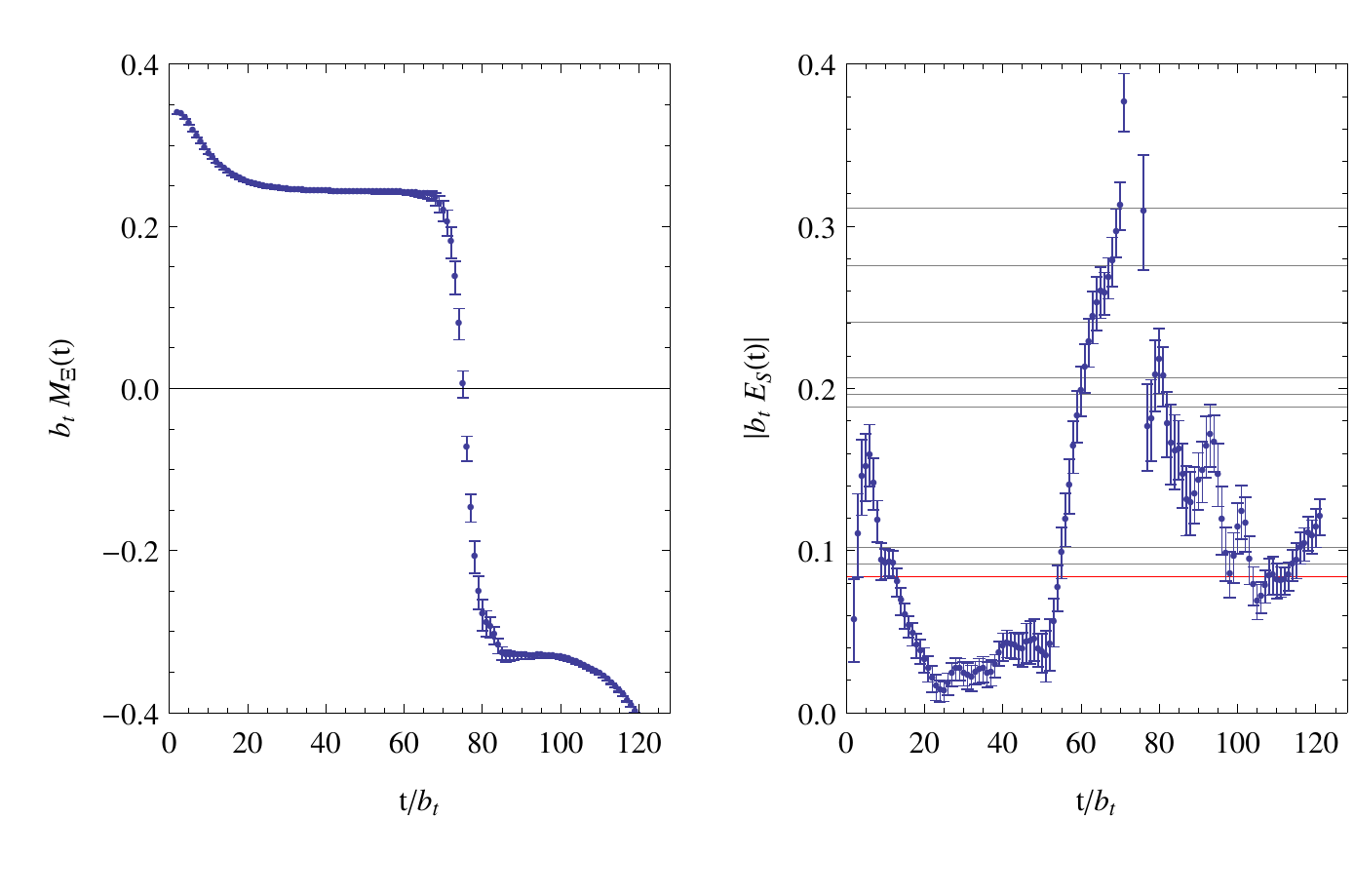}
  \end{tabular}
  \caption{\label{fig:ptshXiStoN} The left panel shows the EM of the
    smeared-point $\Xi$ correlation function formed with $t_J=3$.  The
    right panel shows the energy-scale, $E_{\cal S}$, associated with
    the growth of the noise-to-signal ratio, as defined in
    Eq.~(\protect\ref{eq:Estondefn}).  The horizontal lines correspond
    to the energy scales $m_\Xi-{3\over 2} m_\eta$, $m_\Xi-\mK-{1\over
      2} m_\eta$, $m_\Xi-m_\eta - {1\over 2} \mpi$, $m_\Xi-{1\over 2}
    m_\eta$, $m_\Xi-\mK + {1\over 2} m_\eta$, $m_\Xi-{1\over 2} \mpi$,
    $m_\Xi$, $m_\Xi+{1\over 2} \mpi$, and $m_\Xi+m_\eta -{1\over 2}
    \mpi$ (from lowest energy to highest energy).  }
\end{figure}
%
%%%%%%%%%%%%%%%%%%%%%%%%%%%%%%%%%%%%%%%%%%%%%%%%%%%
%
% FIGURE: Xi sh-sh signal-to-noise
%
%%%%%%%%%%%%%%%%%%%%%%%%%%%%%%%%%%%%%%%%%%%%%%%%%%%
\begin{figure}[!ht]
  \vskip0.5in \center
  \begin{tabular}{c}
    \includegraphics[width=0.99\columnwidth]{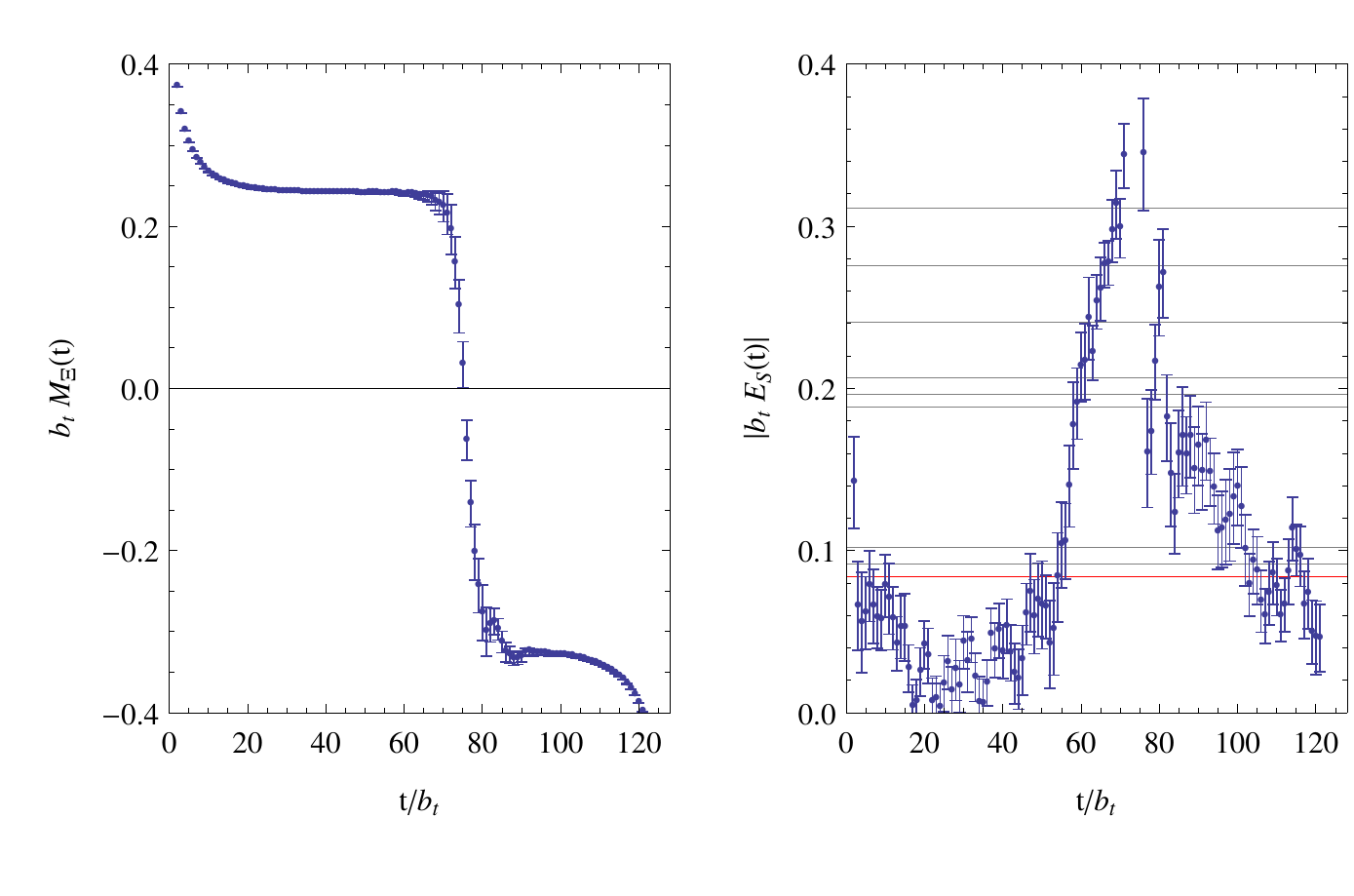}
  \end{tabular}
  \caption{\label{fig:shshXiStoN} The left panel shows the EM of the
    smeared-smeared $\Xi$ correlation function formed with $t_J=3$.
    The right panel shows the energy-scale, $E_{\cal S}$, associated
    with the growth of the noise-to-signal ratio, as defined in
    Eq.~(\protect\ref{eq:Estondefn}).  The horizontal lines correspond
    to the energy scales $m_\Xi-{3\over 2} m_\eta$, $m_\Xi-\mK-{1\over
      2} m_\eta$, $m_\Xi-m_\eta - {1\over 2} \mpi$, $m_\Xi-{1\over 2}
    m_\eta$, $m_\Xi-\mK + {1\over 2} m_\eta$, $m_\Xi-{1\over 2} \mpi$,
    $m_\Xi+{1\over 2} \mpi$, $m_\Xi$, and $m_\Xi+m_\eta -{1\over 2}
    \mpi$ (from lowest energy to highest energy).  }
\end{figure}
The full EMs and the $E_{\cal S}$ plots for the smeared-point and
smeared-smeared $\Xi$ correlation functions are shown in
fig.~\ref{fig:ptshXiStoN} and fig.~\ref{fig:shshXiStoN}.  In both the
smeared-point and smeared-smeared $\Xi$ correlation functions, the
noise-to-signal ratio is growing exponentially slower than naive
expectations, until about time-slice $t\sim 55$.  As the $\Xi$
ground-state dominates the smeared-smeared correlation function beyond
$t\sim 40$, this allows for a extraction of the mass with higher
precision than expected.  This suggests that the noise-source does not
couple to the low-lying mesonic states as strongly as expected, and
that more massive mesonic states are dominating the noise over many
time-slices. However, eventually, for $t\agt 55$, the growth of noise
overshoots the original Lepage expectation (indicated by the lowest
horizontal line in figs.~\ref{fig:ptshXiStoN} and
\ref{fig:shshXiStoN}.

\begin{figure}
  \centering
  \includegraphics[width=0.99\columnwidth]{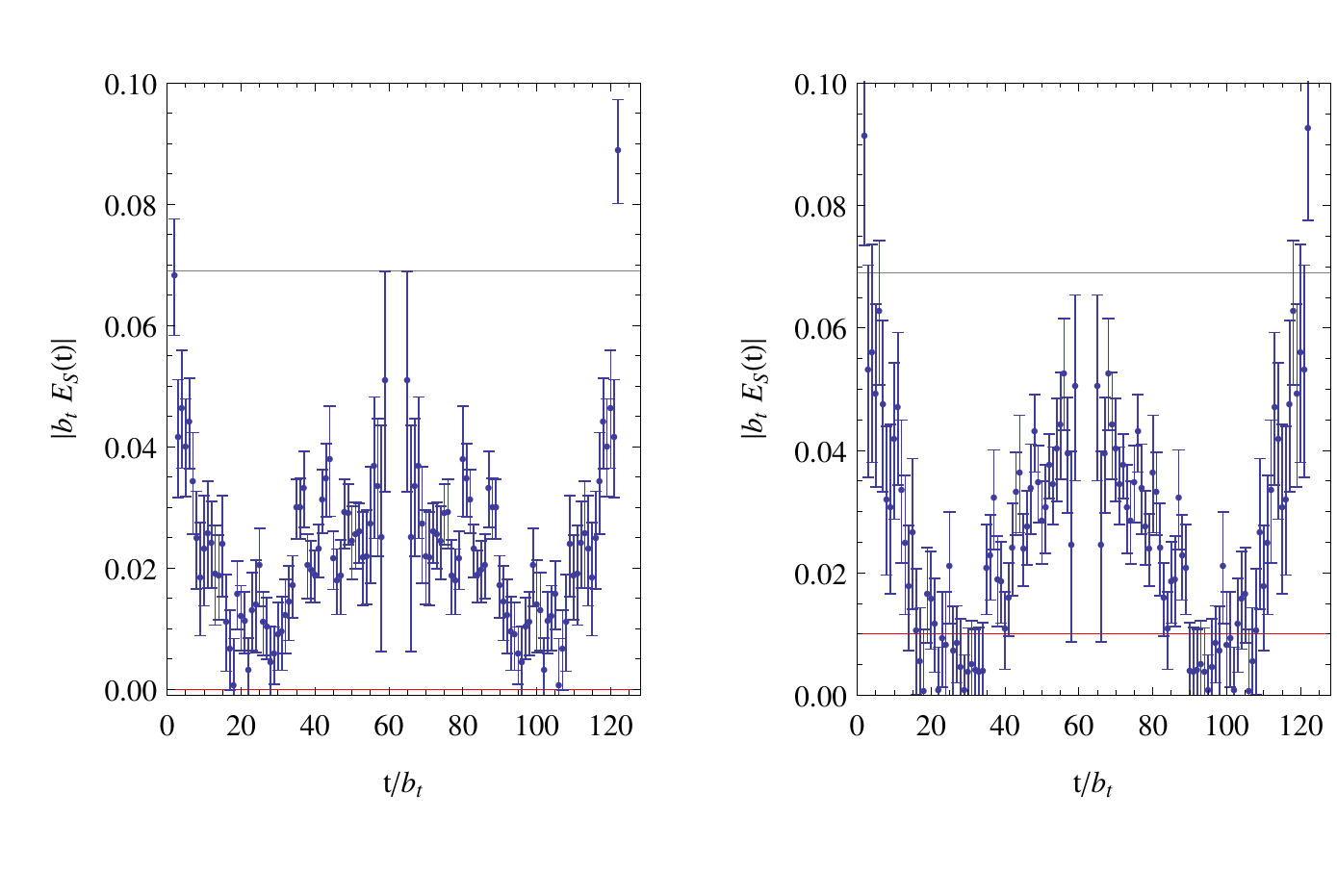}
  \caption{The energy scale, $E_{\cal S}$, associated with the growth
    of the noise-to-signal ratio in the $\pi^+$ (left) and $K^+$
    (right) smeared-smeared correlation functions using $t_J=3$. The
    horizontal lines correspond (from lowest to highest) to 0, $\mpi$
    for the pion and $\mK-\frac{1}{2}m_\eta -\frac{1}{2}m_K$ for the
    kaon.}
  \label{fig:piKnoise}
\end{figure}

Whilst we are primarily interested in noise in the baryonic sector, it
is interesting to note that the mesonic correlation functions also
suffer from similar issues. According to the above arguments, the pion
correlation function on lattices at zero temperature (infinite
temporal extent) will have noise that is independent of time (up to
fluctuations) while the kaon will have noise that grows exponentially
with the small energy difference $m_K-\frac{1}{2}m_\pi
-\frac{1}{2}m_\eta$. However at finite temperature, the noise
correlation functions of both systems receive additional contributions
that grow faster than the above expectations. This is shown in
fig.~\ref{fig:piKnoise}.

%%%%%%%%%%%%%%%%%%%%%%%%
\section{Scaling with Computational Resources}
\label{sec:saturation}
\noindent

An important component of our current work is to address the future
requirements for LQCD calculations in nuclear physics, a field
characterized by small energy scales in heavy systems, for example,
the 2 MeV binding energy of the $\sim 2$~GeV deuteron.
\begin{figure}[!ht]
  \centering
  \includegraphics[width=0.99\columnwidth]{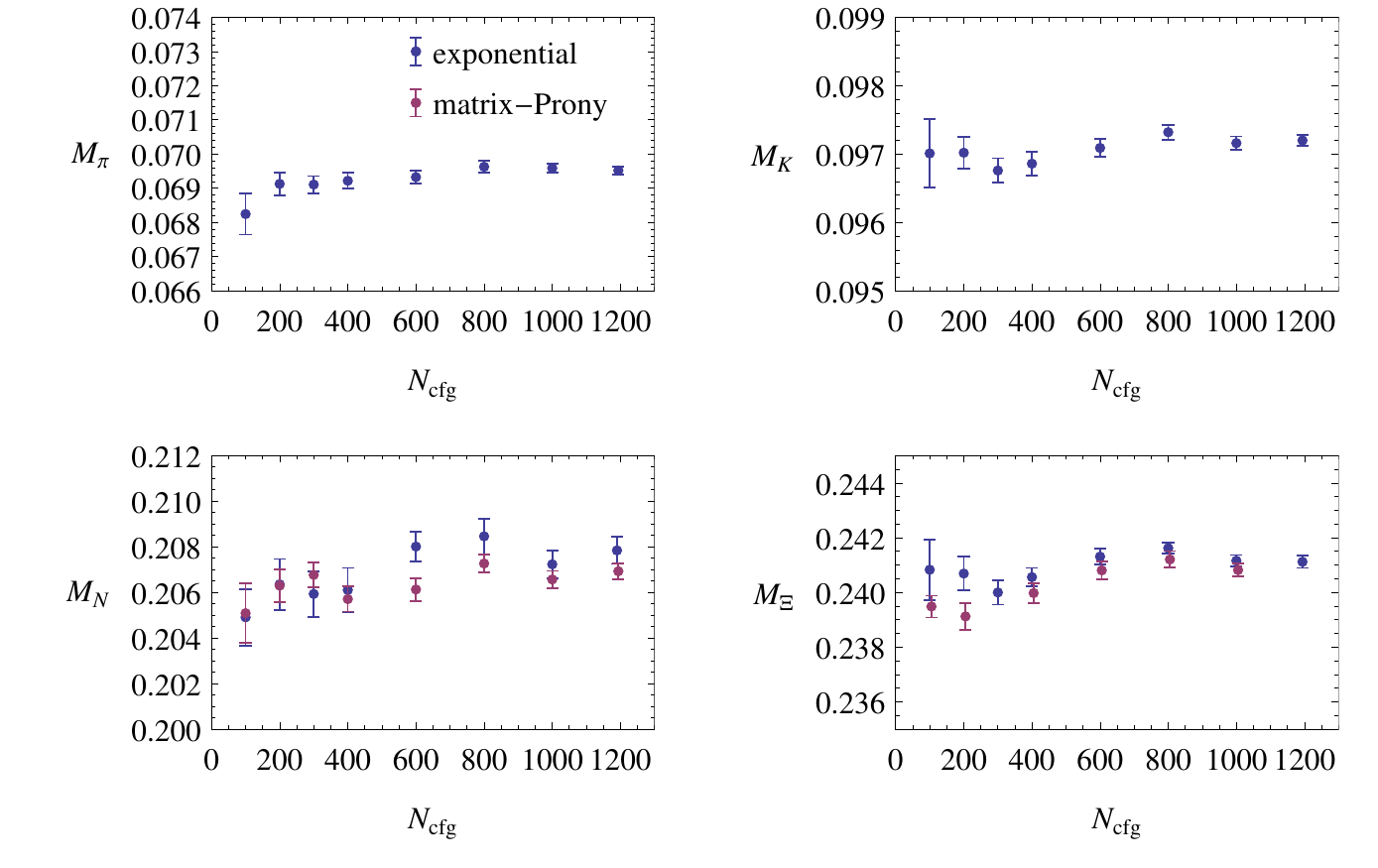}
  \caption{ The extracted masses of the $\pi^+$, $K^+$, N and $\Xi$ as
    a function of the number of configurations (with the full set of
    measurements performed on it).  Statistical and systematic
    uncertainties have been combined in quadrature. For the baryon
    states, both the matrix-Prony and exponential fits are shown.  }
  \label{fig:sdvsNcfgA}
\end{figure}
In fig.~\ref{fig:sdvsNcfgA}, we show the extracted mass of the
$\pi^+$, $K^+$, N and $\Xi$ as a function of the number of
configurations in the ensemble for both the exponential and
matrix-Prony analysis methods.  The full set of measurements performed
on each configuration are included, and the fitting intervals are
chosen to optimize the extraction for each ensemble size.  In each
case, the uncertainty in the mass is reduced, as expected, with
increasing ensemble size, and the mass extracted from the smaller
ensembles tends to be less than that from the larger ensembles.
\begin{figure}[!ht]
  \centering
  \includegraphics[width=0.99\columnwidth]{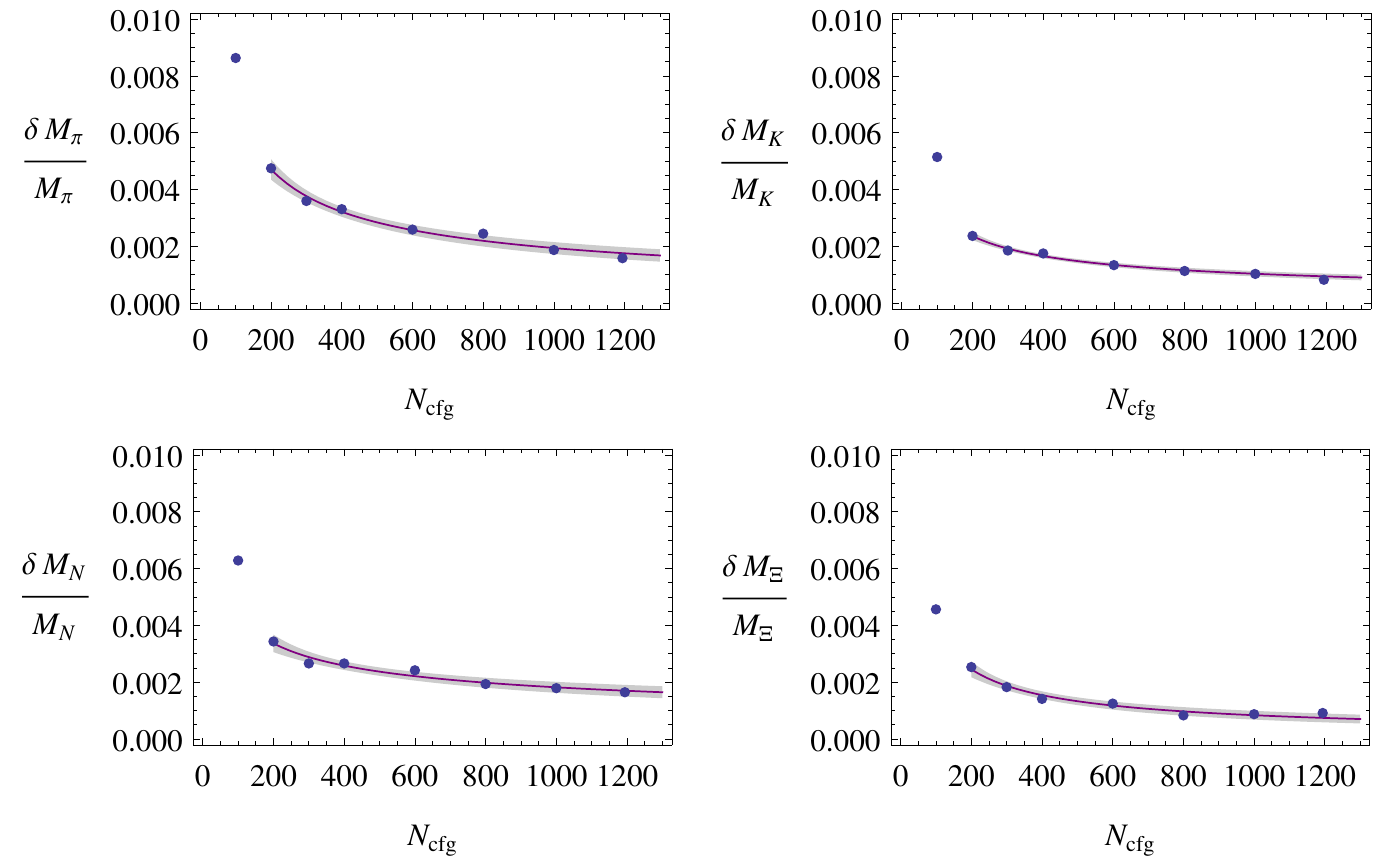}
  \caption{ The fractional uncertainty in the extracted masses of the
    $\pi^+$, $K^+$, N and $\Xi$ as a function of the number of
    configurations (with the full set of measurements performed on it)
    for the exponential analysis. Statistical and systematic
    uncertainties have been combined in quadrature. The curves
    correspond to fits of the form $\delta M/M = A N_{\rm cfg}^b$. The
    exponents extract in these fits are -0.55(4), -0.51(3), -0.38(4),
    -0.67(6) for the $\pi^+$, $K^+$, N and $\Xi$, respectively.
  }  \label{fig:sdvsNcfgB}
\end{figure}
Fig.~\ref{fig:sdvsNcfgB} shows the fractional uncertainty in the mass
of the $\pi^+$, $K^+$, N and $\Xi$, associated with the results in
fig.~\ref{fig:sdvsNcfgA}, as a function of the number of
configurations.  An extrapolation can be performed with a fit to the
uncertainties in fig.~\ref{fig:sdvsNcfgB} of the form $\delta M/M = A
N_{\rm cfg}^b$.  The exponents extract in these fits are -0.55(4),
-0.51(3), -0.38(4), -0.67(6) for the $\pi^+$, $K^+$, N and $\Xi$,
respectively.

\begin{figure}[!ht]
  \centering
  \includegraphics[width=0.99\columnwidth]{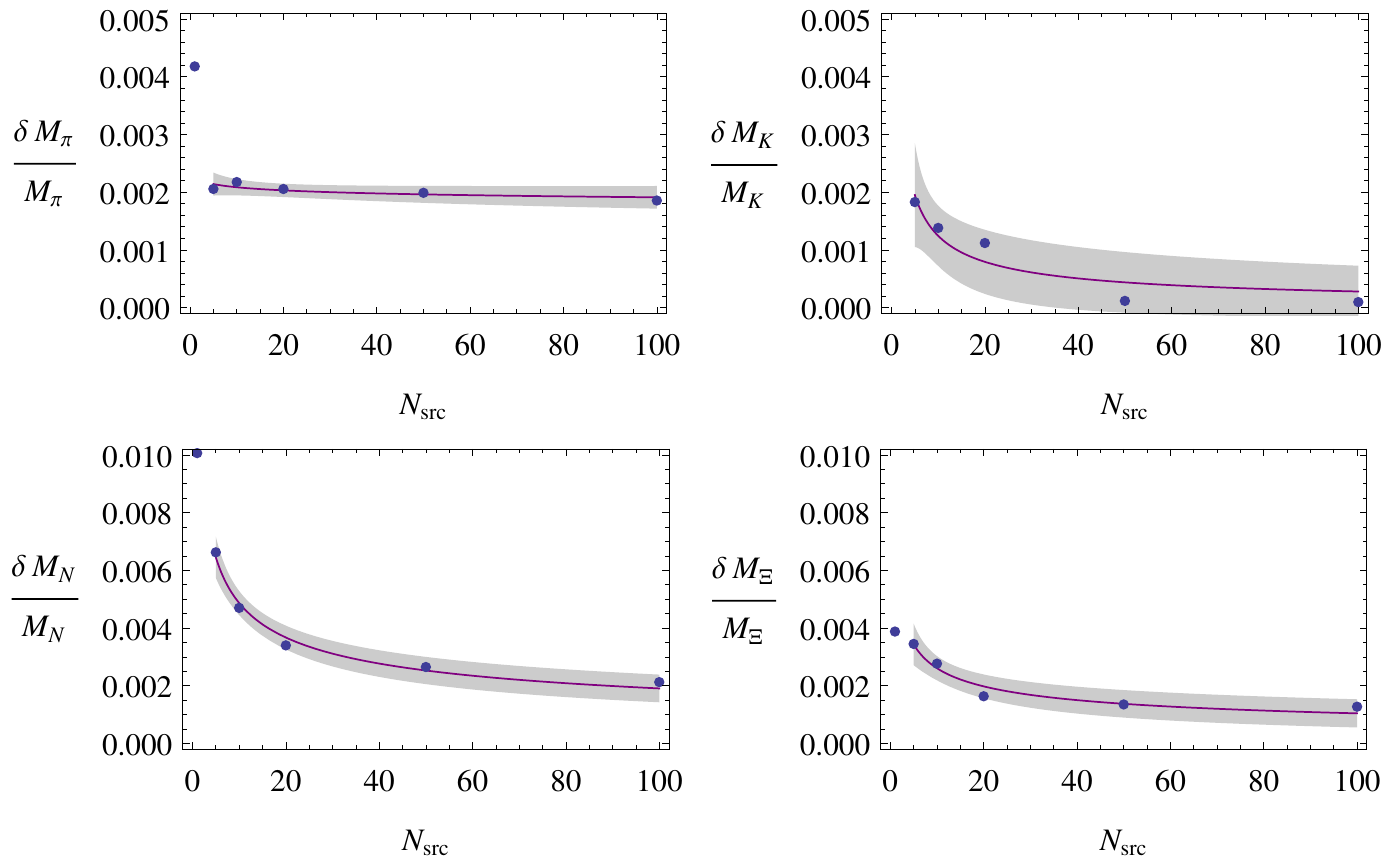}
  \caption{The fractional uncertainty in the extracted masses of the
    $\pi^+$, $K^+$, N and $\Xi$ as a function of the number of sources
    used on each configuration (1012 configurations were used in this
    study) for the exponential analysis. Statistical and systematic
    uncertainties have been combined in quadrature. The curves
    correspond to fits of the form $\delta M/M = A N_{\rm src}^b$. The
    exponents extract in these fits are -0.03(2), -0.65(19), -0.41(3),
    -0.40(6) for the $\pi^+$, $K^+$, N and $\Xi$, respectively.}
  \label{fig:sdvsNsrc}
\end{figure}

The dependence of our results for hadron masses on the number of
sources used in the calculations is explored in
fig.~\ref{fig:sdvsNsrc} where we show the fractional uncertainty in
the mass of the $\pi^+$, $K^+$, N and $\Xi$ as a function of the
number of sources used on each configuration. In this figure we use an
ensemble of 1012 configurations, those on which there are at least 100
measurements. A simple fit of the form $\delta M/M = A N_{\rm src}^b$
returns exponents $b=-0.03(2)$, -0.65(19), -0.41(3), -0.40(6) for the
$\pi^+$, $K^+$, N and $\Xi$, respectively.

The results of this analysis can be simply summarized for baryons
(averaging over the nucleon and $\Xi$) as
\begin{equation}
  \label{eq:scalingform}
  \frac{\delta M_B}{M_B}\sim \frac{1}{N_{\rm src}^{0.4} N_{\rm
      cfg}^{0.5}}\, .
\end{equation}
For mesons, a similar scaling is seen, with a somewhat worse scaling
with the number of measurements per configuration in the case of the
pion, consistent with the saturation seen in
fig.~\ref{fig:saturation}. This functional form enables us to quantify
the relative benefit of increasing the number of sources per
configuration compared to increasing the total number of
configurations. The costs involved in this are as follows:
\begin{itemize}
    \item {\it Gauge configuration generation}: The total cost of
  generating the ensemble of \Ncfgs\ gauge configurations was 2M
  JLab-{\tt 6n} cluster node-hours and the production took a
  significant amount of wall-clock time.  Configuration generation
  costs scale linearly with the number of configurations once a
  Monte-Carlo trajectory has thermalised (in this case the overhead of
  thermalisation was approximately 10\%). In order to generate
  significantly larger ensembles (containing $10^4$ or $10^5$ gauge
  fields) in a reasonable wall-clock time, it will be necessary to run
  multiple trajectories in parallel. Given wall-clock time and memory
  constraints, an individual trajectory will produce ${\cal O}(1000)$
  gauge-field configurations that are useful for
  measurements. Consequently the thermalization overhead will
  conservatively remain at about 10\%. Each configuration requires
  $\sim 2\times 10^3$ JLab-{\tt 6n} node-hours to produce.

    \item {\it Measurement calculations}: The total cost of computing
  all of the measurements performed in this work was \TotalCost\
  Jlab-{\tt 6n} node-hours.  The cost to generate the \PropsperCFG\
  light-quark and strange-quark propagators per configuration on the
  \Ncfgs\ configurations in this ensemble was $\sim 3$M Jlab-{\tt 6n}
  node-hours, while the cost to generate the baryon and meson blocks
  (used at intermediate stages of the calculations) was $\sim 3.5$M
  Jlab-{\tt 6n} node-hours.  Contracting the blocks to accomplish the
  desired measurements (one, two, ...  baryons, one, two,...  mesons
  and so forth) cost $\sim 0.5$M Jlab-{\tt 6n} node-hours.  If
  propagators on a given configuration are computed in sets of 100,
  the initial overhead of constructing deflation vectors in the {\tt
    EigCG} algorithm becomes negligible (at the 1\% level) and can be
  eliminated for further sets of calculations by storing the
  eigenvector information. On typical machines, each set of
  propagators and associated hadron blocks (technically, not an
  efficient way to calculated the single hadron spectrum, but critical
  for two and more hadron calculations) requires 22 JLab-{\tt 6n}
  node-hours to produce.

    \item {\it Anisotropy:} The anisotropy of the lattices used in our
  calculations proved useful in reducing systematic errors in our fits
  (see Table \ref{tab:nExpRange}), providing approximately a
  $1/\sqrt{\xi_{eff}}$ reduction. However, the cost of producing
  gauge-field configurations and propagators scales as approximately
  $\xi^2$ for the same physical extent (one power arises from the
  additional time-slices and one power arises from the worsening
  condition number of the Dirac operator). Comparing these exponents,
  we would conclude that using anisotropic configurations is not
  ideal. However for more complicated multi-hadron systems where
  useful fit ranges are much reduced in physical units, the anisotropy
  will likely prove to be very useful. This remains to be investigated
  further in subsequent studies.
\end{itemize}
Using this information and the scalings in Eq.~(\ref{eq:scalingform}),
we can address the question of how much computation is required to
achieve a particular level of statistical precision. With the current
data this is only possible in the single hadron sector; ongoing
analyses will address the $B>1$ in the near future. To halve the
uncertainty in the determinations of the ground-state baryon masses
(calculating the nucleon mass at the $\sim 1~{\rm MeV}$-level), an
increase in the number of configurations by a factor of four, or of
the number of measurements per configuration by a factor of 5.6 is
required. Achieving this precision by performing more measurements on
the existing set of configurations ($\sim 1100$ additional
measurements on 1200 configurations) would cost 30 M JLab-{\tt 6n}
node-hours. Achieving the same precision by generating an additional
configurations and performing the same number of measurements on them
(3600 configurations with 245 measurements) would require 27 M
JLab-{\tt 6n} node-hours. Both approaches have similar cost at this
level of precision, but for further improvements, the generation of
additional configurations will be more efficient. Additionally, the
second approach will further improve the uncertainty for observables
such as the pion mass that have saturated in terms of the number of
sources per configuration. This approach would clearly  be of more
benefit to the broader community.

%%%%%%%%%%%%%%%%%%%%%%%%
\section{Conclusions \label{sec:conclusions}}
\noindent
The energy-scales that arise in nuclear physics are typically in the
MeV range, and in order for LQCD to have significant impact in this
field, baryon masses (and energy-eigenstates in the volumes relevant
to scattering processes) must be calculable with uncertainties that
are a fraction of an MeV (including isospin-breaking and
electromagnetic interactions, quark mass, lattice volume and lattice
spacing extrapolations).  Current computational resources do not
permit such calculations.  In this work we have performed the first
high-statistics study of baryon correlation functions to better
understand a number of issues that will impact the precision with
which quantities of importance to nuclear physics can be determined
with LQCD. In the future, we will extend our analysis to look at
observables in the $B>1$ baryon sector.

At the single lattice spacing, lattice volume and unphysical light
quark mass used in this work, we find the following set of ground
state masses
\begin{align*}
  &M_{\pi}& &=\ 390.3(0.7)(0.3)(2.5)\  {\rm MeV},&  %\qquad\qquad
  &M_K& &=\ 546.0(0.6)(0.2)(3.6)\  {\rm MeV},& \\
  &M_N& &=\ 1163.9(1.8)(0.6)(7.6)\  {\rm MeV},&  %\qquad\qquad
  &M_\Lambda& &=\ 1252.4(1.6)(0.3)(8.2)\  {\rm MeV},& \\
  &M_\Sigma& &=\ 1283.7(1.6)(1.0)(8.4) \  {\rm MeV},\quad&  %\qquad\qquad
  &M_\Xi& &=\ 1356.1(1.4)(0.2)(8.8)\  {\rm MeV},&\\
  &E_{N(1/2^-)}& &=\  1610(06)(11)(11) \  {\rm MeV},&  %\qquad\qquad
  &E_{\Lambda(1/2^-)}& &=\ 1679(05)(02)(11) \  {\rm MeV},& \\
  &E_{\Sigma(1/2^-)}& &=\ 1727(06)(06)(11) \  {\rm MeV},&  %\qquad\qquad
  &E_{\Xi(1/2^-)}& &=\ 1825(6)(5)(12)\  {\rm MeV},& 
\end{align*}
which we present in physical units. Since the lattice spacing is known
with less precision than the lattice masses presented here, we make
the systematic uncertainty arising from the lattice spacing explicit
(third uncertainty). Given that there are significant ambiguities in
scale setting, the most precise result will be for dimensionless
quantities.

With high precision measurements of baryon correlation functions
obtained from a single type of source for the light-quark and
strange-quarks propagators, we have shown that the number of methods
that can be used to extract the arguments of the contributing
exponentials increases.  This is due to the fact that some methods
become stable when the uncertainties become small, such as the method
of Prony and also the direct fitting of multiple exponentials.
Histograms constructed from the roots found in the Prony method are
found to be useful in identifying mass regions where states may exist,
but we have not yet arrived at a well-defined (rigorous) method with
which to use these histograms directly.  It is likely that a more
refined statistical analysis of these correlation functions using the
most modern statistical tools will further increase the physics that
can be extracted.

The exponential signal-to-noise degradation that plagues baryon
correlation functions is currently a serious limitation for the
calculation of nuclear physics observables, and is one significant
difference between particle and nuclear physics LQCD calculations.
The high statistics calculations we have performed have allowed us to
systematically explore this issue. We find that the issue is more
serious than one would naively expect, due to (what in hindsight is
now obvious) the use of anti-periodic BC's in the time-direction on
the quark-propagators.  The variance of the correlation function is
symmetric about the mid-point of the time-direction of the
configuration.  Therefore, the optimal region in which to determine
the baryon masses (and also their interactions) is in the first half
of the configuration, far from the midpoint.  This significantly
reduces the number of useful time-slices.  Given that most the time
required for these calculations is in the measurements, and not in the
configuration generation, a cure for this problem is to generate
ensembles of configurations that are longer in the time-direction than
those currently being used (as opposed to working with different BC's
on the quark propagators that are less theoretically
``clean'').\footnote{We note that combining quark propagators with
  both periodic and anti-periodic temporal BCs, to effectively double
  the length of the configurations as seen by the valence quarks, will
  not resolve all the noise issues as much of the problem is produced
  by states involving sea quarks which are encoded in the gauge
  configurations.}  The multi-exponential fitting and Prony methods
enable the ground-state to be probed closer to the source where the
statistical uncertainties are exponentially smaller (also one of the
important aspects of the variational method), somewhat reducing the
impact of the exponentially degrading signal-to-noise near the
mid-point of the configuration.  Given that the signal-to-noise
degradation is exponentially more severe in systems containing two or
more baryons, all currently available tools will be required to make
optimal use of the computational resources. For excited states in a
given channel, variational methods seem to be superior to the standard
approach used here.

An important result of this work has been to quantify the statistical
scaling of simple observables in the sub-percent regime of
uncertainty. Scaling with the number of configurations was found to
adhere to the expected $1/\sqrt{N_{\rm cfg}}$ behaviour. We have also
investigated the issue of saturation, asking many measurements can be
performed on a single gauge-field configuration before it becomes more
cost effective to generate another statistically independent
gauge-field configuration?  To address this we have looked for
deviations from $1/\sqrt{N_{\rm src}}$ behavior in the uncertainties
in the correlation functions and extracted masses as a function of the
number of measurements performed on each configuration in the
ensemble.  The measurements of the mesons start to saturate after a
relatively small number of measurements, in this case of order $\sim
10$, while the baryon correlation functions show no signs of
saturation up to $\sim 200$.

A natural question to ask is if the same extractions could have been
performed with fewer computational resources by using the variational
method with a number of different sources for the baryons.  We
estimate that comparable resources would have been required to achieve
comparable uncertainties in the states we have examined.  However, we
have not been able to extract excited states of the nucleon with much
precision because of closely spaced states with the same quantum
numbers. This is likely to be something that the variational method
would better control.  Given that the present work was exploratory in
nature, this is not a concern at present, but it is clear that
high-statistics calculations of correlation functions arising from
multiple interpolating operators will be required in order to explore
the structure and interactions of nuclei.  We are working on
implementing this, but it will require significant computational
resources to perform, even at pion mass $\sim 390~{\rm MeV}$ and for a
relatively small lattice volume and relatively coarse lattice spacing.

It is clear that sub-MeV uncertainties an hadron energies will become
routine with the anticipated increase in computational resources
available to lattice QCD, and that the small energy scales that
characterize nuclear physics are within reach.  However, this program
will require large ensembles of gauge-field configurations that have
large extent in the time-direction, and will require a large fraction
of the computational resources devoted to measurements.

%%%%%%%%%%%%%%%%%%%%%%%%%%%%%%%%%%%
\section{Acknowledgments}

\noindent 
We thank R.~Edwards and B.~Joo for help with the QDP++/Chroma
programming environment~\cite{Edwards:2004sx} with which the
calculations discussed here were performed.  KO would like to thank
A. Stathopoulos useful discussion on numerical linear algebra issues
and for his contribution in the development of the EigCG
algorithm. EigCG development was supported in part by NSF grant
CCF-0728915. We also thank the Hadron Spectrum Collaboration for
permitting us to use the anisotropic gauge-field configurations, and
extending the particular ensemble used herein.  We gratefully
acknowledge the computational time provided by 
NERSC (Office of Science of the U.S. Department of Energy,
No. DE-AC02-05CH11231),, the Institute for Nuclear
Theory, Centro Nacional
de Supercomputaci\'on (Barcelona, Spain), Lawrence
Livermore National Laboratory, and the National Science
Foundation through Teragrid resources provided
by the National Center for Supercomputing
Applications and the Texas Advanced Computing Center. 
Computational support at Thomas Jefferson National Accelerator
Facility  and Fermi National Accelerator Laboratory was provided
by the USQCD collaboration under {\it The
Secret Life of a Quark}, a U.S. Department of Energy SciDAC project 
({\tt http://www.scidac.gov/physics/quarks.html}).
  The work of MJS and WD was supported in part by the
U.S.~Dept.~of Energy under Grant No.~DE-FG03-97ER4014.  The work of KO
and WD was supported in part by the U.S.~Dept.~of Energy contract
No.~DE-AC05-06OR23177 (JSA) and DOE grant DE-FG02-04ER41302. KO and
AWL were supported in part by the Jeffress Memorial Trust, grant
J-813, DOE OJI grant DE-FG02-07ER41527. The work of SRB and AT was
supported in part by the National Science Foundation CAREER grant No.
PHY-0645570.  Part of this work was performed under the auspices of
the US DOE by the University of California, Lawrence Livermore
National Laboratory under Contract No. W-7405-Eng-48.  The work of AP
is partly supported by the Spanish Consolider-Ingenio 2010 Programme
CPAN CSD2007-00042, by grants Nos. FIS2008-01661 from MICINN (Spain) and
FEDER and 2005SGR-00343 from Generalitat de Catalunya, and by the EU
contract FLAVIAnet MRTN-CT-2006-035482.

%%%%%%%%%%%%%%%%%%%%%%%%%%%%%%%%%%%%%%%%%%%%%%%%%%
%
% BIBLIOGRAPHY
%
%%%%%%%%%%%%%%%%%%%%%%%%%%%%%%%%%%%%%%%%%%%%%%%%%%

\end{document}